\documentclass[vecphys]{svmult}

\usepackage{makeidx}         
\usepackage{graphicx}            
\usepackage{multicol}        
\usepackage[bottom]{footmisc}

\makeindex             

\begin{document}

\title*{Electromagnetic probes}

\author{Rupa Chatterjee\inst{1} \and Lusaka Bhattacharya\inst{2} 
\and Dinesh K. Srivastava\inst{1}}

\institute{Variable Energy Cyclotron Centre, 1/AF, Bidhan Nagar, 
Kolkata 700064, India, \texttt{rupa@veccal.ernet.in, 
dinesh@veccal.ernet.in}\and Saha Institute of 
Nuclear Physics, 1/AF, Bidhan Nagar, Kolkata 700064, India, 
\texttt{lusaka.bhattacharya@saha.ac.in}}
\maketitle
\begin{abstract}
We introduce the seminal developments in the theory and experiments 
of electromagnetic probes for the study of the dynamics 
of relativistic heavy ion collisions and quark gluon plasma.
\end{abstract}

\section{Introduction}

Collision of heavy nuclei at relativistic energies is expected 
to lead to formation of a deconfined state of matter known as 
Quark-Gluon Plasma (QGP)~\cite{muller}, where quarks and gluons 
are the effective degrees of freedom rather than nucleons or 
hadrons~\cite{wong, hwa}. It is now well accepted that a few 
microseconds after the `Big Bang' the whole universe was in 
the state of QGP~\cite{uli}.

Several experiments performed at the Super Proton Synchrotron 
(SPS) at CERN and Relativistic Heavy Ion Collider (RHIC) at 
Brookhaven National Laboratory, New York, have provided a 
significant evidence of the formation of this novel state
of matter. A giant accelerator known as Large Hadron Collider 
(LHC) at CERN will be in operation very soon and will provide 
many new insights about the properties of QGP and the theory of
strong interactions.

Heavy ion collisions at relativistic energies produce extremely high 
temperatures and energy densities within a very small volume. As a 
result, quarks and gluons (also known as partons) no longer remain 
confined within the nucleonic volume and create a deconfined state 
of partons due to multiple scatterings  and production of secondaries 
due to gluon multiplications. The system (or the fireball) may reach 
a state of local thermal equilibrium. It cools  by expansion and 
below a certain critical temperature ($T_c \sim 180$ MeV) or energy 
density, the partons are confined to form hadrons and the system 
reaches a hadronic state. It may undergo a further expansion and 
cooling before the freeze-out takes place. 

Radiation of photons and dileptons has been proposed as the most 
promising and efficient tool to characterize the initial state of 
heavy ion collisions. Unlike hadrons, which are emitted from the 
freeze-out surface  after undergoing  intense re-scatterings, 
photons come out from each and every phase of the expanding fireball. 
Being electromagnetic in nature, they interact only weakly and their 
mean free path is larger than the typical system size ($\sim 10 $ fm). 
As a result once produced, they do not suffer further interaction 
with the medium ($\alpha \ll \alpha_s$) and carry undistorted 
information about the circumstances of their production to the 
detector~\cite{feinberg}. 

Initially, photons (real as well as virtual), were studied in 
order to get only the temperature of the plasma. Several other 
possibilities, e.g., (i) evolution of the system size by intensity 
interferometry~\cite{prl_inter}, (ii) momentum anisotropy of the 
initial partons~\cite{CFHS} as well as formation time of quark-gluon 
plasma~\cite{cs} using  elliptic flow of thermal photons, (iii) an 
accurate check on jet quenching and other aspects of the collision 
dynamics by photons due to passage of high energy jets through 
plasma~\cite{fms_phot} etc. have come to the fore. Of-course 
dileptons are considered as the most reliable messengers of the 
medium modification of vector mesons~\cite{KKMM}.
\begin{figure}
\centering
\includegraphics[height=4.8cm, width=5.50cm]{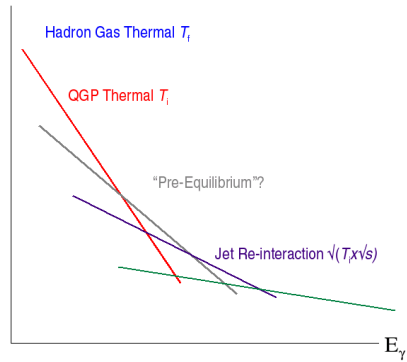}
\caption{Schematic diagram of different sources of photons and 
their relative $p_T$ spectra.}
\label{phot_mass}      
\end{figure}
\section{Sources of photons}

In order to proceed, it is useful to identify various sources of 
photons from relativistic heavy ion collisions. Their production 
is a result of convolution of the emissions from the entire history 
of the nuclear collision. Photons are emitted from the 
pre-equilibrium stage, from QGP phase, from hadronic phase and also 
from the decay of hadrons produced at the time of freeze-out. An 
ideal situation would ensue if the contributions from different 
stages dominate different parts of the $p_T$ spectrum. A schematic 
diagram of the different sources of photons and their slopes is 
shown in Fig.~\ref{phot_mass}. We, thus need rates and models to 
study the evolution from different sources. Hydrodynamics, Cascade, 
Fire-ball, Cascade+Hydrodynamics are the vastly used models for 
this purpose.

\subsection{Direct photons}

The term `direct photons' stands for the photons which emerge 
directly from a particle collision. In a heavy ion collision 
experiment, the detector captures all the emitted photons 
including those from decay of final state hadrons. The resultant 
spectrum is the inclusive photon spectrum. However, more than 90\% 
of the photons in this spectrum are from hadron decay. One can 
subdivide this broad category of `direct photons' into `prompt', 
pre-equilibrium, `thermal' (from QGP as well as hadronic phase) 
and `jet conversion' depending on their origin. Before we come 
to the different sub-categories of direct photons, we start our 
discussion with decay photons and their subtraction from the 
inclusive photon spectrum.
 
\subsection{Decay Photons}

As mentioned earlier, most of the decay  photons are from 
2-$\gamma$ decay of $\pi^0$ and $\eta$ mesons. $\omega$, 
$\eta'$ etc. also contribute to the decay photon spectrum, 
marginally. Subtraction of the decay background from inclusive 
photon spectrum is a very challenging task. WA98 Collaboration
~\cite{subtraction} used the subtraction method using invariant 
mass analysis for decay background and later PHENIX Collaboration
~\cite{rhic} has developed this method to a much higher level 
of sophistication.

\subsubsection{Invariant mass analysis}
\begin{figure}
\centering
\includegraphics[height=7.2cm, width=10.0cm]{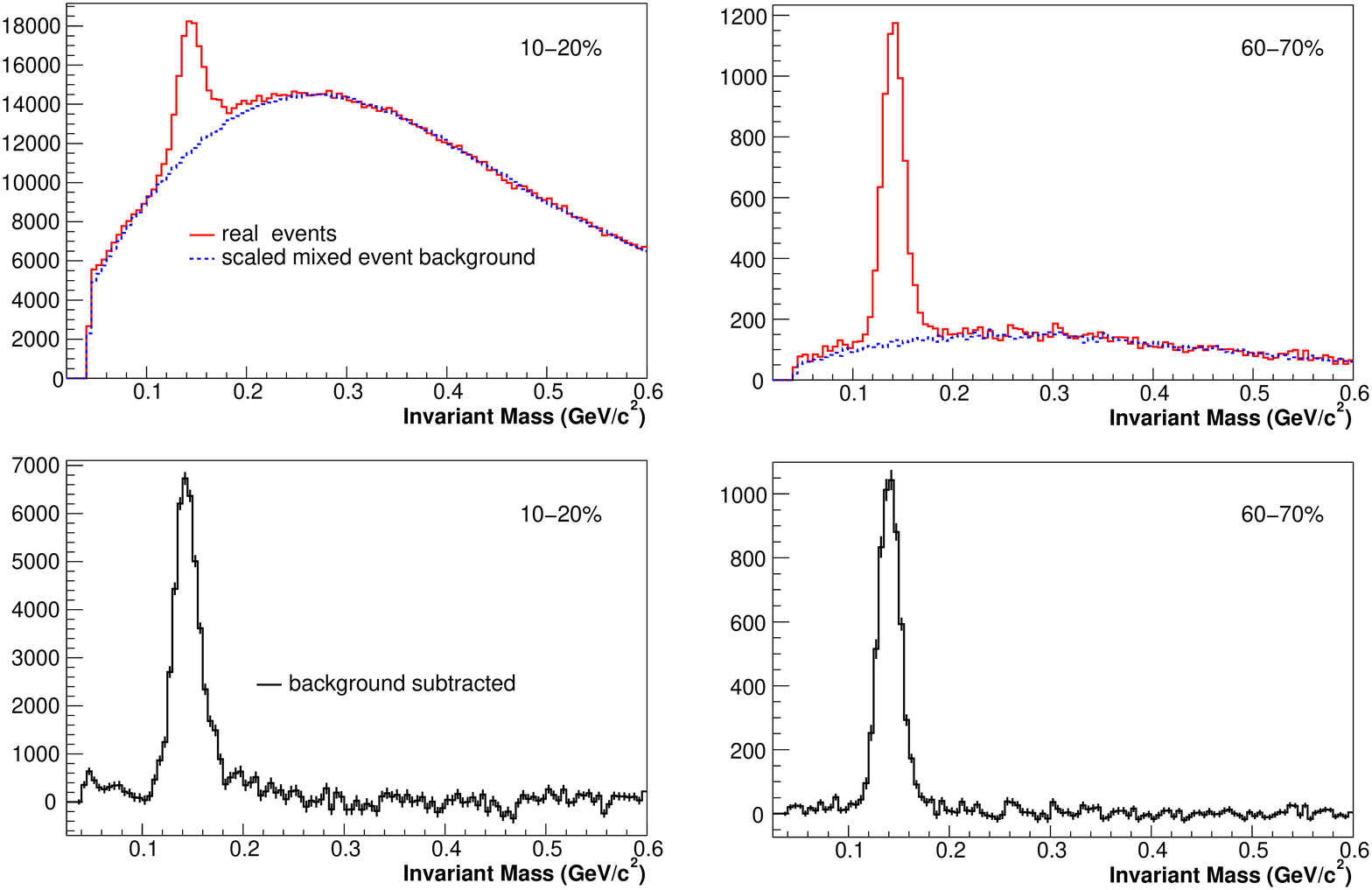}
\caption{ Invariant mass distributions of pairs of electromagnetic 
clusters passing photon selection cuts for pair transverse momenta 
satisfying $3.0 < 3.5 $ GeV. Top panels: $m_{\gamma \gamma}$ 
distributions in $Au+Au$ events compared to a normalized mixed-event 
sample representing the combinatoric background. Bottom panels: 
The $m_{\gamma \gamma}$ distributions after subtraction of the 
combinatoric background for 10-20\% and 60-70\% centrality 
bins~\cite{pion_supp}.}
\label{decay_subtraction}    
\end{figure}
Extraction of direct photon spectrum from the inclusive photon spectrum 
is done using invariant mass analysis as the primary step. First of all, 
all the detected photons are listed on an event by event basis. Then by 
selecting two photons randomly from an event, invariant mass of the
 pair is calculated. If $E_1$, $E_2$ are the energies and $\bf{p_1}$, 
$\bf{p_2}$ are the three momentum of the photons respectively, then 
the invariant mass of the pair is,
\begin{equation}
M_{\gamma\gamma} = [ (E_1 + E_2)^2 - ({\bf{p_1}} +{\bf{p_2}})^2 ]^{1/2}.
\end{equation}
If the value of $M_{\gamma\gamma}$ is close to $m_{\pi^0}$,  
it is assumed that they are the decay products from the same 
pion and a $\pi^0$ spectrum is obtained. Similarly the $\eta$ 
spectrum is obtained. These two spectra are then used to 
determine the decay photon spectrum using kinematics and the 
subtraction of this spectrum from  the inclusive spectrum gives 
the direct photon spectrum. 

However, in an event having N number of photons, the total number 
of photon pairs that can be formed is $^NC_2$. For large values 
of N, this number gets very large. Thus, there is a very high 
probability of getting a pair of photons having an invariant mass 
$m_{\pi^0}$ or $m_\eta$, which are not decay product of the same pion 
or eta meson. 

Several other issues also need to be accounted for: (i) the 
detector resolution is finite, (ii) the opening angle between 
the decay photons can get very small, specially if the energy 
of the pion is large, (iii) the photons may not deposit all their 
energy in the detector, and (iv) one of the photons may be 
outside the coverage of the detector.

\subsubsection{Mixed event analysis}

Thus, the major problem of invariant mass analysis is that, the 
accidental (false) photon pairs can also give rise to pion mass 
and it is not possible to distinguish them from the correlated 
pairs. To overcome this problem, a mixed event analysis
~\cite{pion_supp} procedure has been used successfully. The 
basic idea of mixed event technique is to compare particle 
spectrum from one event to the result for particle combinations 
from different events, which are {\it a priori} not correlated. 
As a first step,  properly normalized mixed events are constructed 
by randomly sampling photons from different events. The difference 
of the invariant mass spectra (see Fig.~\ref{decay_subtraction})
of the real events and the mixed events then gives the pion and $\eta$ 
distributions. Once again, the decay photon spectrum is subtracted 
from the inclusive photon spectrum to get the direct photons. 

\subsubsection{Internal conversion and tagging of decay photons}

\begin{figure}
\centering
\includegraphics[height=3.2cm, width=3.20cm]{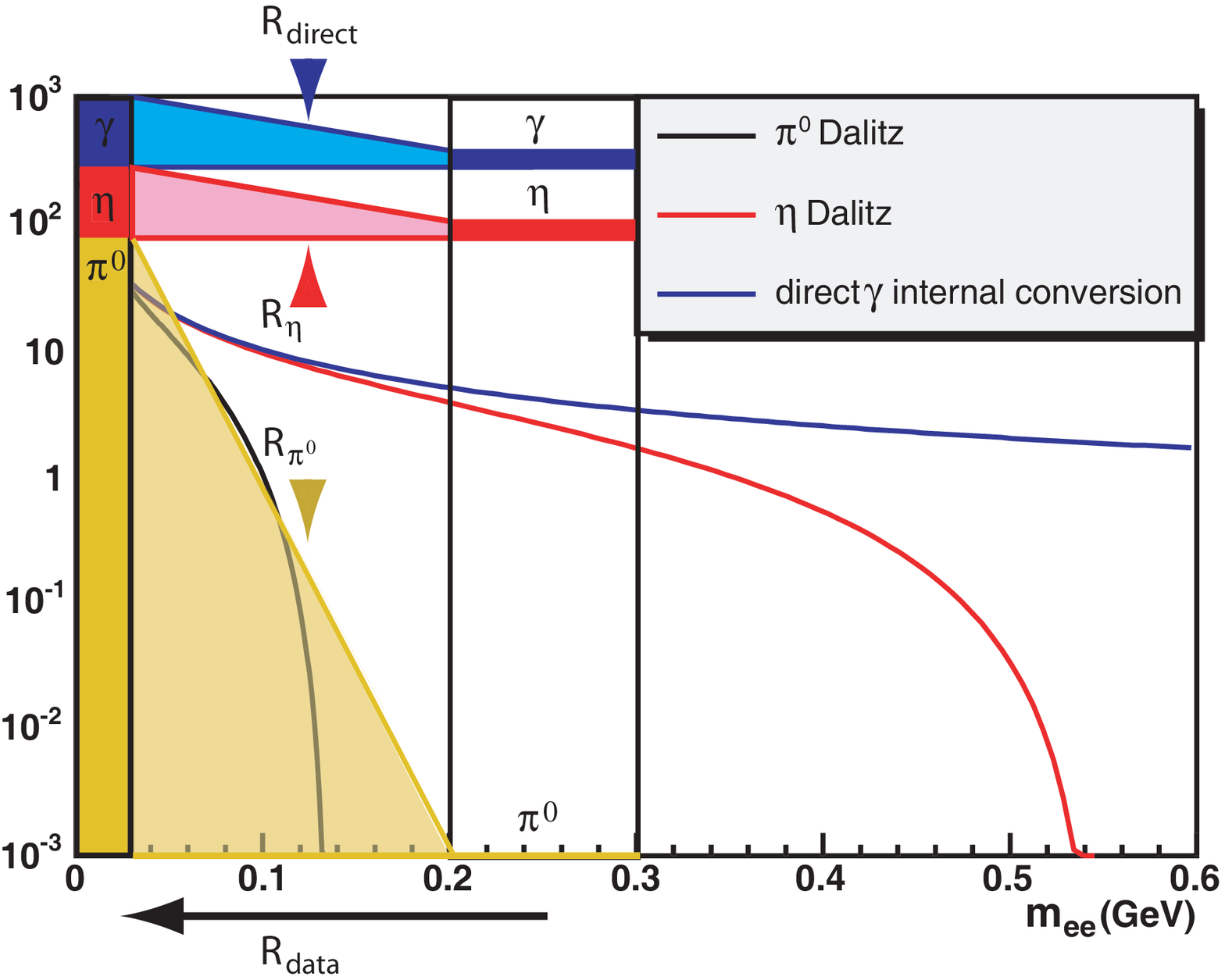}
\includegraphics[height=3.2cm, width=7.60cm]{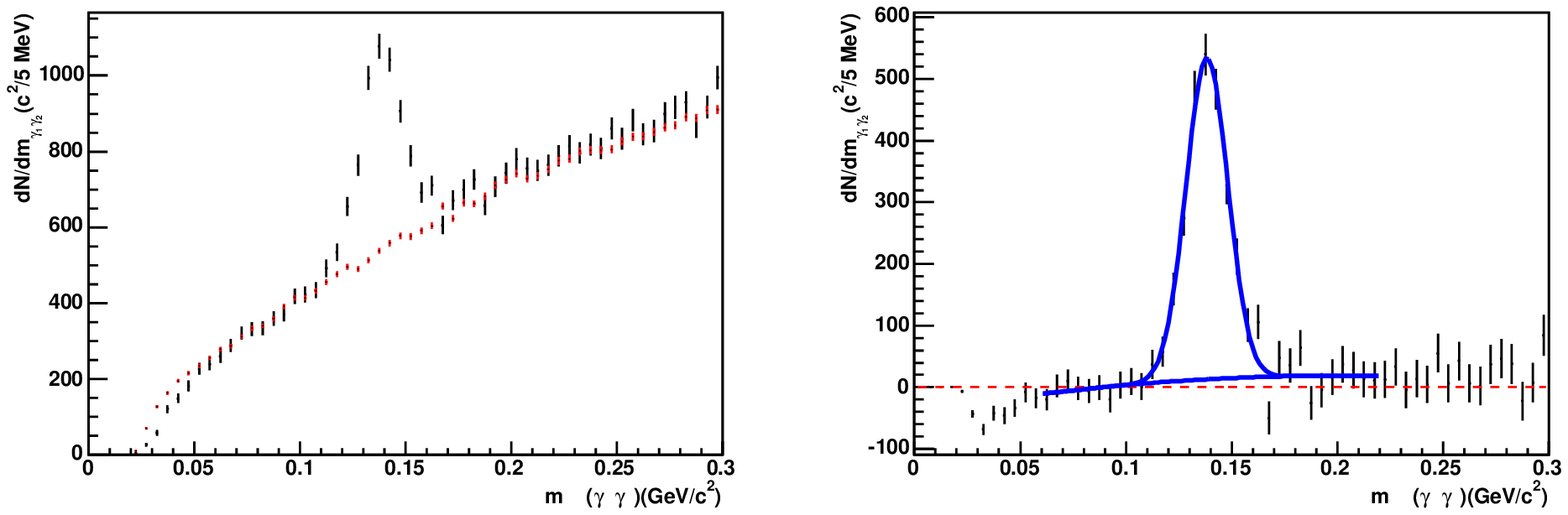}
\caption{Left Panel: Invariant mass distribution of virtual 
photons from $\pi^0$ and $\eta$ Dalitz decay as well as from 
direct photons~\cite{nucl_ex_0804.4168}. Middle and right panels: 
Invariant mass spectrum in `tagging' of decay photon 
method~\cite{gong}.}
\label{decay_phot_tag}   
\end{figure}

An alternative approach of separating direct photons from decay 
background is by measuring the `quasi-real' virtual photons which 
appear as low mass electron positron pair. It is assumed that, any 
source of real photons also produces low mass virtual photons which 
decay  into  $e^+e^-$ pair. This method is known as internal 
conversion method~\cite{nucl_ex_0804.4168} and is based on two 
assumptions. The first assumption is that the ratio of direct to 
inclusive photons is the same for real as well as virtual photons 
having $m_{\gamma} < 30$ MeV, i.e, $ \gamma^*_{\rm dir}/\gamma^*_{\rm incl} 
= \gamma_{\rm {dir}}/\gamma_{\rm {incl}}$. Secondly, the mass 
distribution follows the Kroll-Wada formula~\cite{kroll_wada}:
\begin{equation}
\frac{d^2n_{ee}}{dm_{ee}} = \frac {2 \alpha} {3 \pi} \frac {1} {m_{ee}} 
\sqrt{ 1- \frac {4m_e^2}{m_{ee}^2}} \left ( 1+ \frac {2m_e^2}{m_{ee}^2} 
\right ) S dn_{\gamma}.
\label{kroll}
\end{equation}
Here, $m_{e}$ and $m_{ee}$ are the masses of electron and $e^+e^-$ 
pair respectively and $\alpha$ is the fine structure constant. This 
method is used for Compton scattering ($ q + g \rightarrow q + \gamma^*  
\rightarrow q + e^+ + e^- $), Dalitz decay ( $\pi^0, \eta \rightarrow 
e^+e^- \gamma, \ \omega \rightarrow e^+ e^- \pi^0$) and also for two 
$\gamma$ decay of several other hadrons. The factor S in Eqn.~(\ref{kroll})
is process dependent and  for $\pi^0$ ($\rightarrow \gamma \gamma^* 
\rightarrow \gamma e^+e^-$) decay it is expressed as~\cite{S_decay}: 
\begin{eqnarray}
S= | F(m_{ee}^2)|^2 \left (1- \frac {m_{ee}^2} {M_h^2} \right )^3,
\end{eqnarray}
where, $M_h$ is the hadron mass and $F(m_{ee}^2)$ is the form factor. 
The factor S is 0 for $m_{ee} > M_h$ and goes to 1 as $m_{ee} 
\rightarrow 0$ or $m_{ee} \ll p_T$.  The key advantage of this method 
is the greatly improved signal to background ratio which is achieved 
by elimination of the contribution of Dalitz ($\pi^0$) decay. The 
experimentally measured quantity is the ratio of $e^+ e^-$ pairs 
in a particular invariant mass bin and the direct photon spectrum 
is obtained by multiplying $ \gamma^*_{\rm{dir}}/\gamma^*_{\rm incl}$ 
to the measured inclusive photon spectrum (left panel of 
Fig~\ref{decay_phot_tag}).

Due to the excellent resolution of the PHENIX detector to measure 
charged particles at low momenta, another powerful  technique 
known as `tagging of decay photons'~\cite{gong} is used to eliminate 
the $\pi^0$ decay background. This method is very useful in the low 
and intermediate $p_T$ range ( $1 < p_T < 5$ GeV) as the systematic 
uncertainties introduced by detector efficiency, acceptance etc. 
cancel out for measuring  $\gamma/\gamma_{\pi^0}$ directly in-place 
of conventional double ratio ($R= (\gamma/\pi^0)_{\rm measured}/(\gamma/
\pi^0)_{\rm decay}$) technique. In this method, the invariant mass 
($m_{\gamma e^+ e^-}$) distribution  of the $\pi^0$ ( $\rightarrow 
\gamma \gamma^* \rightarrow \gamma e^+ e^- $) decay products is 
constructed and then mixed event analysis is used for the final 
subtraction (middle and right panel of Fig~\ref{decay_phot_tag}).
\subsection{Sources of direct photons}

Direct photons can be classified into different categories depending 
on their origin from different stages of the expanding fireball formed 
after the collision. These are: (1) prompt photons, which originate 
from initial hard scatterings, (2) pre-equilibrium photons, produced 
before the medium gets thermalized, (3) thermal photons from quark-gluon 
plasma as well as by hadronic reactions in the hadronic phase, and (4) 
photons from passage of jets through plasma. It is not possible 
experimentally to distinguish between the different sources. Thus,
theory can be used with a great advantage to identify these sources 
of direct photons and their relative importance in the 
spectrum~\cite{dks_qm}. 

\subsubsection{Partonic processes for production of prompt photons}

\begin{figure}
\centering
\includegraphics[height=4.cm, width=12 cm]{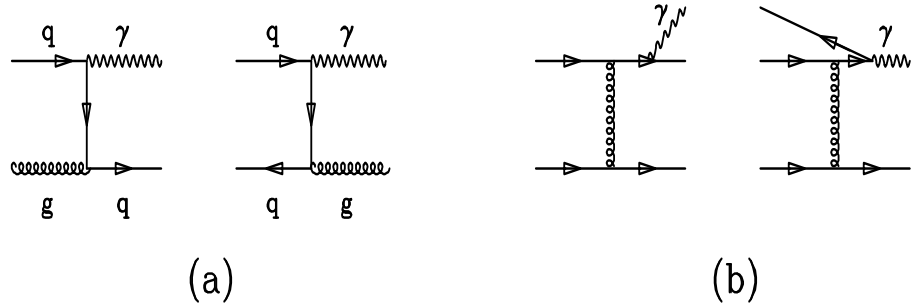}
\caption{Partonic processes for production of photons from (a) quark gluon 
Compton scattering, quark anti-quark annihilation process and (b) quark 
fragmentation.}
\label{partonic}   
\end{figure}

In relativistic heavy ion collisions, prompt photons are produced 
due to quark gluon  Compton scattering ($ q + g \rightarrow g + 
\gamma$), quark anti-quark annihilation process ($q + \bar q \rightarrow 
g + \gamma$), and quark fragmentation ( $ q \rightarrow q + \gamma$) 
following scattering of partons of the nucleons in the colliding nuclei 
(see Fig~\ref{partonic}). 
At lowest order in $\alpha \alpha_s$, quark gluon Compton scattering 
and quark anti-quark annihilation processes dominate the photon 
production. In Next to Leading Order (NLO) calculation, many more 
complicated scattering processes appear 
in the photon production cross-section and the total contribution 
can be written as addition of two different terms as~\cite{Aurenche},
\begin{equation}
\frac {d \sigma} {d \overrightarrow p_T d \eta} = {\frac {d \sigma^{(D)}} 
{d \overrightarrow p_T d \eta}} + {\frac {d \sigma^{(F)}} {d 
\overrightarrow p_T d \eta}}
\label{prompt}
\end{equation}
In the above equation `D' stands for `Direct' or the total Compton 
scattering and annihilation contribution while the photons from 
fragmentation are denoted by `F'. It is clear that the `D' photons
are well separated from hadrons. The `F' photons on the other hand
have their origin in collinear fragmentation of coloured high $p_T$ 
partons and are accompanied by hadrons. This can be used with 
advantage as the produced photons can be separated out by employing 
useful isolation cuts. The two terms in Eqn.~(\ref{prompt}) can be 
written explicitly 
as~\cite{old_aurenche}:
\begin{eqnarray}
\frac {d \sigma^{(D)}} {d \overrightarrow p_T d \eta} &=& \sum_{i,j=q,
\bar{q},g} \int dx_{1} dx_{2} \ F_{i/h_1}(x_{1},M)\ F_{j/h_2}(x_{2},M) 
{\frac { \alpha_s (\mu_R)}{2 \pi}}
\nonumber\\
& \times& \left 
({\frac {d \hat{\sigma}_{ij}}{d \overrightarrow p_T d \eta}} +{\frac 
{ \alpha_s (\mu_R)} {2 \pi}} K_{ij}^{(D)}  (\mu_{R},M,M_{_F})\right )
\end{eqnarray}
and
\begin{eqnarray}
\frac {d \sigma^{(F)}} {d \overrightarrow p_T d \eta} &=& \sum_{i,j,k=q,
\bar{q},g} \int dx_{1} dx_{2} {\frac{dz}{z^2}}\ F_{i/h_1}(x_{1},M)
\ F_{j/h_2}(x_{2},M) \ D_{\gamma/k}(z, M_{_F})
\nonumber \\
&\times &({\frac { \alpha_s (\mu_R)}{2 \pi}})^2 \left( {\frac {d 
\hat\sigma_{ij}^k}{d \overrightarrow p_T d \eta}} +{\frac { \alpha_s 
(\mu_R)}{2 \pi}} K_{ij,k}^{(F)} (\mu_{R},M,M_{_F}) \right).
\end{eqnarray}
Here $F_{i/h_{1,2}}(x,M)$ are the parton distribution functions 
and $\alpha_s(\mu_R)$ is the strong coupling defined in the 
$\overline{MS}$ renormalization scheme at the renormalization 
scale $\mu_R$. For details see Ref.~\cite{Aurenche}. As mentioned 
earlier, in a complete and consistent NLO pQCD $[O (\alpha 
\alpha_s^2)]$ calculation important contribution to prompt 
photon result arises from various possible $ 2 \rightarrow 
3 $ process like $ ab \rightarrow \gamma cd$ for the Direct 
as well as for the Fragmentation processes~\cite{old_aurenche,
Vogel}. 
\begin{figure}[b]
\centering
\includegraphics[height=7.5cm, width=8.cm]{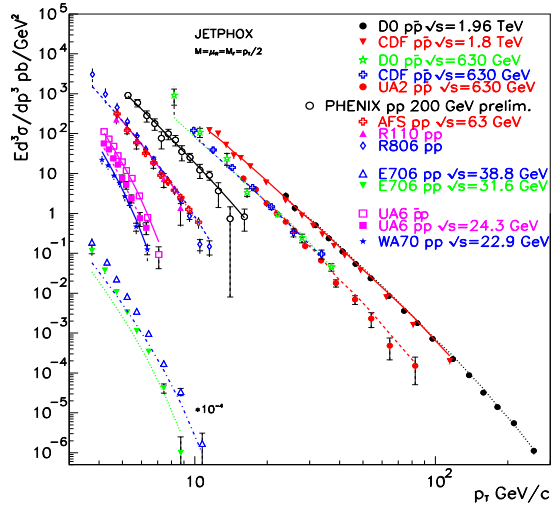}
\caption{World's inclusive and isolated direct photon 
production cross-sections measured in $pp$ and $p\bar p$ 
collisions compared to JETPHOX NLO predictions, using BFG 
II(CTEQ6M) for fragmentation (structure) functions and a 
common scale $p_T/2$~\cite{Aurenche}.}
\label{all_phot}    
\end{figure}

In case of photons from fragmentation, the higher order correction 
is very important at low $x_T\,( =\, 2 p_T/\sqrt{s})$. By choosing a 
scale for factorization, renormalization, and fragmentation, all 
equal to $p_T/2$, a very good quantitative description is obtained 
for all the available $pp$ and $p \overline{p}$ data without 
introduction of any intrinsic $k_T$. Results spanning over two 
orders of magnitude in energy and over nine orders of magnitude 
in cross sections, are shown in Fig.~\ref{all_phot}.

It should be mentioned though, that a similar exercise for pions 
requires a scale of $p_T/3$~\cite{pi_aurenche}.

In the early calculations, the results for nucleus-nucleus scatterings 
were often obtained by multiplying the $pp$ results for some $\sqrt{s}$ 
with corresponding scaling factor $N_{\rm {coll}}$ ($ = \sigma_{NN} 
T_{AB}$; where, $\sigma_{NN}$ is the nucleon-nucleon cross-section 
and $T_{AB}$ is the nuclear overlapping function for nuclei A and B) 
or number of binary collisions.  In actual practice, often enough,
$\sigma_{NN}$ was replaced by $\sigma_{pp}$. However, as the valence 
quark structures of protons $(uud)$ and neutrons $(udd)$ are different, 
one needs to correctly account for the iso-spin of the nucleons to 
calculate the prompt contribution. This correction will strongly 
affect the results in the $p_T$ range, where the valence quark 
contribution is significant. Also, remember that there is no direct 
measurement for $pn$ and $nn$ cross-sections, though they can be 
estimated by comparing results of scatterings involving deuterons. 
The effect of shadowing on structure function and energy loss of 
final state quarks before they fragment into hadrons are two other 
corrections~\cite{JMS} which need to be accounted for.

\begin{figure}[t]
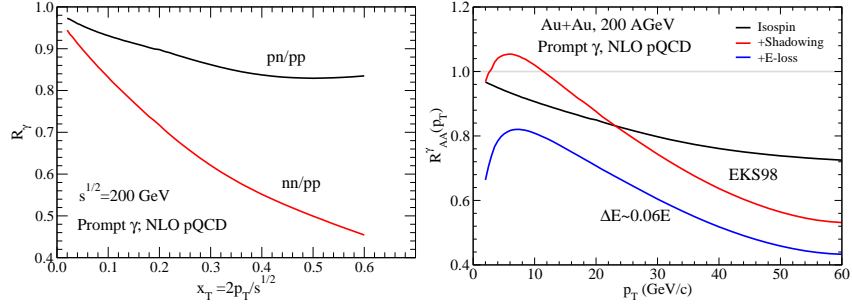

\centering
\includegraphics[height=4.0cm]{1.eps}
\includegraphics[height=4.00cm]{2.eps}
\caption{Left panel: Prompt photons for $pn$ and $nn$ collisions 
normalized to those for $pp$ collisions at $\sqrt{s}=200$ GeV. 
Right panel: Iso-spin, shadowing and energy loss corrected $R_{AA}$ 
for prompt photons at $200A$ GeV $Au+Au$ collisions using NLO pQCD.}
\label{ifs1}       
\end{figure}

In order to clearly see this point we show the results for prompt 
photons for $nn$ and $pn$ collisions normalized to those for $pp$ 
collisions at $\sqrt{s}= 200$ GeV in the left panel of Fig.~\ref{ifs1}. 
It is clear that at low $x_T$ (where processes involving gluons 
dominate) the effect is not very significant, however with larger 
values of $x_T$ (where the processes involving valence quarks 
dominate), the production of photons decreases by a large factor 
for the case of $pn$ and $nn$ collisions. 

\begin{figure}[ht]
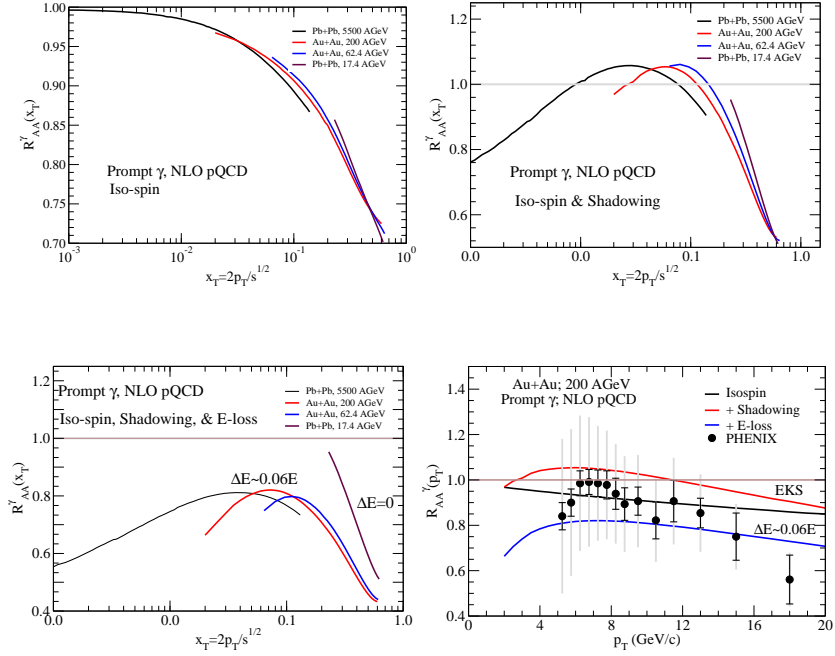

\centering
\includegraphics[height=3.8cm]{6.eps}
\includegraphics[height=3.8cm]{7.eps}

\vspace{1cm}
\includegraphics[height=3.8cm]{8.eps}
\includegraphics[height=3.8cm]{3.eps}
\caption{ Iso-spin, shadowing and fragmentation energy-loss 
corrected NLO pQCD results at different collider energies with 
different target projectile combination and comparison of the 
RHIC results with PHENIX~\cite{Phenix} data.}
\label{ifs4}       
\end{figure}

The transverse momentum dependent  nuclear modification factor 
$R_{AA}$, defined as,

\begin{equation}
R_{AA}(p_T)= \frac {1} {N_{\rm {coll}}} \frac{d \sigma^{AA}_{\gamma}
(p_T)/dy d^2{p_T}} {d \sigma^{pp}_{\gamma}(p_T)/dy d^2{p_T}}
\end{equation}
for prompt photons using NLO pQCD and considering iso-spin, shadowing 
and energy-loss effects for $200A$ GeV $Au+Au$ collisions at RHIC  is 
shown in the right panel of Fig.~\ref{ifs1}. We note that, for $p_T 
< 10$ GeV, the iso-spin and shadowing corrected result shows an 
enhancement in the prompt photon production compared to the situation 
when only the iso-spin correction is incorporated~\cite{DKS_ifs}. This
is due to anti-shadowing for larger values of Bjorken $x$ or large $x_T$. 
The inclusion of the energy-loss pushes down the value of $R_{AA}$ to less 
than one for all $p_T$. We also give a comparison with the 
PHENIX~\cite{Phenix} experimental data in the bottom lower panel of 
Fig.~\ref{ifs4}.

\begin{figure}[t]
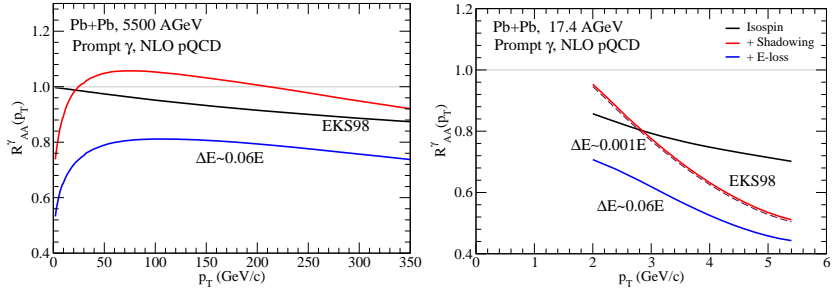

\centering
\vspace{0.3cm}
\includegraphics[height=3.80cm]{4.eps}
\includegraphics[height=3.80cm]{5.eps}
\caption{Iso-spin, shadowing and energy-loss corrected $R_{AA}$ for 
prompt photons at LHC (left panel) and SPS (right panel) energies 
using NLO pQCD.
}
\label{ifs2}   
\end{figure}
NLO results at LHC ($Pb+Pb@5.5A$ TeV) and SPS ($Pb+Pb@17.4A$ GeV) are 
shown in Fig.~\ref{ifs2}. To clearly demonstrate the relative features 
of iso-spin, shadowing and energy-loss corrected prompt photons at 
different collider energies, results for $R_{AA}$ as function of 
$x_T$ are shown in Fig.~\ref{ifs4}.
%

\subsection{Photon production from Quark-Gluon Plasma} 

The thermal emission rate of photons with energy $E$ and momentum 
$p$ from a small system (compared to the photon mean free path) 
is related to the imaginary part of photon self energy by the 
following relation 
\begin{eqnarray}
E\frac{dR}{d^3p}=\frac{-2}{(2\pi)^3}Im{\Pi_{\mu}}^{R,\mu}\frac{1}
{e^{E/T}-1}
\label{phorate}
\end{eqnarray}
where, ${\Pi_{\mu}}^{ R,\mu}$ is the retarded photon self-energy 
at a finite temperature $T$. This relation is valid in the 
perturbative~\cite{Mc_Toi_&} as well as non-perturbative~\cite{gale_kapu} 
limits. It is also valid to all orders in the strong interactions
and to order $e^2$ in the electromagnetic interactions. If the photon 
self-energy is approximated by carrying out a loop expansion to some 
finite order, then the formulation of Eqn.~(\ref{phorate}) is equivalent 
to relativistic kinetic theory. In order to illustrate this, we closely 
follow the treatment of Ref.~\cite{KLS}.

Thus using relativistic kinetic theory formulation, the contribution 
of these processes to the rate can be written as~\cite{KLS,BNNR}: 
\begin{eqnarray}
{\mathcal R}_i &=&{\mathcal N}\int \frac{d^3p_1}{2E_1(2\pi)^3}
\frac{d^3p_2}{2E_2(2\pi)^3}f_1(E_1)f_2(E_2)(2\pi)^4
\delta({p_1}^{\mu}+{p_2}^{\mu}-{p_3}^{\mu}-p^{\mu}) \nonumber\\
&\times& |{\mathcal M}_i|^2\frac{d^3p_3}{2E_3(2\pi)^3}
\frac{d^3p}{2E(2\pi)^3}[1 \pm f_3(E_3)]
\label{rate1}
\end{eqnarray}   
where ${\mathcal M}_i$ represents the amplitude for one of these 
processes and the $f's$ are the Fermi-Dirac or Bose-Einstein 
distribution functions as appropriate. Positive and negative signs 
in the last part of Eqn.~(\ref{rate1}) correspond to Bose enhancement 
and Pauli suppression respectively.  

The integral above can be simplified by introducing Mandelstam 
variables $s=(p_1+p_2)^2$, $t=(p_1-p_3)^2$ and $u=(p_1-p)^2$. Now  
the differential 
photon rate can be written as: 
\begin{eqnarray}
E\frac{d{\mathcal R}_i}{d^3p}&=&\frac{{\mathcal N}}{(2\pi)^7}
\frac{1}{16E}\int ds dt ~|{\mathcal M}_i(s,t)|^2\int
dE_1 dE_2 f_1(E_1)f_2(E_2)\nonumber\\ & \times &
[1\pm f_3(E_1+E_2-E)]\theta(E_1+E_2-E)
(a{E_1}^2+bE_1+c)^{-1/2} 
\end{eqnarray}
where,
\begin{eqnarray}
a &=&-(s+t)^2, \nonumber\\
b &=&2(s+t)(Es-E_2t),\nonumber\\
c &=&st(s+t)-(Es+E_2t)^2.
\end{eqnarray}    
Considering the photon energy to be large, one can consider,  
 $f_1(E_1)f_2(E_2) \approx e^{-(E_1+E_2)/T}$ and simplify the 
above as
\begin{eqnarray}
E\frac{d{\mathcal R}_i}{d^3p}=\frac{{\mathcal
N}}{(2\pi)^6}\frac{T}{32E}e^{-E/T} \int \frac{ds}{s}~\ln(1\pm
e^{-s/4ET})^{\pm 1}\int dt |{\mathcal M}_i(s,t)|^2
\end{eqnarray}
In the above equation, positive and negative signs stand for 
fermions ($q$) and bosons ($g$) in the final state respectively. 

The relation between the amplitude and differential cross-section for 
massless particles can be written as,
\begin{eqnarray}
\frac{d\sigma}{dt}=\frac{|{\mathcal M}|^2}{16\pi s^2}
\end{eqnarray}
And thus, the differential cross-sections for annihilation process and 
Compton scattering are:
\begin{eqnarray}
\frac{d\sigma^{\rm {annihilation}}}{dt}=\frac{8\pi \alpha \alpha_s}
{9s^2}\frac{u^2+t^2}{ut}
\end{eqnarray}
and
\begin{eqnarray}
\frac{d\sigma^{\rm {Compton}}}{dt}=\frac{-\pi \alpha \alpha_s}{3s^2}
\frac{u^2+s^2}{us}
\end{eqnarray}
For annihilation process, ${\mathcal N}=20$ when summing over 
$u$ and $d$ quarks and for Compton scattering ${\mathcal N}=320/3$. 
The total cross-section can be obtained after integrating over $t$.
These differential cross sections have a singularity at $t$ and/or 
$u=0$ and the total cross-section is infinite as the processes 
involve exchange of mass-less particle.

To screen this divergence many-body effects are necessary. This 
approach will be discussed later. As a first step let us isolate the 
region of phase space causing the divergences. The integration is 
done over 
\begin{eqnarray}
&&-s+{k_c}^2 \le t\le -{k_c}^2, \hspace{0.2cm} 2{k_c}^2\le s \le \infty, 
\end{eqnarray}
where $T^2 \gg {k_c}^2>0$ is an infrared cut-off. 

This treats $u$ and $t$ symmetrically and maintains the
identity $s+t+u=0$ appropriate for all massless particles.

In the limit that ${k_c}^2\to 0$,
\begin{eqnarray}
E\frac{d{\mathcal R}^{\rm {Compton}}}{d^3p}= \frac{5}{9}\frac{\alpha
  \alpha_s}{6{\pi} ^2} T^2 e^{-E/T}[\ln (4ET/{k_c}^2)+C_F]\\
E\frac{d{\mathcal R}^{\rm {annihilation}}}{d^3p}= \frac{5}{9}\frac{\alpha
  \alpha_s}{3{\pi} ^2} T^2 e^{-E/T}[\ln (4ET/{k_c}^2)+C_B]
\end{eqnarray}
where
\begin{eqnarray}
C_F=\frac{1}{2}-C_{\rm {Euler}}+\frac{12}{{\pi}^2}\sum_{n=2}^{\infty} 
\frac{(-1)^n}{n^2}\ln~ n= 0.0460.........,
\label{cf}
\end{eqnarray}
\begin{eqnarray}
C_B=-1-C_{\rm{ Euler}}-\frac{6}{{\pi}^2}\sum_{n=2}^{\infty} 
\frac{1}{n^2} \ln~n= -2.1472.......  .
\label{bf}
\end{eqnarray} 
These expressions use the full Fermi-Dirac or Bose-Einstein
distribution functions in the final state. 

These results have a very interesting structure. Thus, the
factor $5/9$ arises from the sum of the squares of the electric  
charges of the $u$ and $d$ quarks, the factor $\alpha \alpha_s$ 
comes from the topological structure of the diagrams, a factor $T^2$  
comes from phase space which gives the overall dimension of the rate, 
and we have the Boltzmann factor $e^{-E/T}$ for photons of energy $E$. 
The logarithm arises due to the infrared behavior.

\subsubsection{Infrared contribution}

The infrared divergence in the photon production rate \cite{KLS} 
discussed above is caused by a diverging differential cross-section 
when the momentum transfer goes to zero. Often-times long-ranged 
forces can be screened by many-body effects at finite temperatures. 
Braaten and Pisarski have analyzed problems such as this one in QCD 
\cite{wel13}. They have argued that a cure can be found in reordering 
perturbation theory by expanding correlation functions in terms 
of effective propagators and vertices instead of bare ones. These 
effective propagators and vertices are just the bare ones plus 
one-loop corrections, with the caveat that the one-loop corrections 
are evaluated in the high temperature limit. This makes them relatively 
simple functions. 

The analysis of Braaten and Pisarski shows that a propagator must be
dressed if the momentum flowing through it is soft (small 
compared to $T$). This is because propagation of soft momenta is 
connected with infrared divergences in loops. Dressing of propagators 
are necessary, otherwise corrections due to these are also infinite.

Using the one loop corrected propagators and vertices and the 
contribution to the rate coming from the infrared sensitive part 
of phase space can be written as,
\begin{eqnarray}
E\frac{d{\mathcal R}^{\rm BP}}{d^3p}=\frac{5}{9}\frac{\alpha \alpha_s}
{2 {\pi}^2}T^2e^{-E/T}\ln(\frac{{k_c}^2}{2{m_q}^2})
\end{eqnarray}
where, $2{m_q}^2=\frac{1}{3}g^2 T^2$. 

Adding this contribution to those given by Eqns.~(\ref{cf}) and 
~(\ref{bf}), the final result can be written as;

\begin{eqnarray}
E\frac{d{\mathcal R}}{d^3p}=\frac{5}{9}\frac{\alpha \alpha_s}{2
  {\pi}^2}T^2e^{-E/T}\ln \left (\frac{2.912}{{g}^2} {\frac{E}{T}} \right).
\end{eqnarray}

This is independent of cut-off $k_c$. Thus, the Braaten-Pisarski 
method has worked beautifully to shield the singularity encountered 
above. We also note that in kinetic theory calculation, the effective 
infrared cutoff is ${k_c}^2=2{m_q}^2$. In earlier works an infrared 
cut-off was often imposed by giving the exchanged quark an effective
temperature-dependent mass. 

These early results have been brought to a high degree of 
sophistication and results complete to leading order in $\alpha_s$ 
with inclusion of LPM effects are now available, which should be 
used for detailed calculation~\cite{AMY}.  

\subsubsection{Photons from passage of jets through QGP}

The relativistic heavy ion collisions at RHIC (and LHC) energies 
are marked by a large production of high energy quark and gluon 
jets which lose energy while passing through the QGP due to 
collision and radiation of gluons. This is the celebrated phenomenon 
of jet-quenching. A quark jet having a transverse momentum $p_T$ 
would be formed within $\tau \sim 1/p_T$ which can be much smaller 
than $\tau_0$, when QGP is formed, for large $p_T$. This quark (or 
anti-quark) jet, while passing through QGP may annihilate with a 
thermal anti-quark (or quark) or undergo a Compton scattering with 
a thermal gluon and lead to production of a high energy photon. This 
process is called the jet-photon conversion \cite{fms_phot}. 

We have seen that the  kinematics of the annihilation of a quark 
anti-quark pair ($q + {\bar q}\to \gamma+g$) is expressed in terms 
of the Mandelstam variables $s=(p_q+p_{\bar q})^2$, $t=(p_q-p_{\gamma})^2$ 
and $u=(p_{\bar q}-p_{\gamma})^2$. We also recall that the largest 
contribution to the production of photons arises from small values 
of $t$ or $u$, corresponding to $p_{\gamma}\sim p_q$ or $p_{\gamma}
\sim p_{\bar q}$~\cite{wong,fms_phot,nad}. 

The phase-space distribution of the quarks and gluons produced in a
nuclear collision can be approximately decomposed into two components,
a thermal component $f_{\rm th}$ characterized by a temperature $T$ and a
hard component $f_{\rm jet}$ given by hard scattering of the partons and
limited to transverse momenta $p_T \gg 1$ GeV: $f({\bf p})=f_{\rm th}
({\bf p})+f_{\rm jet}({\bf p})$. $f_{\rm jet}$ dominates for large momenta, 
while at small momenta $f$ is completely given by the thermal part.

The phase space distribution for the quark jets propagating through
the QGP is given by the perturbative QCD result for the jet yield 
\cite{fms_phot}:
\begin{eqnarray}
f_{\rm jet}({\bf p})= \frac{1}{g_q}\frac{(2\pi)^3}{\pi R_{\perp}^2 
\tau p_T}\frac{dN_{\rm jet}}{d^2p_T dy}R(r)\delta(\eta-y)
\Theta(\tau_{\rm max}-\tau_i)\Theta(R_{\perp}-r)
\label{dks31}
\end{eqnarray}
where $g_q=2\times 3$ is the spin and colour degeneracy of the quarks,
$R_{\perp}$ is the transverse dimension of the system, 
and the $\eta$ is the
space-time rapidity. $R(r)$ is a transverse profile function. 
$\tau_{\rm max}$ is the smaller of the lifetime $\tau_f$ of the QGP
and the time $\tau_d$ taken by the jet produced at position {\bf r} to
reach the surface of the plasma.

One can approximate the invariant 
photon differential cross sections for the annihilation process and 
Compton scattering as~\cite{wong,fms_phot},
\begin{eqnarray}
E_{\gamma}\frac{d\sigma^{(a)}}{d^3p_{\gamma}}\sim \sigma^{(a)}(s)\frac{1}
{2}E_{\gamma}[\delta({\bf p}_{\gamma}-{\bf p}_q)+\delta({\bf p}_{\gamma}-
{\bf p}_{\bar q})],
\label{dks27}
\end{eqnarray}
and 
\begin{eqnarray}
E_{\gamma}\frac{d\sigma^{(C)}}{d^3p_{\gamma}}\sim \sigma^{(C)}
(s)E_{\gamma}\delta({\bf p}_{\gamma}-{\bf p}_q)
\label{dks28}
\end{eqnarray} 
Here $\sigma^{(a)}(s)$ and $\sigma^{(C)}(s)$ are the corresponding total
cross sections. 

Using Eqn.~(\ref{dks27}) and Eqn.~(\ref{dks28}), the rate of production 
of photons due to annihilation and Compton scattering are 
given by~\cite{wong}:
\begin{eqnarray}
E_{\gamma}\frac{dN^{(a)}}{d^4xd^3p_{\gamma}}&=&\frac{16E_{\gamma}}
{2(2\pi)^6}\sum_{q=1}^{N_f}f_q({\bf p_{\gamma}})\int d^3pf_{\bar q}
({\bf p})[1+f_g({\bf p})]\sigma^{(a)}(s) \nonumber\\
&\times& \frac{\sqrt{s(s-4m^2)}}{2E_{\gamma}E}+(q\leftrightarrow{\bar q})
\label{dks29}
\end{eqnarray}
\begin{eqnarray}
E_{\gamma}\frac{dN^{(C)}}{d^4xd^3p_{\gamma}}&=&\frac{16E_{\gamma}}
{(2\pi)^6}\sum_{q=1}^{N_f}f_q({\bf p_{\gamma}})\int d^3pf_g({\bf p})
[1-f_q({\bf p})]\sigma^{(C)}(s)\nonumber\\
&\times& \frac{(s-m^2)}{2EE_{\gamma}}+(q\to{\bar q})
\label{dks30} 
\end{eqnarray}
The $f's$ are the distribution functions for the quarks, anti-quarks
and gluons. Inserting thermal distributions for the gluons and quarks
one can obtain an analytical expression for these emission rates for
an equilibrated medium~\cite{wong,KLS,BNNR}.

In order to estimate the jet photon conversion contribution, we first 
note that, the integrals over {\bf p} in Eqns.~(\ref{dks29}) and 
(\ref{dks30}) are dominated by small momenta. Therefore dropping 
the jet part in the distributions, $f({\bf p})$ in the integrands is 
approximated by the thermal part. Now performing the integrals and 
identifying the quark and anti-quark distributions outside the 
integrals with the jet distributions, results for Compton and 
annihilation scatterings due to jet-conversions are given as,
\begin{eqnarray}
E_{\gamma}\frac{d{N_{\gamma}}^{(a)}}{d^3p_{\gamma}d^4x}&=&E_{\gamma}
\frac{d{N_{\gamma}}^{(C)}}{d^3p_{\gamma}d^4x}\nonumber\\
&=&\frac{\alpha \alpha_s}{8{\pi}^2}{\sum^{{\mathcal N}_f}_{f=1}}
(\frac{e_{q_f}}{e})^2 [f_q({\bf p}_{\gamma})+f_{\bar q}({\bf p}_{\gamma})
]T^2 [\ln(\frac{4E_{\gamma}T}{m^2})+C]
\label{phjet}
\end{eqnarray} 
Here, $C=-1.916$. If we include  three lightest quark flavours, 
then $\sum_f {e_{q_f}}^2/e^2=2/3$. We also assume that the mass $m$ 
introduced here to shield the infrared divergence can be identified
 with the thermal quark mass $m_{th}$.

These pioneering works have now been corrected for energy loss and flavour 
change suffered by the jets, as they pass through the plasma~\cite{tur} 
as well as bremsstrahlung induced by the passage of the jets through the 
plasma~\cite{zakha}.

\subsubsection{Equilibration time for photons}

It is of interest to get an idea about the equilibration  
time of photons in the medium. Following the treatment of 
Ref.~\cite{KLS},  we note that the six-dimensional phase 
space distribution $dn/d^3p$($=dN/d^3xd^3p$), for the photons 
satisfies the rate equation:
\begin{equation}
\frac {d} {dt} \left ( {\frac {dn}{d^3p}} \right ) = {\frac {dR} 
{d^3p}} \left ( 1 - {\frac {dn/d^3p}{dn_{\rm eq}/d^3p}} \right ).
\label{eq}
\end{equation}
In the above equation, $dn_{\rm eq}/d^3p$ is the equilibrium 
distribution and is expressed as Planck's distribution,
\begin{equation}
\frac {dn^{\rm eq}} {d^3 p} = {\frac {2} {(2 \pi)^3}} {\frac {1} 
{e^{E/T}-1} }.
\end{equation}
Considering $\tau_{\rm eq}$ ($ = \frac 
{dn_{\gamma}^{\rm eq}} {d^3p}/ \frac {dR} {d^3p}$) as the time for 
equilibration and assuming zero photons at the beginning we can 
write, 
\begin{equation}
\frac {dn} {d{^3}p} = \frac {dn^{\rm eq}} {d^3p} 
\left(1-e^{t/\tau_{\rm eq}}\right).
\end{equation}
Using the rate equation, the thermalization time can be expressed 
in a simplified form (considering $ E > 2T $) as:
\begin{equation}
\tau_{\rm eq} = {\frac {9E} {10 \pi \alpha \alpha_s T^2 }} {\frac {1} 
{\ln( {\frac {2.9} {g^2}} {\frac {E} {T}}+1)}}.
\end{equation}
For energy values E =0.5, 1, 2, 3 GeV, corresponding values for 
$\tau_{\rm eq}$  will be about 270, 356, 505, 639  fm/$c$ respectively, 
when the temperature is 200 MeV. As the life time of the system is 
very short, of the order of few tens of fm/$c$, it is very clear that  
the high energy photons will never reach the equilibrium state in heavy 
ion collisions. This is an important confirmation for the validity of the 
assumption made in all such studies that photons, once produced in the 
collision, leave the system with out any further re-interaction.

\begin{figure}[t]
\centering
\includegraphics[height=5.50cm]{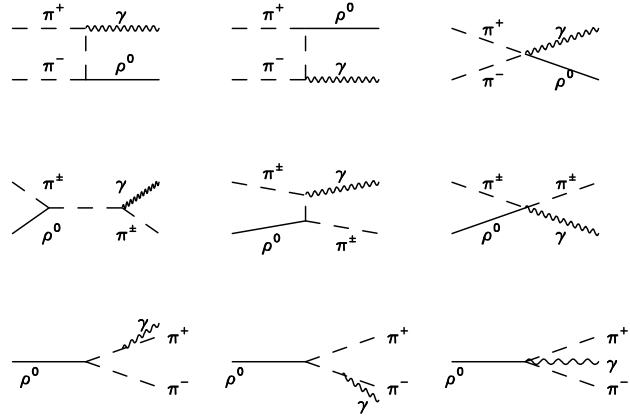}
\caption{ Typical hadronic reactions for photon production.}
\label{phot_fms+had_yield}   
\end{figure}  

\subsection{Photons from hot hadronic matter}

Hot hadronic matter produced after the hadronization of the quark-gluon 
plasma will also lead to production of photons due to hadronic reactions.
These photons will dominate the spectrum at lower $p_T$ ( $< 1 $ GeV ).  
The first ever calculation of production of thermal photons from 
hadronic matter was performed by Kapusta {\it et al.}~\cite{KLS}.

In a hot hadronic gas (having temperature of the order of pion mass), 
the most important hadronic constituents for photon production are 
$\pi$ and $\rho$ mesons (see Fig.~\ref{phot_fms+had_yield}). 
The low mass of pions and the large spin iso-spin degeneracy of 
$\rho$ mesons, make them the most easily accessible particles in 
the medium. In order to illustrate the photon production from these 
two mesons, we closely follow the treatment of KLS~\cite{KLS}. In a 
hadronic reaction involving charged $\pi$ and  $\rho$ meson, the 
typical Lagrangian describing the interaction can be written as:
\begin{eqnarray}
\mathcal L &=& \mid D_\mu \Phi \mid^2 -m_\pi^2 \mid \Phi \mid^2 - 
\frac {1} {4} \rho_{\mu \nu} \rho^{\mu\nu} + \frac {1} {2} m_\rho^2 
\rho_\mu \rho^\mu - \frac {1} {4} F_{\mu\nu}F^{\mu\nu}
\end{eqnarray}
where,
\begin{eqnarray}
 D_\mu &=&\partial_\mu -ieA_\mu - ig_\rho \rho_\mu, \nonumber
\end{eqnarray}
$\Phi$ is the complex pion field, $\rho_{\mu\nu}$ is the $\rho$ field 
strength and $F_{\mu\nu}$ is the photon field tensor. The differential 
cross-sections for the dominating photon producing processes ($\pi \rho 
\rightarrow \pi \gamma$) in the hadronic phase is expressed as~\cite{KLS},
\begin{figure}[th]
\centering
\includegraphics[height=9.0cm, width=7.0cm]{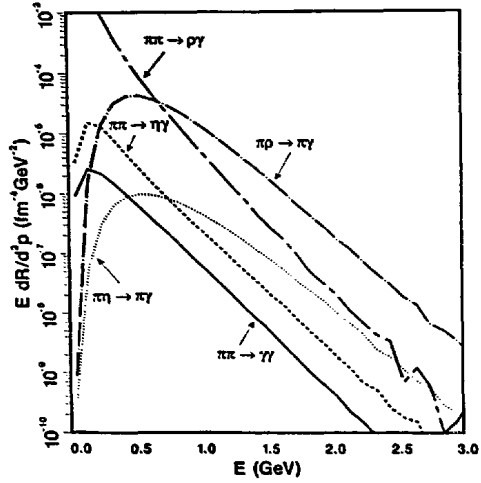}
\caption{Left Panel: Yield of photons from different  hadronic 
channels (taken from~\cite{KLS}).}
\label{check}   
\end{figure}  
\begin{eqnarray}
\frac{d \sigma} {dt} (\pi^+ \rho^0 \longrightarrow \pi^+ \gamma)&=&\frac {d 
\sigma} {dt} (\pi^- \rho^0 \longrightarrow \pi^- \gamma)\nonumber\\
& =& {\frac {\alpha g_\rho^2} {12sp_{\rm c.m.}^{2}}} 
 \left [ 2 - {\frac {(m_\rho^2-4m_\pi^2)s}{(s-m_\pi^2)^2}} - 
(m_\rho^2-4m_\pi^2) \right.\nonumber\\ &\times & \left.
 \left (\frac {s-m_\rho^2 + m_\pi^2} {s-m_\pi^2} \frac{1} {u-m_\pi^2} + 
\frac {m_\pi^2}{(u-m_\pi^2)^2} \right ) \right ].
\end{eqnarray}
Similarly for
\begin{eqnarray}
\frac{d \sigma} {dt} (\pi^- \rho^+ \longrightarrow \pi^0 \gamma)&=&\frac {d 
\sigma} {dt} (\pi^+ \rho^- \longrightarrow \pi^0 \gamma)\nonumber\\
&=& - {\frac {\alpha g_\rho^2} {48sp_{\rm c.m.}^2}}\left [ 4(m_\rho^2-4m_\pi^2)
 \left [ {\frac {u} {(u-m_\pi^2)^2}} + {\frac {t} {(t-m_\rho^2)^2}} 
\right.\right.\nonumber\\ &-& \left.{\frac {m_\rho^2}{s-m_\pi^2}} 
\left( {\frac {1} {u-m_\pi^2}} + {\frac {1}{t-m_\rho^2}} \right)\right ] 
+ \left [\left(3+ {\frac {s-m_\pi^2}{m_\rho^2}}\right) \right.
\nonumber\\ &\times & \left. {\frac {s-m_\pi^2} {t-m_\rho^2}}\right] - 
\left. {\frac {1} {2}} + {\frac {s} {m_\rho^2}} - \left( {\frac 
{s-m_\pi^2} {t-m_\rho^2}} \right)^2 \right ]
\end{eqnarray}
and also,
\begin{eqnarray}
\frac{d \sigma} {dt} (\pi^0 \rho^+ \longrightarrow \pi^+ \gamma)&=&\frac {d 
\sigma} {dt} (\pi^0 \rho^- \longrightarrow \pi^- \gamma)\nonumber\\
&=& \frac {\alpha g_\rho^2} {48sp_{\rm c.m.}^2} \left [ \frac {9} {2} - 
\frac {s}{m_\rho^2} - \frac {4(m_\rho^2-4m_\pi^2)s} {(s-m_\pi^2)^2} 
\right. \nonumber\\ &+& \frac {(s-m_\pi^2)^2 - 4m_\rho^2(m_\rho^2-4m_\pi^2)} 
{(t-m_\rho^2)^2} \nonumber\\
&+& \frac {1} {t-m_\rho^2} \left (5(s-m_\pi^2)- \frac {(s-m_\pi^2)^2} 
{m_\rho^2} \right.\nonumber\\
 & - & \left. \left. {\frac {4(m_\rho^2 - 4 m_\pi^2)}{s-m_\pi^2} } 
(s-m_\pi^2+m_\rho^2) \right ) \right ]
\end{eqnarray}
In the above set of equations, $s, \ t, \ u$ are the Mandelstam variables 
and $p_{\rm c.m.}$ is the three momentum of the interacting particles 
in their centre of mass frame. Typical results are shown in 
Fig.~\ref{check}.

\begin{figure}[t]
\centering
\includegraphics[height=5.0cm, width=6.0cm]{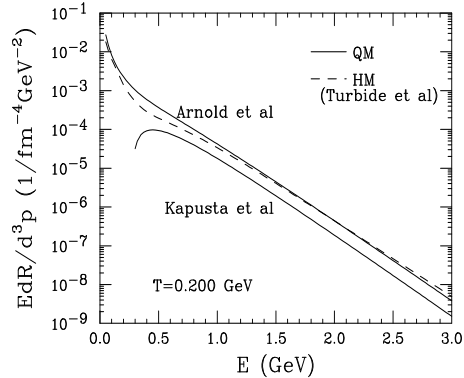}
\caption{Complete leading order rates from QGP and 
exhaustive reactions in hadronic matter~\cite{KLS,AMY,TRG}.}
\label{phot_rate}   
\end{figure}

Xiong {\it et al.}~\cite{XIONG} and Song~\cite{SONG} first introduced 
the $\pi \rho \rightarrow a_1 \rightarrow \pi \gamma$ channel for 
photon production in hadronic phase, whereas baryonic processes and 
medium modification were included by Alam {\it et al.}~\cite{alam1,alam2}. 
Several refinements, e.g., inclusion of strange sector, use of massive 
Yang-Mills theory and t-channel exchange of $\omega$ mesons, were 
incorporated by Turbide {\it et al.}~\cite{TRG}. The last calculation 
is essentially the state of art result at the moment. Photon spectra 
considering a complete leading rate from QGP~\cite{AMY} and exhaustive 
reactions in hadronic matter~\cite{KLS,TRG} are shown in Fig.~\ref{phot_rate}.

\section{Photons from Pb+Pb@SPS to Au+Au@RHIC}

\subsection{SPS}

In order to get an idea of greatly increased insights  provided by 
single photon production, let us briefly recall some of the important 
results from the SPS era, many of which preceded the large strides 
made in our theoretical understanding mentioned above. The first hint 
of single photon production, which later turned out to be the upper 
limit of their production, came from the $S+Au$ collisions studied at the 
SPS energies~\cite{wa80}.

\begin{figure}[t]
\centering
\includegraphics[height=4.50cm, width=5.50cm]{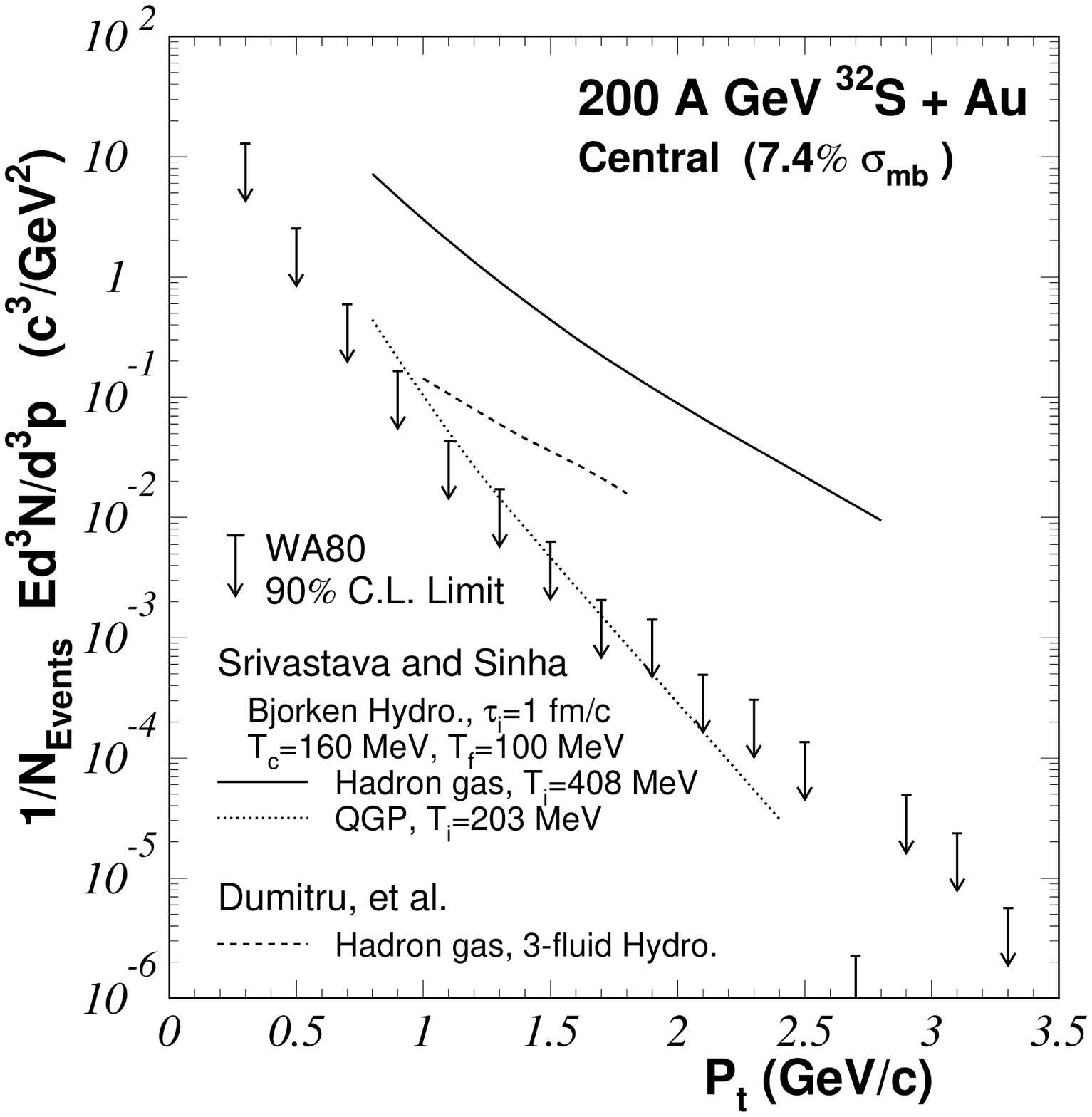}
\includegraphics[height=4.20cm, width=5.50cm]{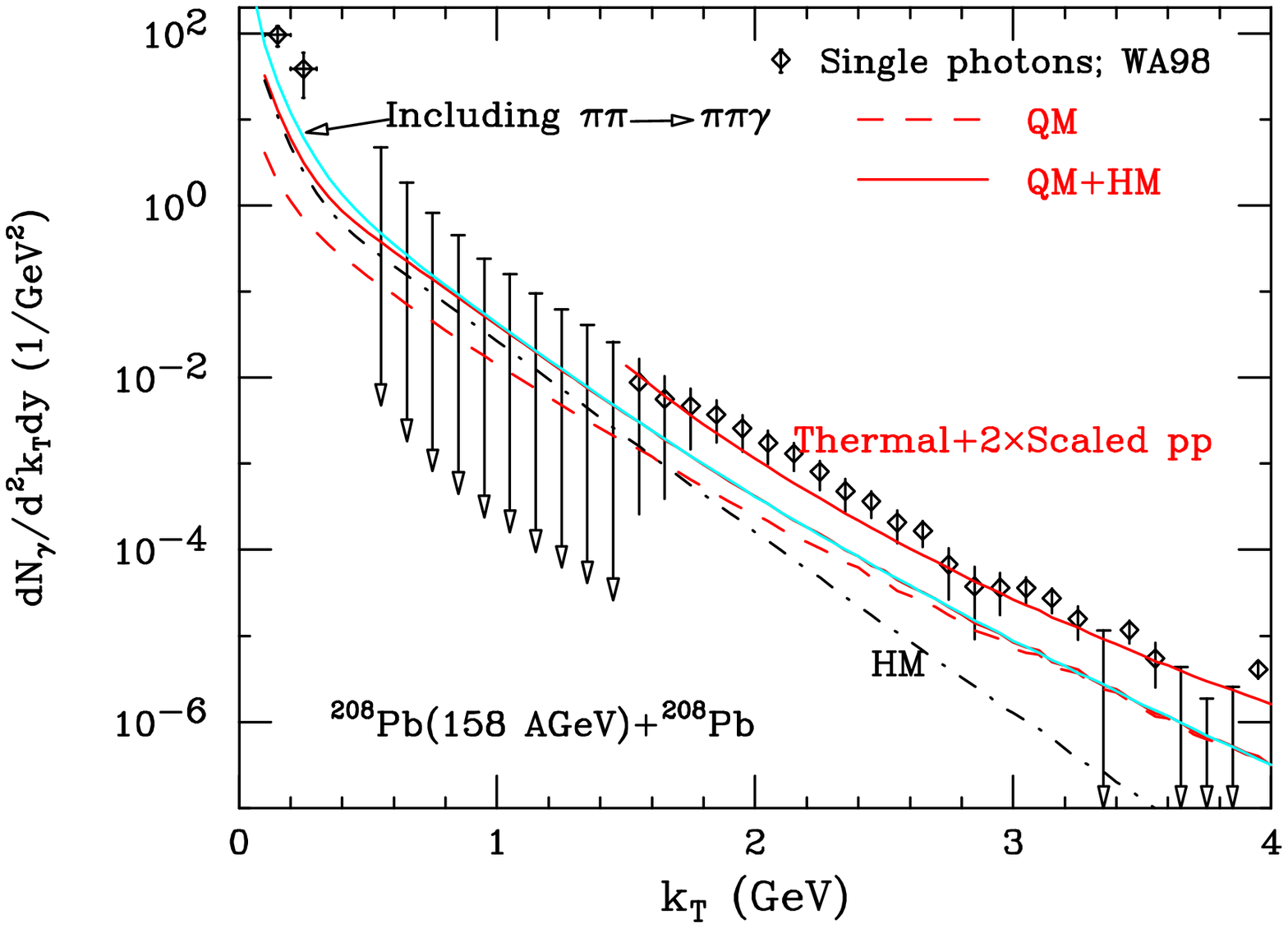}
\caption{ Left panel: Upper limits at the 90 \% confidence level on 
the invariant excess photon yield for the 7.4 \% $\sigma_{mb}$ most 
central collisions of $200A$ GeV $^{32}S+Au$~\cite{wa80}. The solid 
and the dashed curves give the thermal photon production expected 
from hot hadron gas calculation~\cite{ss}, while the dotted curve 
is the calculated thermal photon production expected in the case 
of a QGP formation. Right panel: Single photon production in $Pb+Pb$ 
collision at CERN SPS. Prompt photons are estimated using pQCD 
(with a K-factor estimated using NLO calculation) and intrinsic 
$k_T$ of partons~\cite{DKS_WA98}.}
\label{fig_wa98}     
\end{figure}

These results were analyzed in two different scenarios by authors 
of Ref.~\cite{ss}. In the first scenario, a thermally and chemically 
equilibrated quark-gluon plasma was assumed to be formed at some 
initial time ($\tau_0\approx$ 1 fm/$c$), which expanded~\cite{assb}, 
cooled, and converted into a mixed phase of hadrons and QGP at a 
phase-transition temperature, $T_C \approx$ 160 MeV. When all the 
quark-matter was converted into a hadronic matter, the hot hadronic 
gas continued to cool and expand, and underwent a freeze-out at a 
temperature of about 140 MeV. The hadronic gas was assumed to consist 
of $\pi$, $\rho$, $\omega$, and $\eta$ mesons, again in a thermal and 
chemical equilibrium. This was motivated by the fact that the included 
hadronic reactions involved~\cite{KLS} these mesons. This was already 
a considerable improvement over a gas of mass-less pions used in the 
literature at that time. In the second scenario, the collision was 
assumed to lead to a hot hadronic gas of the same composition. The 
initial temperature was determined by demanding that the entropy of 
the system be determined from the measured particle rapidity 
density~\cite{HK}. It was found that the scenario which did not 
involve a formation of QGP led to a much larger initial temperature 
and a production of photons which was considerably larger than the 
upper limit of the photon production, and could be ruled out. The 
calculation assuming a quark-hadron phase transition yielded results 
which were consistent with the upper limit of the photon production. 
These results were confirmed~\cite{wa80_ana} by several calculations 
exploring different models of expansion (see left panel of 
Fig.\ref{fig_wa98}).

It was soon realized that one may not limit the hadronic gas to contain
just $\pi$, $\rho$, $\omega$, and $\eta$ mesons, as there was increasing
evidence that perhaps all the mesons and baryons were being produced in
a thermal and chemical equilibrium in such collisions. Thus authors of 
Ref.~\cite{Clay} explored the consequences of using a hadronic gas 
consisting of essentially all the hadrons in the Particle Data Book, 
in a thermal and chemical equilibrium. This led to an interesting 
result for the $Pb+Pb$ collision at SPS energies, for which experiments 
were in progress. It was found that with the rich hadronic gas, the 
results for the production of photons in the phase-transition and no 
phase transition models discussed above were quite similar, suggesting 
that measurement of photons at the SPS energy could perhaps not 
distinguish between the two cases. However, in a very important 
observation, it was also noted that the calculations involving hot 
hadronic gas at the initial time would lead to hadronic densities of 
several hadrons/fm$^3$, and while those involving a quark gluon plasma 
in the initial state would be free from this malady. Thus, it was 
concluded that the calculations involving a phase transition to 
QGP offered a more natural description.

The WA98 experiment~\cite{subtraction}, reported the first observation 
of direct photons in central $158A$ GeV $Pb+Pb$ collisions studied at the 
CERN SPS. This was explained~\cite{ss1} in terms of formation of quark 
gluon plasma in the initial state (at $\tau_0\approx$ 0.2 fm/$c$), which 
expanded, cooled and hadronized as in Ref.~\cite{Clay} (see right panel 
of Fig.~\ref{fig_wa98}). An independent confirmation of this approach was 
provided by an accurate description~\cite{dks_na50} of excess dilepton 
spectrum measured by the NA60 experiment for the same system.

Once again the results for single photons  were analyzed by several 
authors using varying models of expansion as well as rates for production 
of photons; viz., with or with-out medium modification of hadronic properties 
(see, e.g., Ref.~\cite{TRG,sps1,sps2}). The outcome of all these efforts can 
be summarized as follows: the single photon production in $Pb+Pb$ collisions 
at SPS energies can be described either by assuming a formation of QGP in 
the initial state or by assuming the formation of a hot hadronic gas whose 
constituents have massively modified properties. The later description, 
however, involved a hadronic density of several hadrons/fm$^3$, which 
raises doubts about the applicability of a description in terms of hadrons, 
as suggested by Ref.~\cite{Clay}.

\subsection{RHIC}
\begin{figure}[b]
\centering
\includegraphics[height=6.2cm]{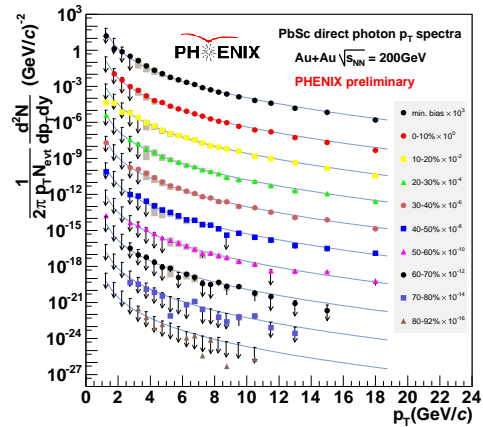}
\caption{Direct photon results for $Au+Au$ collisions (left panel) and 
$p+p$ collisions (right panel) at $\sqrt{s_{NN}} = 200A$ GeV~\cite{rhic}.} 
\label{rhic_fig}
\end{figure}      

The first experimental photon result obtained from the Relativistic 
Heavy Ion Collider was the centrality dependent single photon data  
for $200A$ GeV $Au+Au$ collisions by PHENIX collaboration~\cite{rhic}, 
where single photons were identified clearly for  $p_T > 4$ GeV. For 
central collisions, the lower bound on photon transverse momenta was 
even much lower, upto 2 GeV (Fig.~\ref{rhic_fig}). This data was in 
good agreement with NLO pQCD results for $pp$ collisions, scaled by 
number of binary collisions ( without considering the iso-spin effect).

Thermal radiation dominates the direct photon spectrum at lower values 
of $p_T$ ($ \le 3$ GeV) and the measured slope of the thermal photon 
spectrum can be related to the temperature of the system. 

\begin{figure}
\centering
\includegraphics[height=4.5cm, width=5.5cm]{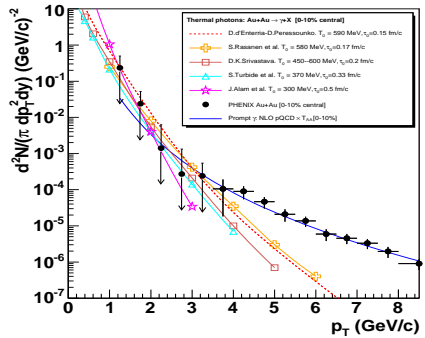}
\caption{Thermal photon spectra for central $200A$ GeV $Au+Au$ collision 
computed within different models ~\cite{Ente_Press} compared to the 
expected pQCD prompt photon yield and to the experimental total direct 
photon spectrum measured by PHENIX~\cite{Phenix}.}
\label{ther_phot_rhic} 
\end{figure}

In Fig.~\ref{ther_phot_rhic} results from several theoretical models 
based on hydrodynamics for thermal photon production at RHIC are shown. 
The initial temperature and thermalization time for $Au+Au$ collision at 
$200A$ GeV quoted by several theoretical groups~\cite{TRG, ss1, sps1, sps2,
anrds, Ente_Press} are in the range of 450 - 650 MeV  and about 0.2 
fm/$c$ respectively.

All these calculations are comparable to the experimental data and 
with each others within a factor of 2 and also confirm the dominance 
of thermal radiation in the direct photon spectrum in low and 
intermediate $p_T$ range. 

\subsubsection{Indications for jet conversion photons}

We have already discussed the procedure for calculating the high 
energy photons due to passage of jets through the quark gluon plasma. 

The  parametrized $p_T$ distribution of jets (quarks, anti-quarks, and 
gluons) obtained by using CTEQ5L parton distribution function 
and EKS98 nuclear modification factor is given by:
\begin{eqnarray}
\frac{dN^{\rm jet}}{d^2p_Tdy}|_{y=0}=
T_{AA}\frac{d\sigma^{\rm jet}}{d^2p_Tdy}|_{y=0}=K \frac{a}
{(1+p_T/b)^c}
\end{eqnarray}
Here $T_{\rm AA}=9A^2/8\pi {R_{\perp}}^2$ is the nuclear thickness 
for a head-on collision and to include the higher order effects 
a $K$ factor of $2.5$ is introduced. Numerical values for the 
parameters for quarks, anti-quarks and gluons
can be found in the Ref.~\cite{fms_phot}.

Now as a first step, let us ignore the transverse expansion of the 
plasma and assume that a thermally and chemically equilibrated plasma 
is produced in the collision at an initial time $\tau_0$ at temperature 
$T_0$. In an isentropic longitudinal expansion, ${T_0}$ and ${\tau_0}$ 
are related by the observed particle rapidity density ($dN/dy$) by:
\begin{eqnarray}
{T_0}^3{\tau_0}= \frac{2 {\pi}^4}{ 45 \zeta(3)}\frac{1}{ 4 a \pi 
R_\perp^2}\frac{dN}{dy}
\end{eqnarray}   
where $a=42.25 \pi^2/90$, for QGP consisting of $u, d$ and $s$ quarks and 
gluons. $dN/dy$ can be taken as $\simeq 1260$, based on the charge particle 
pseudo-rapidity density measured by PHOBOS experiment~\cite{Phobos_coll} 
for central collision of $Au$ nuclei at ${\sqrt s_{NN}}=200$ GeV. For 
central collision of $Pb$ nuclei at LHC energies $dN/dy \simeq 5625$ 
was used by Fries {\it et al.}~\cite{fms_phot} as in Ref.~\cite{prl25}. 

Further assuming a rapid thermalization, initial conditions can then 
be estimated as~\cite{prl25} $T_0=446$ MeV and $\tau_0=0.147$ fm/$c$ 
for the RHIC and $T_0=897$ MeV and $\tau_0= 0.073$ fm/$c$ for the LHC. 
Now, taking the nuclei to be  uniform spheres, the  transverse profile 
for initial temperature is given by $T(r)=T_0[2(1-r^2/{R_{\perp}}^2)
]^{1/4}$ where $R_{\perp}=1.2A^{1/3}$ fm. The same profile 
$R(r)=2(1-r^2/{R_{\perp}}^2)$ is used for the jet 
production.

The jet traverses the QGP medium until it reaches the surface or
until the temperature drops to the transition temperature $T_c$ 
($\sim 160$ MeV).
\begin{figure}
\centering
\includegraphics[height=4.6cm, width=5.50cm]{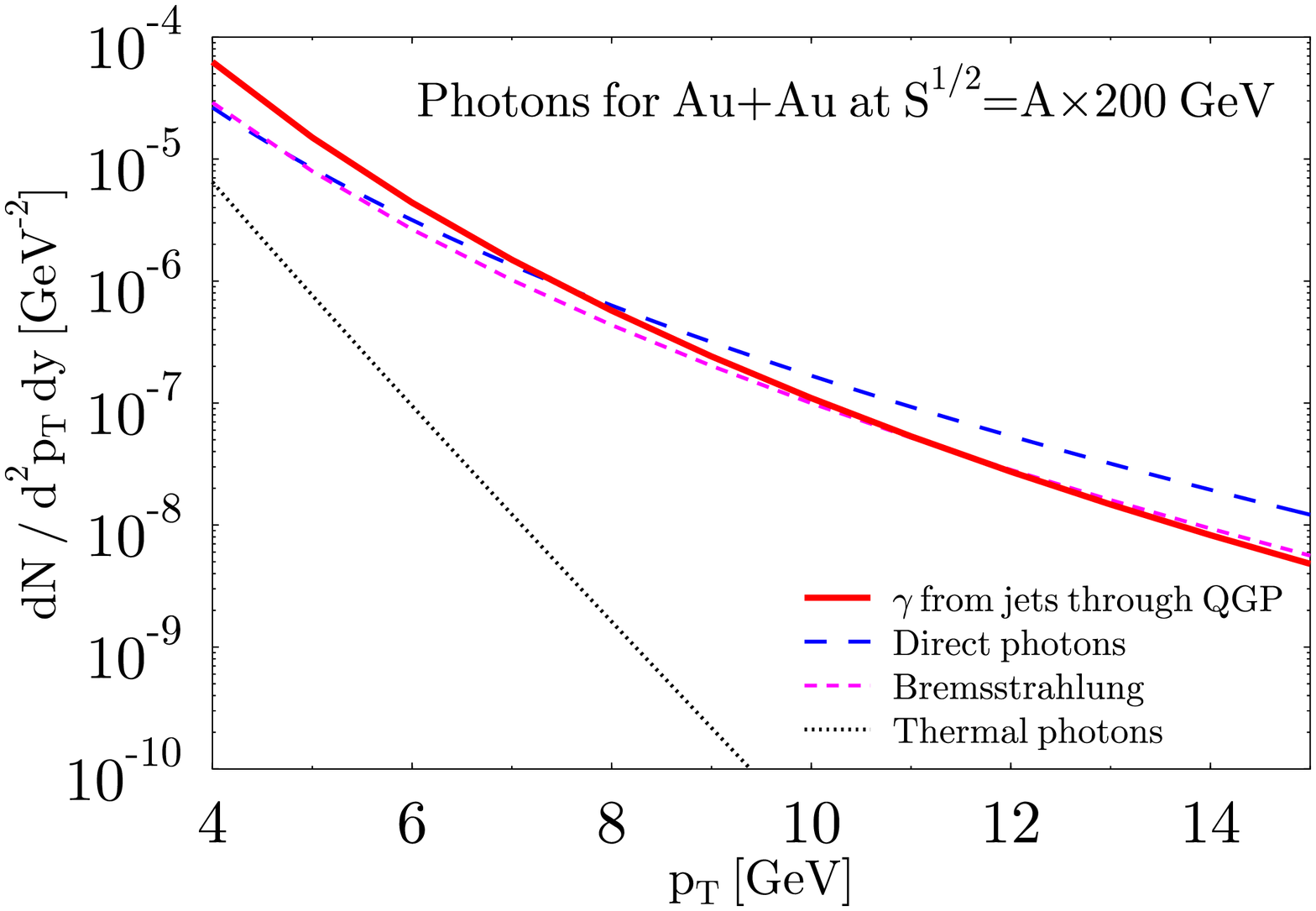}
\includegraphics[height=4.6cm, width=5.50cm]{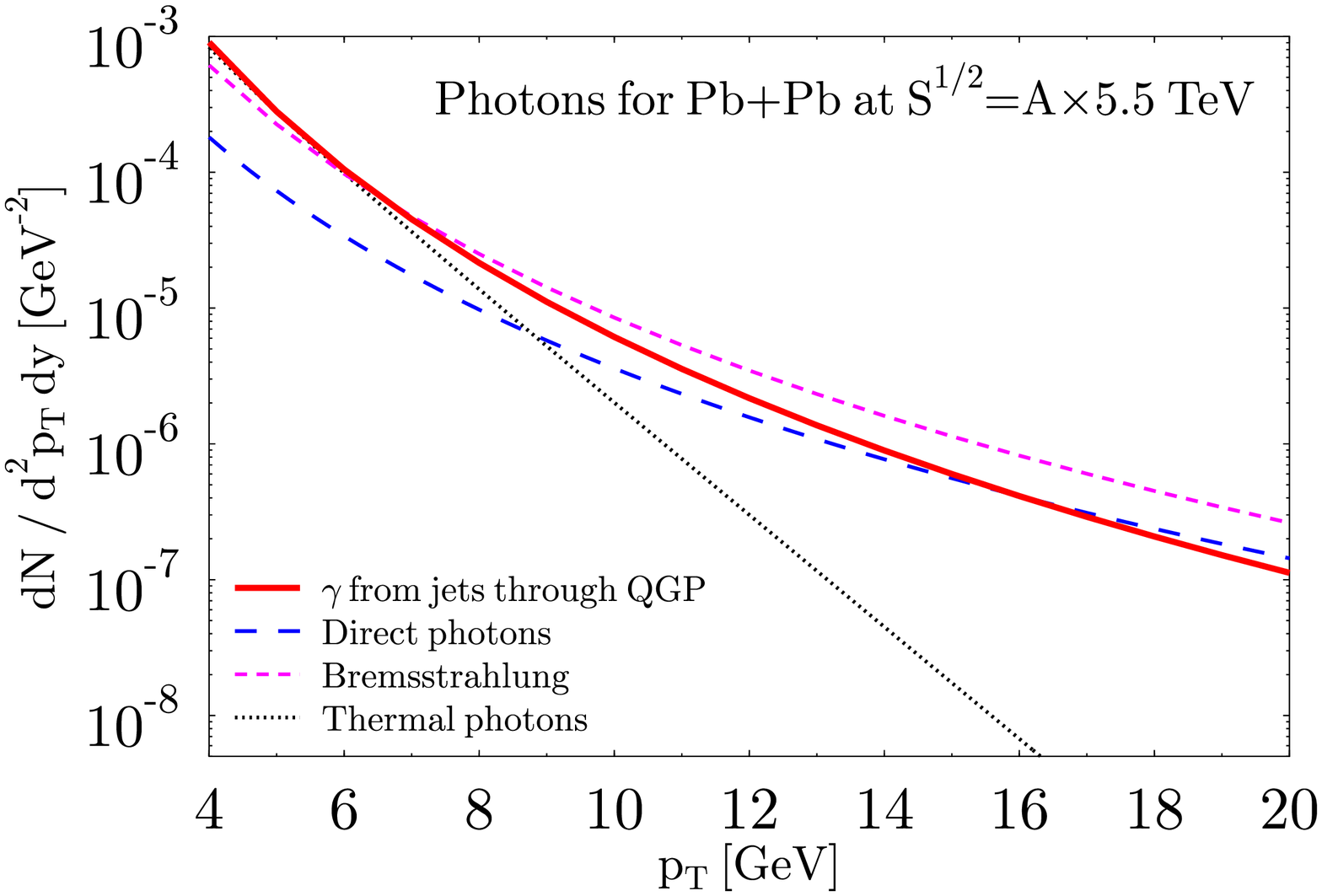}
\caption{Jet conversion photons at RHIC [left panel] and LHC 
[right panel] energies (taken from Ref.~\cite{fms_phot}).}
\label{FMS_rhic_lhc}   
\end{figure}  

First estimates for the jet conversion photons at RHIC and LHC energies 
along with other sources of photons having large transverse momentum are 
given in Fig.~\ref{FMS_rhic_lhc}. We see that the jet conversion photons 
make a fairly large contribution in the $p_T$ range of 4-10 GeV. 

\begin{figure}[t]
\centering
\includegraphics[height=4.7cm]{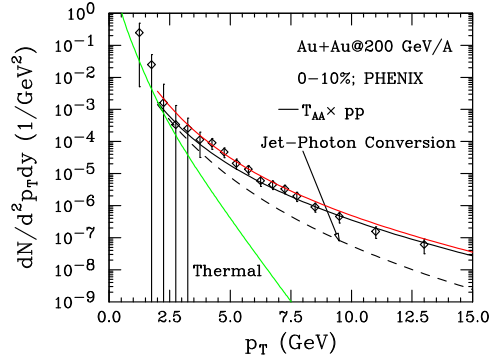}
\caption{Photon yield as a function of $p_T$ in central (0-10\%) 
$Au+Au$ collisions at $200A$ GeV. Results of jet-photon conversion 
(solid and labeled), primary hard photons (dashed) and the sum of 
the two (upper-most solid curve) along with thermal photons are 
shown separately. Data are from the PHENIX collaboration~\cite{Phenix}.} 
\label{FMS_phot}   
\end{figure} 

\begin{figure}
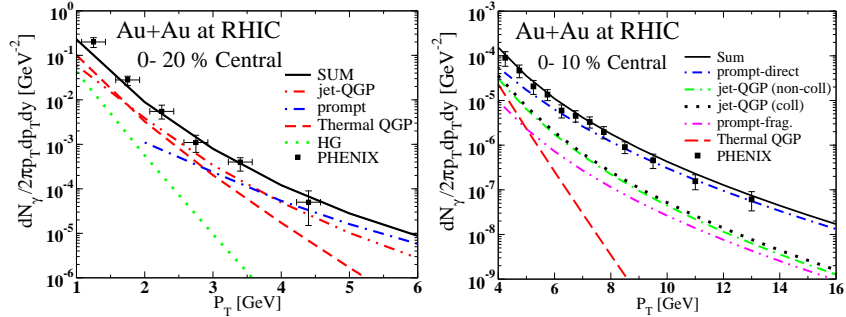

\centering
\includegraphics[height=4.2cm, width=5.50cm]{gale1.eps}
\includegraphics[height=4.2cm, width=5.50cm]{gale2.eps}
\caption{Yield of photons in $Au+Au$ collisions at RHIC, for centrality 
classes $0-10\%$ (left panel) and $0-10\%$ (right panel). See~\cite{TGFH} 
for details. The data sets are from Refs.~\cite{HB} and ~\cite{Phenix} 
respectively.}
\label{gale_evan_1_2} 
\end{figure}

The centrality dependence of jet conversion photons was also studied by 
Fries {\it et al.}~\cite{fms_phot2} (see figure~\ref{FMS_phot}), which 
indicates a small but clear contribution of photons due to passages of 
jets through the plasma. These early calculations have been brought to 
a high degree of sophistication by the McGill group~\cite{TGFH} (see 
figure~\ref{gale_evan_1_2}), where jet quenching and jet-photon conversion 
along with bremsstrahlung is treated in a single frame work. The details 
can be seen in Ref.~\cite{TGFH}.

In Fig.{\ref{gale_evan_1_2}} the results for thermal photons, direct 
photons due to primary processes, bremsstrahlung photons and the 
photons coming from jets passing through the QGP in central collision 
of gold nuclei at RHIC energies is plotted. The quark jets passing 
through the QGP give rise to a large yield of high-energy photons. For 
RHIC this contribution is dominant source of photons up to $p_T \simeq 
6$ GeV. Due to multiple scattering suffered by the fragmenting 
partons a suppression of the bremsstrahlung contribution is found. This 
will further enhance the importance of the jet-photon conversion process.

Obviously a high statistics data at several centralities and energies 
will go a long way in clearly establishing the presence of jet-conversion 
photons at RHIC and LHC energies. This is also of importance, as there are 
indications that these photons measure the initial spatial anisotropy of 
the system~\cite{TGFH}.

\section{Dileptons}

\begin{figure}
\centering
\includegraphics[height=3.50cm ]{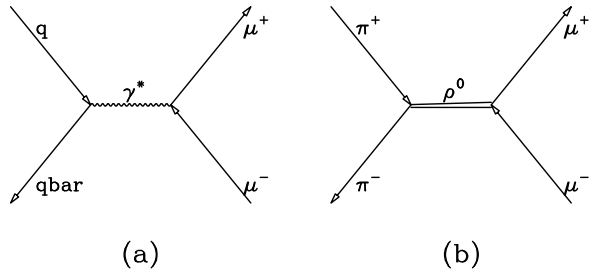}
\caption{ Dilepton production from (a) quark anti-quark annihilation and 
(b) $\pi^+ \pi^-$ annihilation through $\rho$ channel.}
\label{dil_dia}      
\end{figure}

Virtual photons or dileptons are also very powerful and efficient 
probes like the real photons, to study the dynamics of heavy ion 
collisions and the properties of the medium created in the 
collisions. Real photons are massless, whereas dileptons are massive. 
Thus, invariant mass $M$ and the transverse momentum $p_T$ are the 
two parameters available for dileptons, which can be tuned 
to investigate the different stages of the expanding fireball. Dileptons 
having large invariant mass and high $p_T$, are emitted very early, 
soon after the collision when the temperature of the system is very 
high. On the other hand, those having lower invariant masses come out 
later from a relatively cooler stages and at low temperatures. 
\begin{figure}
\centering
\includegraphics[height=5.50cm, width= 6.0cm]{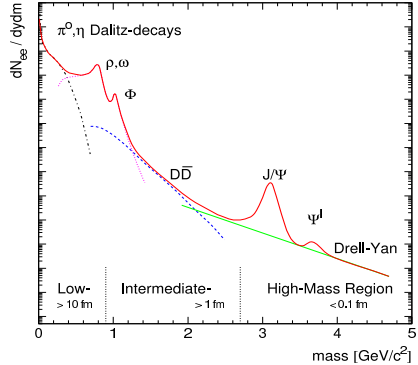}
\caption{Expected sources of dilepton production as a function of 
invariant mass in relativistic heavy ion collisions (schematic).}
\label{dil_mass}      
\end{figure}

Similar to the real photons, dileptons  are also emitted from every 
stage of heavy ion collisions~\cite{wong}. In the QGP phase, a quark 
can interact with an anti-quark to form a virtual photon $\gamma^*$, 
which subsequently decays in to a lepton ($l^-$) and an anti-lepton 
($l^+$) pair, together known as dilepton. In the hadronic 
phase, dileptons are produced from interactions of charged hadrons 
with their anti-particles by processes like ($\pi^+ + \pi^- \rightarrow 
l^+ + l^-$) (see Fig.~\ref{dil_dia}). They are also produced from decay 
of hadronic resonances 
like $\rho$, $\omega$, $\phi$, and $J/\Psi$ as well as from Drell-Yan 
process. In the Drell-Yan process, a valence quark from a nucleon in 
the projectile nucleus interacts with a sea anti-quark from a nucleon 
in the target nucleus to form a virtual photon, which decays into a 
lepton pair. The dilepton 
emissions can be classified into three distinct regimes in a rough 
estimation, depending on the invariant mass $M$ of the emitted lepton 
pairs. These are:

a) Low Mass Region (LMR): $M \le M_{\phi}$(= 1.024 MeV): In this 
mass range, vector meson decays are the dominating source of dilepton 
production and medium modified spectral density is one of the key issues 
which needs to be addressed. 

b) Intermediate Mass Region (IMR): $ M_{\phi}<  M < M_{J/\Psi}$(= 
3.1 GeV): In intermediate mass region, continuum radiation from QGP 
dominates the dilepton mass spectrum and thus this region is important 
for getting a pure QGP signature.

c) High Mass Region (HMR): $M \ge M_{J/\Psi}$: In the HMR, the most 
interesting phenomenon are the primordial emission and heavy quarkonia 
like $J/\Psi$  and $\Upsilon$ suppression.

A schematic diagram of continuous dilepton mass spectrum is shown in 
Fig.~\ref{dil_mass}.

\subsection{Dileptons from QGP and hadronic phase}

Dilepton production by quark anti-quark annihilation  process or by 
$\pi^+$ $\pi^-$ annihilation process can be expressed in a general 
form as,
\begin{equation}
{a^+ + a^- \rightarrow  l^+ + l^-}
\end{equation}
where, the particle `a' can be either a quark or a $\pi$ meson. We 
closely follow the treatment of Ref.~\cite{KKMM} for calculation of 
dilepton spectra in this section. From Quantum Electrodynamics (QED), 
the cross section for $e^+ e^- \rightarrow \mu^+ \mu^-$ can be 
written as,
\begin{eqnarray}
\tilde\sigma(M)= \frac {4 \pi } {3} \frac {\alpha^2} {M^2} \left [1+ \frac 
{2m_l^2}{M^2} \right] \left [1- \frac {4m_l^2} {M^2} \right]^{1/2}
\label{cross}
\end{eqnarray}
In the above equation, $\alpha$ is the fine structure constant 
($\approx 1/137$), $m_l$ is the mass of $\mu$, and M is the dilepton invariant 
mass. For the quark anti-quark annihilation process, this cross-section 
is modified by inclusion of a colour factor $N_c (= 3)$, and the fractional 
charges of the up(u) and down(d) quarks. Thus, the new cross-section takes 
the form as,
\begin{eqnarray}
\sigma_q(M) & = & F_q \tilde \sigma(M),\\
 F_q  & = & N_c(2s+1)^2 {\sum_{f} e_f^2}
\end{eqnarray}
where s is the spin of quarks, $e_f$ is the fractional charge, and the 
sum is over different quark flavour `f'. If we consider only u and d 
quarks having fractional charges 1/3 and 2/3 respectively, then the factor 
$F_q$ is 20/3. 

In the hadronic phase, the $\pi^+\pi^-$ annihilation from vector 
meson dominance model can be expressed as,
\begin{equation}
\pi^+ + \pi^- \rightarrow \rho \rightarrow l^+ + l^-
\end{equation}
For this case, the QED cross-section is multiplied by form factor 
$F_\pi(M)$ which is of Breit-Wigner form as;
\begin{equation}
F_\pi(M)= {\frac {m_\rho^4} {(m_\rho^2-M^2)^2+m_\rho^2\Gamma_\rho^2}}
\end{equation}
Here, $M_\rho$ is the mass of the $\rho$ meson ($\sim 770$ MeV) and 
$\Gamma_\rho$ ( $\sim 155$ MeV ) is the decay width. Thus, the total pion 
cross-section as a function of M becomes,
\begin{equation}
\sigma_\pi(M)=F_\pi \tilde \sigma(M)(1-4m_\pi^2/M^2)^{1/2},
\end{equation}
The reaction rate `R' can be obtained from the kinetic theory:
\begin{equation}
R(a^+a^- \rightarrow l^+ l^-) = \int \frac {d^3p_1} {(2\pi)^3} f({\bf p_1}) 
\int \frac {d^3p_2} {(2\pi)^3} f({\bf p_2})\sigma(a^+a^- \rightarrow 
l^+l^-;{\bf p_1p_2})v_{\rm rel}
\label{rate}
\end{equation}
where,
\begin{equation}
v_{rel}= {\frac {[(p_1.p_2)^2 - m_a^4]^{1/2}} {E_1E_2}}
\nonumber\\
\end{equation}
$f({\bf p})$, is the occupation probability at momentum ${\bf p}$ and 
energy $E=\sqrt{{\bf p}^2 + m_a^2}$. Now, using the distribution 
function $f({\bf p})\sim e^{-E/T}$, and integrating over five out 
of the six variables, the reaction rate takes a simplified form as,
\begin{equation}
R(T)= \frac {T^6} {(2\pi)^4} \int_{z_0}^{\infty} \sigma(z)z^2(z^2-4z_a^2)
K_1(z)dz
\end{equation}
Here, $z=M/T$, $z_a=m_a/T$, and $K_1$ is the modified Bessel function of 
the first kind. The value of the parameter $z_0$ is taken as the larger 
of $2m_a/T$ and $2m_l/T$. Now, for massless $u$ and $d$ quarks ( as $m_u$, 
$m_d$ $\ll T$), the $e^+e^-$ emission rate takes a simple form of $T^4$ 
law, 
\begin{equation}
R= \frac {10} {9\pi^3} \alpha^2T^4 .
\end{equation}
One can also estimate the relaxation time for lepton pairs to come
to equilibrium with the QGP:
\begin{equation}
{t_{\rm rel}= {\frac {n_{\rm eq}^l} {2R}} = {\frac {9 \pi} 
{10 \alpha^2 T}}}  .
\end{equation} 
For a temperature range of 200 to 500 MeV, the value of $t_{\rm rel}$ 
varies from $20 - 60 \times 10^3$ fm/$c$. We know that the life time of 
the QGP phase is only $\sim 10$ fm/$c$, which is more than three order 
of magnitude smaller than $t_{\rm rel}$. Thus, like the real photons, 
the produced lepton pairs also escape the system without suffering 
significant absorption in the medium.

Now, the dilepton emission rate $R$ is actually defined as the total number 
of lepton pairs emitted from a 4-volume element $\ d^4x \ (= \ d^2x_T \ 
d\eta \ \tau \ d\tau)$ at a particular temperature $T$ is given by:
\begin{equation}
R=dN/d^4x .
\end{equation}
Thus, the rate of production of dileptons having invariant mass $M$ 
can be expressed using Eqn.(\ref{rate}) as,
\begin{eqnarray}
\frac{dN} {d^4xdM^2}= \frac {\sigma(M)}{2(2\pi)^4}M^3TK_1(M/T) \left[1- 
\frac {4m_a^2}{M^2} \right].
\end{eqnarray}

From the last equation and using properties of the modified Bessel's 
functions, the production rate per unit 4-volume for total energy 
$E$, momentum $p$ and invariant mass $M$ (where, $E = \sqrt{ p^2 + 
M^2}$ ) can be written as,
\begin{eqnarray}
E \frac {dN}{d^4xdM^2d^3p} = \frac {\sigma(M)} {4(2 \pi)^5} M^2 e^{-E/T} 
\left[1- \frac{4m_a^2}{M^2}\right] .
\end{eqnarray}
Integrating the rate of emission over the entire 4-volume from QGP and 
hadronic phase, one can obtain the $p_T$ spectra at a particular $M$ as,
\begin{eqnarray}
\frac {dN} {dM^2 d^2p_T dy} & = & \int \ \tau \ d\tau \ r \ dr \ d\eta \ 
d\phi  \left [ \left [E \frac {dR} {dM^2 d^3p} \right ]_{\rm QGP}  
\it{f}_{\rm QGP} (r, \tau)  \right. \nonumber\\ & + & \left. \left [E 
\frac {dR} {dM^2 d^3p} \right ]_{\rm HM}  \it{f}_{\rm HM}(r, \tau) \right]
\label{dil_spec}
\end{eqnarray}
where, $\eta $ is the space time rapidity. In the above equation, 
the temperature $T$, QGP and hadronic matter (HM) distribution 
functions ($\it{f}_{\rm QGP}, \it{f}_{\rm HM}$), all are functions 
of space $\bf{r}$ (x, y), transverse velocity $v_T$, and proper time 
$\tau$. Eqn.~(\ref{dil_spec}) can be solved 
numerically by using any appropriate model [e.g., Ref.~\cite{KKMR}] 
with a proper equation 
of state (EOS). One can also get the invariant mass spectrum by 
integrating out the variable $p_T$ from Eqn.~(\ref{dil_spec}). 
The invariant mass spectrum of thermal dilepton is dominated 
by QGP radiation above $\phi$ mass and hadronic radiation 
outshines the QGP contribution for $M \le M_{\phi}$.

\subsection{Medium modification}

As mentioned earlier, in the low mass region, dilepton emission is 
largely mediated by $\rho(770)$, a broad vector meson, as a result 
of its strong coupling to the $\pi \pi$ channel and a short lifetime, 
which is about 1.3 fm/$c$. In-medium properties of vector mesons, like 
change in medium mass and width have long been considered as prime signatures 
of a hot and dense hadronic medium. CERES/NA45~\cite{ceres, NA45} started the 
pioneering experiment on dilepton measurement during the period 1989-1992 at 
CERN SPS. 
For $p+Be$ and $p+Au$ collisions, the theoretical 
estimates, considering only hadron decay (as source of dilepton 
production), were in excellent agreement with the experimental data 
points. However, the CERES/NA45 experimental data for $S+Au$ and $Pb+Au$ 
at $200A$ GeV and $158A$ GeV respectively exhibit excess radiation 
beyond the electromagnetic final state decay of produced hadrons below 
the $\phi$ mass. From these observations, it was concluded that, the 
theoretical models based on $\pi \pi$ annihilation can reproduce the 
experimental data, only if the properties of the intermediate $\rho$ 
meson is modified in the medium. This was an exciting discovery and 
a genuine consequence of many body physics. However, the resolution and 
statistical accuracy of the data was insufficient to distinguish between 
models suggesting a drop in the mass of $\rho$ mesons~\cite{BR} and those 
which suggest an increase in their decay width~\cite{RW} and thus, 
determine the in-medium spectral properties of the $\rho$ meson. 

Shortly after this, an excess dimuon (over the sources expected from 
$pA$  measurement)  was identified experimentally by Helios/3~\cite{heli} 
(measured both $e^+ e^-$ and $\mu^+ \mu^-$) and NA38/NA50~\cite{na50} 
collaborations. 
\begin{figure}[ht]
\centering
\includegraphics[height=4.2cm, width=5.50cm]{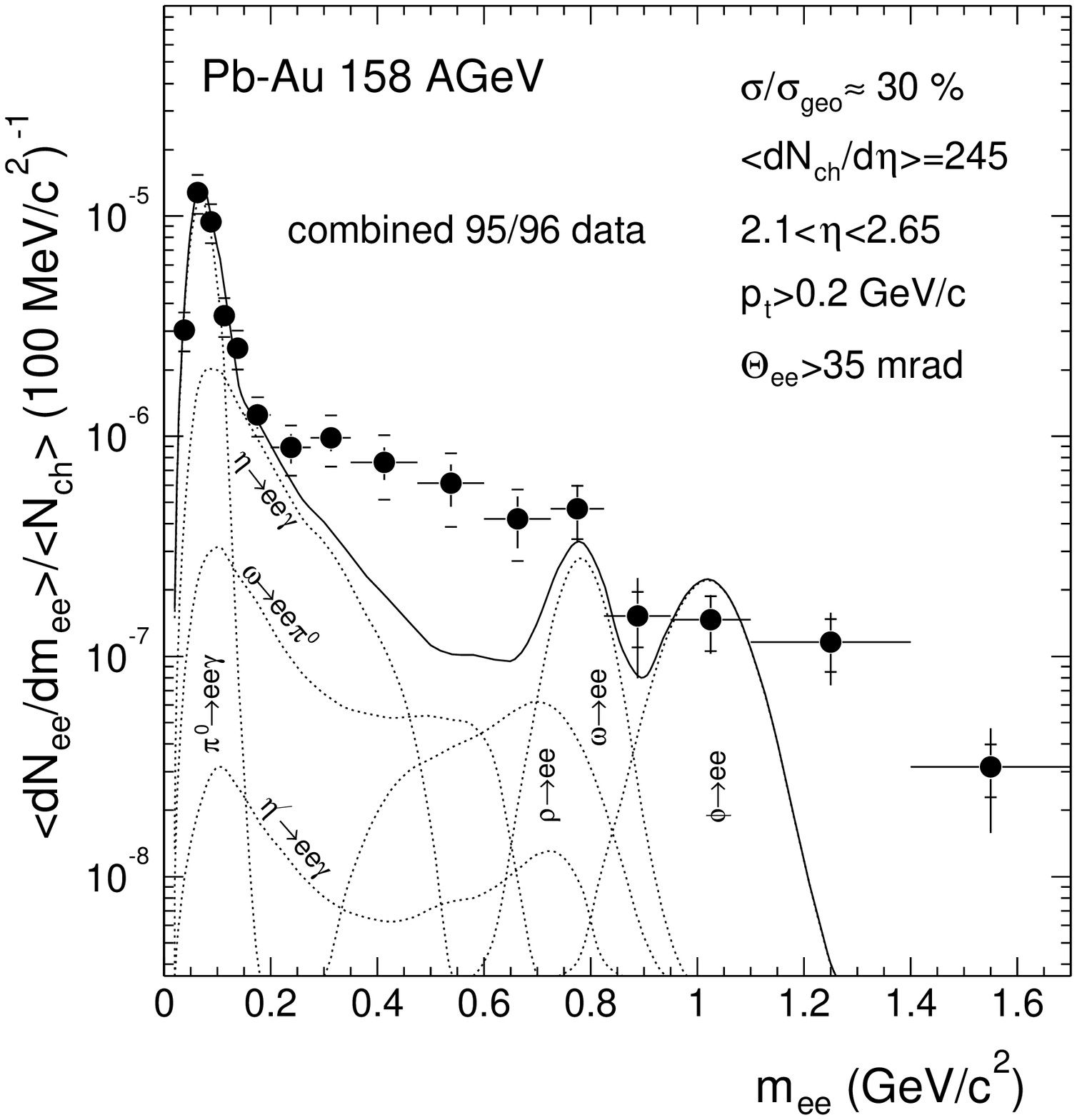}
\includegraphics[height=4.2cm, width=5.50cm]{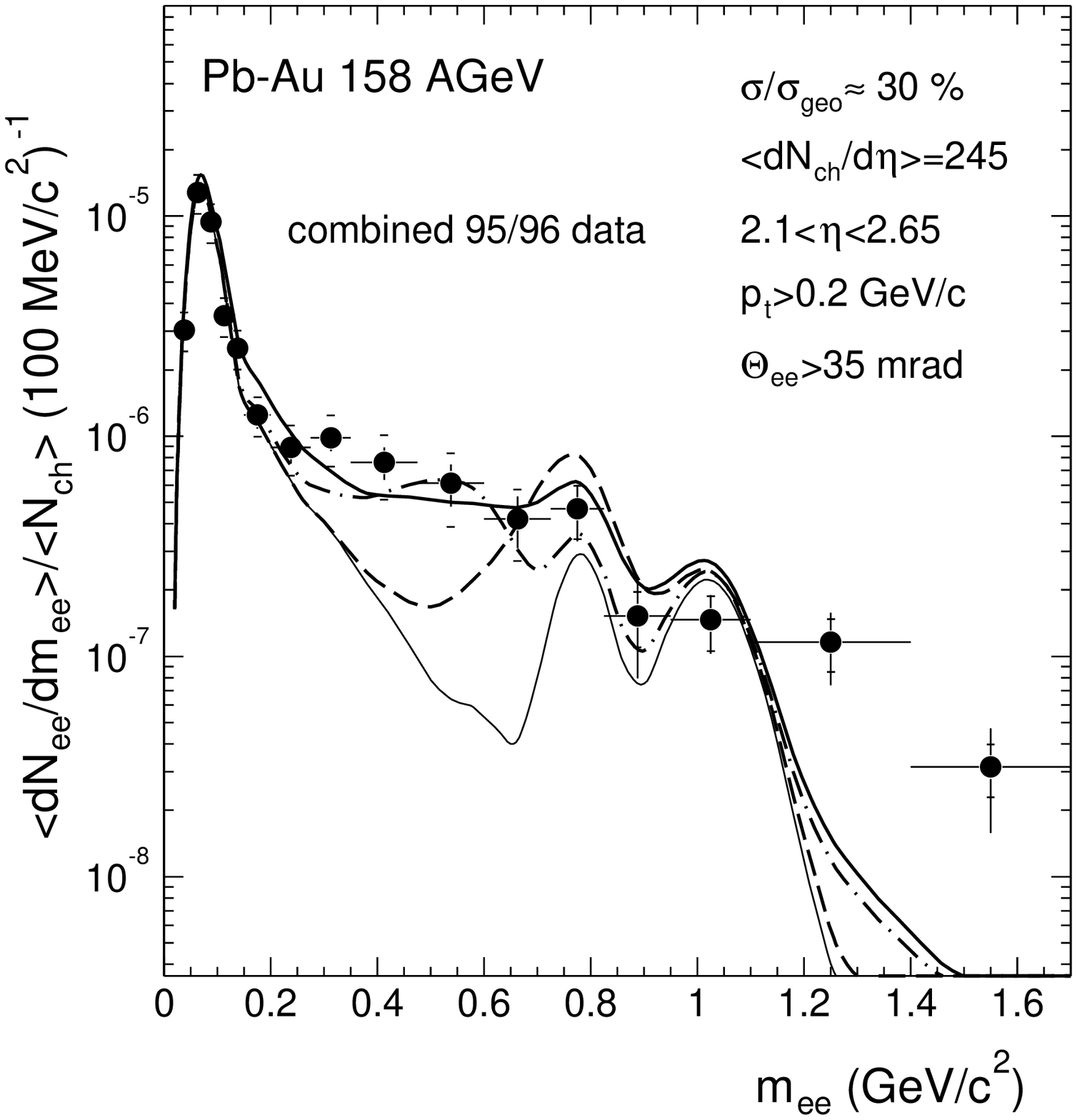}
\caption{ Left panel: Invariant mass spectrum of $e^+ e^-$ pairs 
emitted in $158 A$ GeV $Pb+Au$ collisions from the combined analysis 
of 1995 and 1996 data by CERES/NA45~\cite{ceres,NA45}. The solid line 
shows expected yield from hadron decay and dashed lines indicate the 
individual contribution to the total yield. Right panel: Comparison 
of the experimental data to i) free hadron decays without $\rho$ 
decays (thin solid line), ii) model calculations with a vacuum $\rho$ 
spectral function (thick dashed line), iii) with dropping in-medium 
$\rho$ mass (thick dash-dotted line), iv) with a medium modified 
$\rho$ spectral function (thick solid line).}
\label{ceres_1_2}  
\end{figure}
%

\begin{figure}
\centering
\includegraphics[height=3.2cm, width=3.60cm]{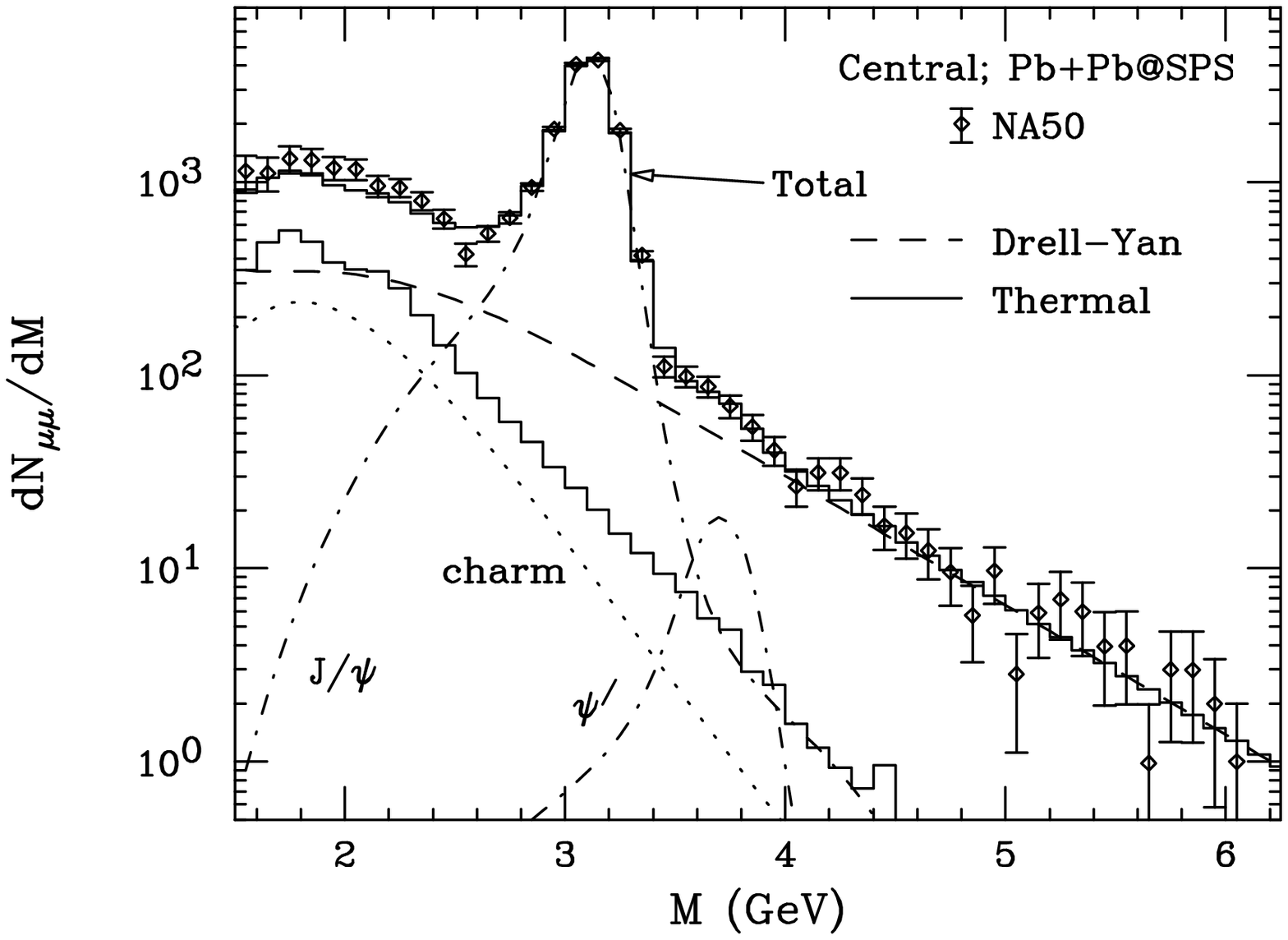}
\includegraphics[height=3.2cm, width=3.60cm]{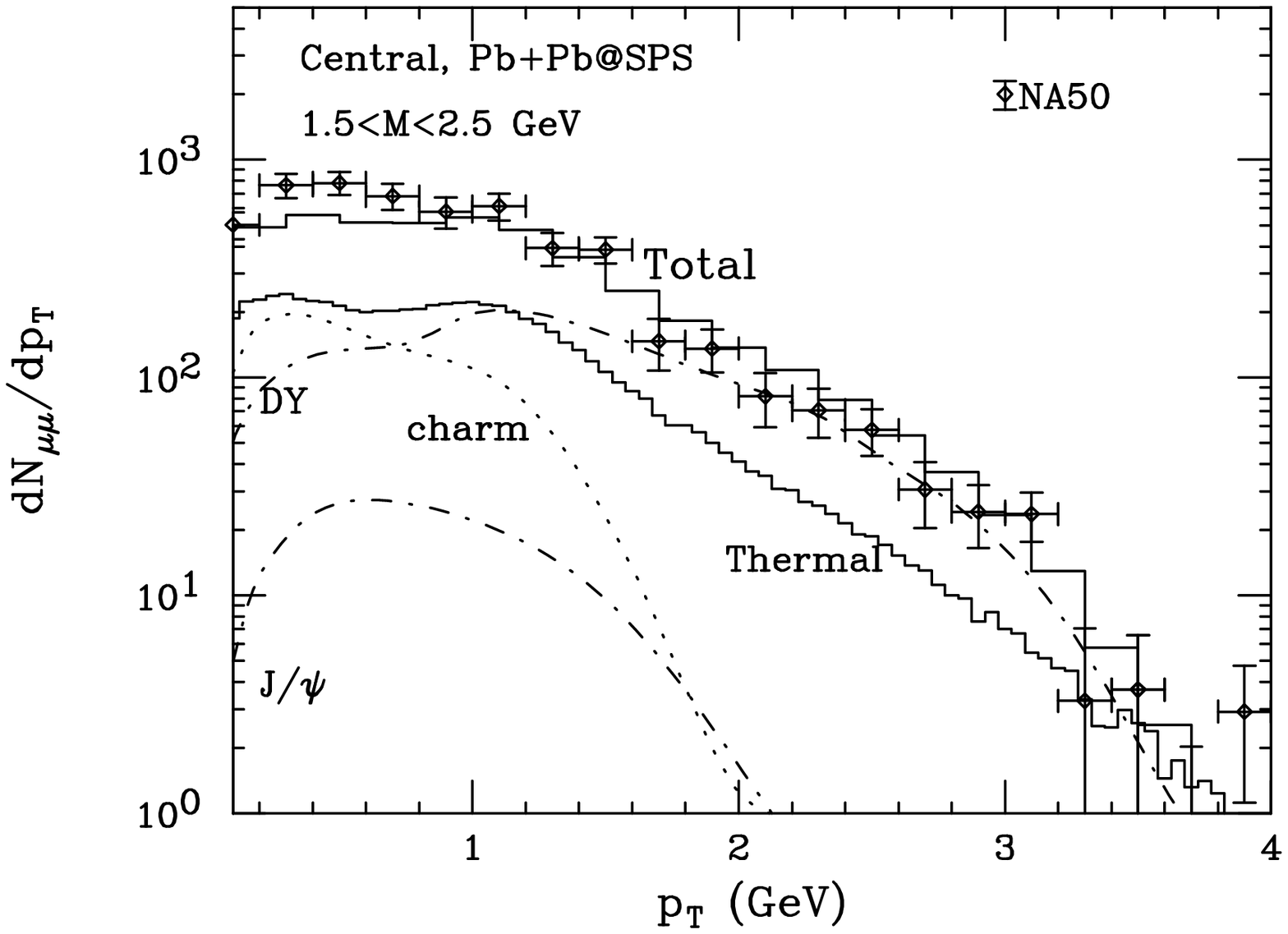}
\includegraphics[height=3.6cm, width=3.20cm, angle=90]{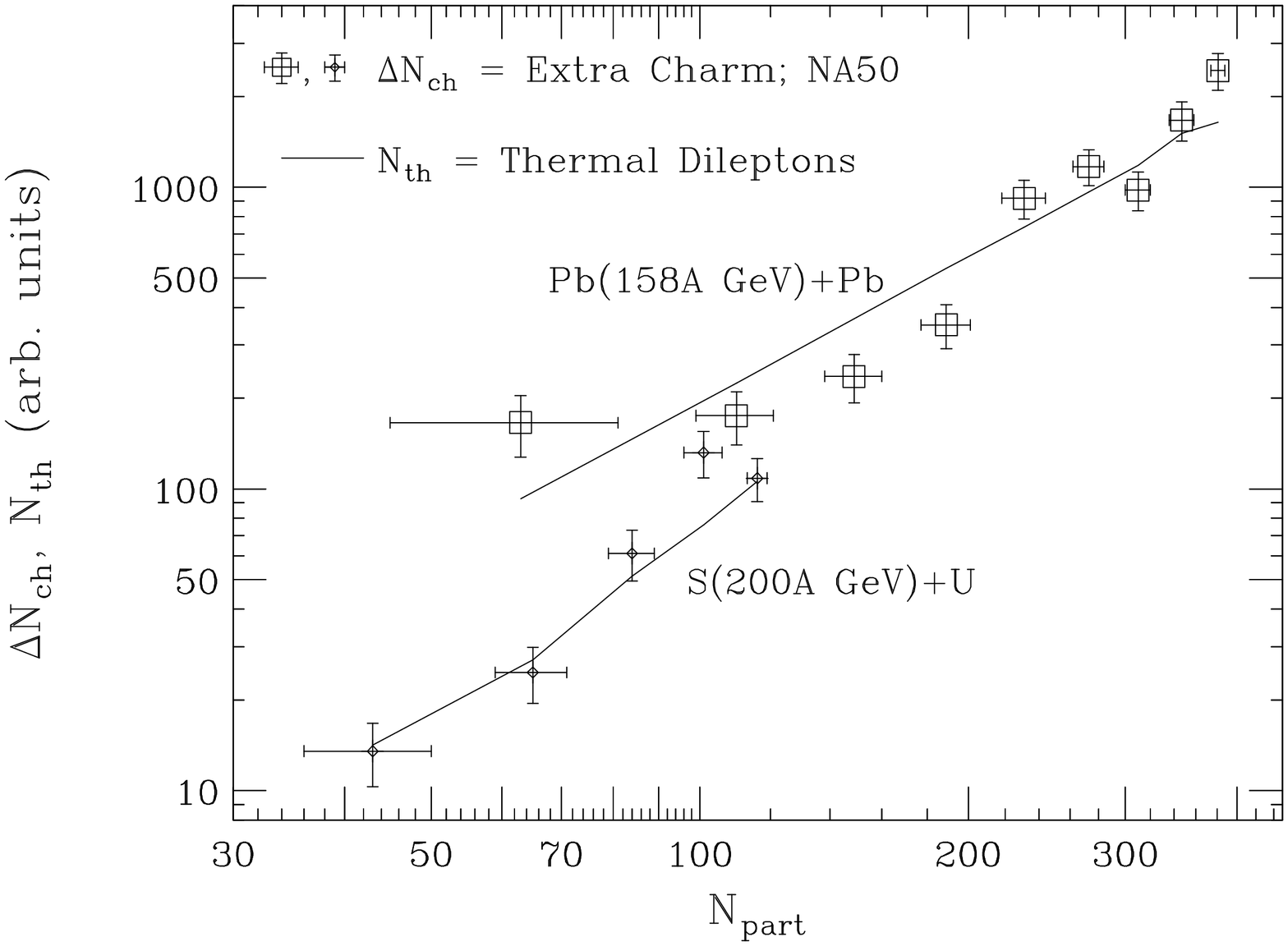}
\caption{(a) The dimuon invariant mass spectrum after correcting for 
detector acceptance and resolution and NA50 data. The Drell-Yan and 
thermal contributions are shown separately along with correlated charm 
decay and direct decays of the $J/\Psi$ and $\bar{J}/\Psi$. (b) The 
dimuon transverse momentum spectrum. (c) Centrality dependent 
results from Ref.~\cite{dks_na50} and NA50 data~\cite{na50}.}
\label{DKS_na50}   
\end{figure}

Rapp and Shuryak~\cite{rapp_sur} and Kvasnikova {\it et al.}~\cite{dks_na50}
showed that, the excess dimuon observed by NA50 in the mass region $ 1.5  
< M  < 2.5$ GeV can be explained by thermal signal without invoking any 
anomalous enhancement in the charm production. Kvasnikova {\it et al.} 
studied the NA50 intermediate mass dimuon result using hydrodynamic 
model and a detailed analysis of the rates of dilepton production from 
hadronic phase~\cite{dks_na50}. Detector resolution and acceptance were 
accurately modeled in their calculation and the normalization was 
determined by a fit to the Drell-Yan data 
using the MRSA parton distribution function as done in the experimental 
NA50 analysis. The results are shown in Fig.~\ref{DKS_na50}. They also 
studied the centrality dependence of the NA50 data using hydrodynamics 
by incorporating azimuthal anisotropy in the calculation. The results 
were in fairly good agreement with the measured excess dilepton data, 
as can be seen in the third panel of Fig.~\ref{DKS_na50}.

\subsubsection{Dropping $m_{\rho}$ vs. increasing $\Gamma_{\rho}$}

\begin{figure}
\centering
\includegraphics[height=4.25cm, width=5.0cm]{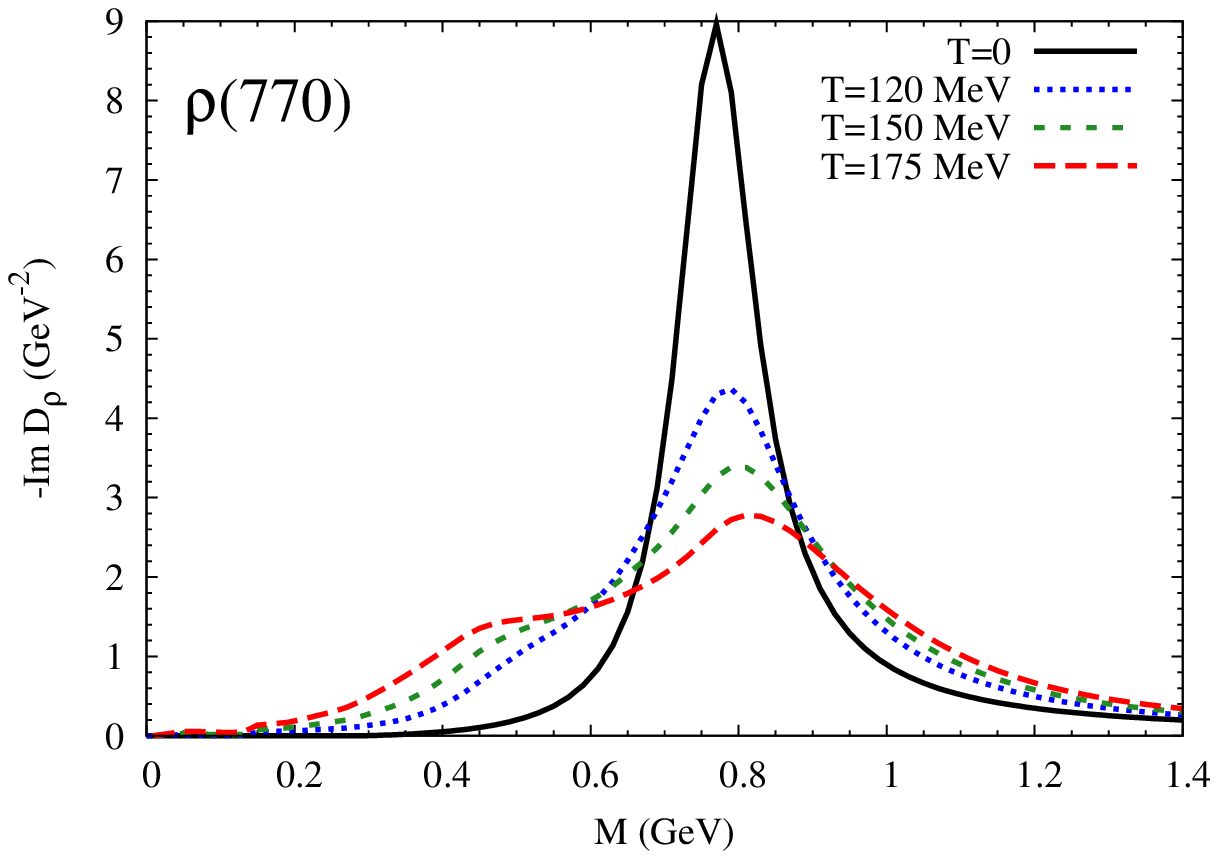}
\includegraphics[height=4.2cm, width=5.50cm]{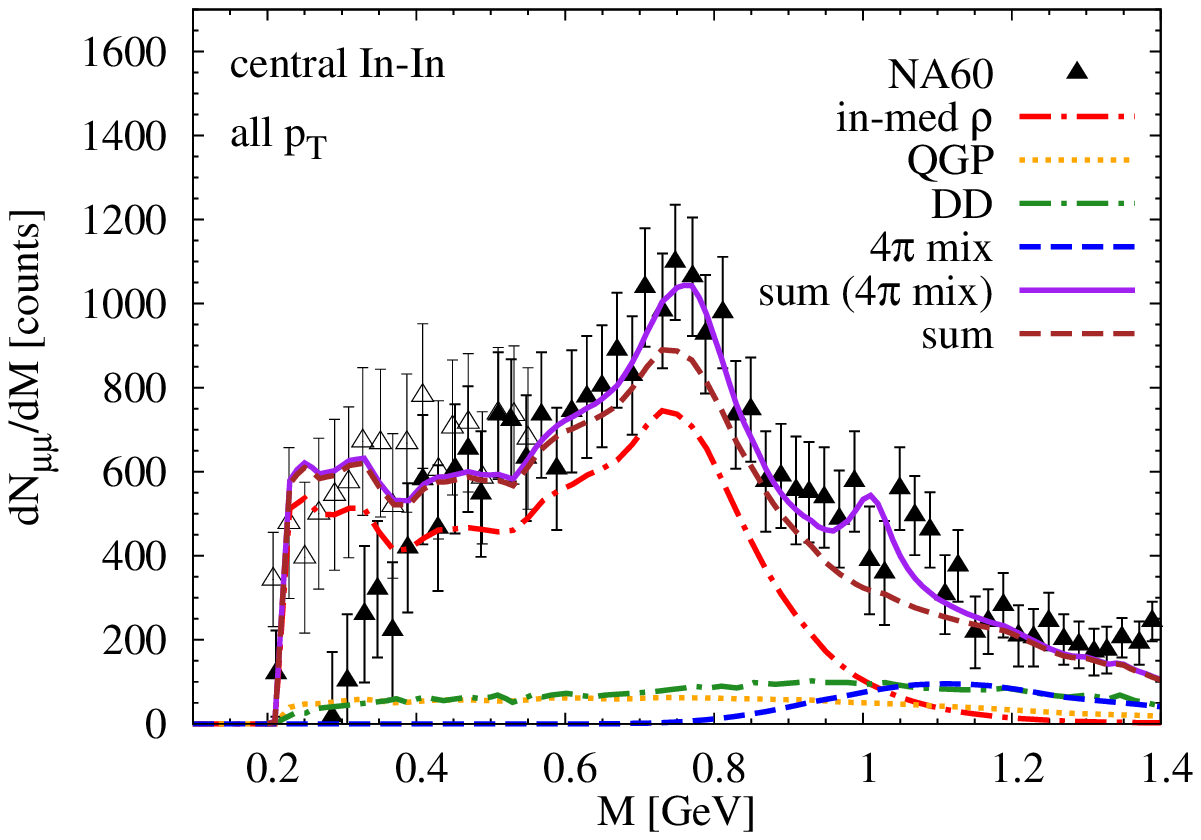}
\caption{ Left panel: In medium vector meson spectral functions 
from hadronic many body theory. The $\rho$ in cold nuclear matter 
at several densities proportional to the saturation density $\rho_0 
=0.16$ fm$^{-3}$. Right panel: NA60~\cite{NA60} excess dimuons in 
central $In+In$ collisions at SPS compared to thermal dimuon radiation 
using in-medium electromagnetic rates~\cite{Hees_Rapp}.}
\label{rapp_1}      
\end{figure}

The CERES data~\cite{ceres,NA45} was unable to distinguish between 
the models suggesting drop in the mass of $\rho$ meson and those 
suggesting an increase in its width, as seen in Fig.~\ref{ceres_1_2}. 
The high statistics data obtained by the NA60 Collaboration for $In+In$ 
collisions at SPS energies for the 
dimuon excess rather clearly ruled out the models advocating the 
dropping mass scenarios. They rather firmly establish the models 
advocating the substantial broadening of the decay width of the 
$\rho$ meson due to many body effects (see Fig~\ref{rapp_1}).

\section{Elliptic Flow}

Elliptic flow is one of the key observables in relativistic collisions 
of heavy nuclei, which confirms the collectivity and early thermalization 
in the created hot and dense matter. For a non-central collision (impact 
parameter $b \ne 0$ ) of two spherical nuclei, the overlapping zone 
between the nuclei no longer remains circular in shape, rather it takes 
an almond shape. This initial spatial anisotropy of the overlapping zone 
is converted into momentum space anisotropy of particle distribution via 
the action of azimuthally anisotropic pressure gradient, which gives rise 
to elliptic flow~\cite{olli}. Note that, this anisotropic flow or elliptic 
flow can also be produced in central collisions of deformed nuclei, such 
as $U+U$ collisions~\cite{uli_khul}. The driving force for momentum 
anisotropy is the initial spatial eccentricity $\epsilon_x$ ($ = <y^2 - 
x^2>/<y^2 + x^2>$) and the momentum anisotropy can grow as long as 
$\epsilon_x > 0$~\cite{QGP3}. Elliptic flow coefficient $v_2$ is quantified 
as the second Fourier co-efficient of particle distribution in the $p_T$ 
space, which is of the form:
\begin{eqnarray}
\frac{dN(b)}{p_T dp_T \, dy \,d\phi} & = & \frac{dN(b)}{ 2 \pi p_T \, 
dp_T \, dy} (b)\times [1+ 2 v_1(p_T,b) \cos( \phi) 
\nonumber\\ 
& + & 2 v_2(p_T,b) \cos(2 \phi)+ 2 v_3(p_T,b) \cos(3 \phi)+ \dots ].
\end{eqnarray}
At mid-rapidity ($y=0$) and for collisions of identical nuclei, only 
the even cosine terms survive in the above Fourier series and $v_2$ 
is the lowest non-vanishing anisotropic flow co-efficient. Now, the 
value of $v_2$ depends on the impact parameter $b$, transverse 
momentum $p_T$ as well as on the particle species through their 
rest masses $m$. For massless real photons, $v_2$ depends on the 
elliptic flow of parent particles.

\subsection{Thermal Photon $v_2$}

Elliptic flow co-efficient $v_2$ for photons is a much more powerful 
tool than the $v_2$ of hadrons, to study the evolution history of 
the system depending on the different emission time and production 
mechanism of photons compared to hadrons. Photons are emitted from 
all stages and throughout the evolution of the system, whereas, 
hadrons are emitted only from the freeze-out surface at a relatively 
much cooler temperature ($\sim 100$ MeV). The interplay of the photon 
contributions from fluid elements at different temperatures with 
varying radial flow pattern, gives the photon $v_2$ a richness, 
which is not possible for hadrons. Also, photons emitted from QGP 
phase as a result of $q \bar q$ annihilation and quark gluon Compton 
scattering, carry the momenta of the parent quarks or anti-quarks, 
which makes them an unique tool for QGP study. 
%
\begin{figure}[t]
\centering
\includegraphics[height=4.2cm, width=5.50cm]{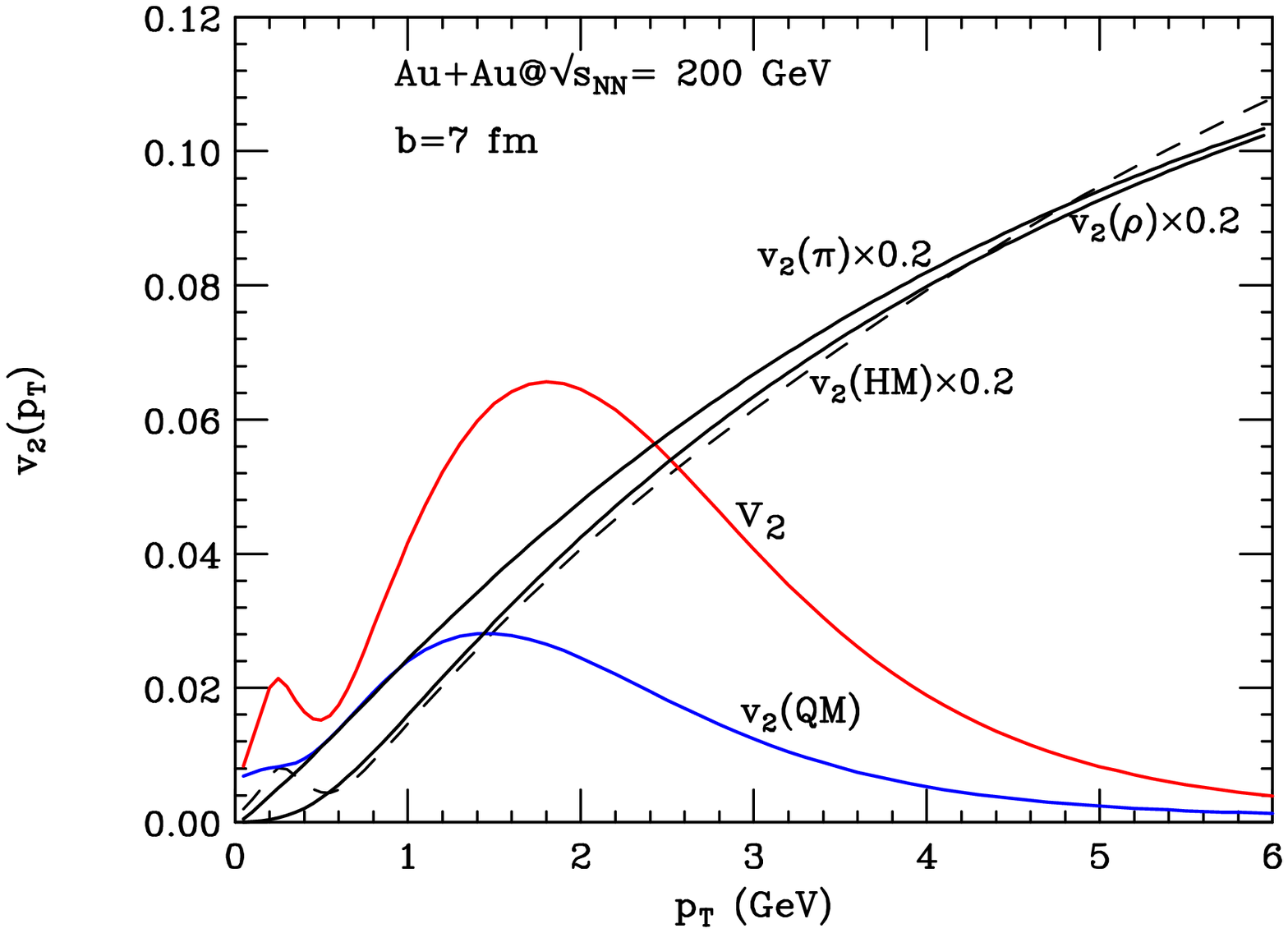}
\includegraphics[height=4.2cm, width=5.50cm]{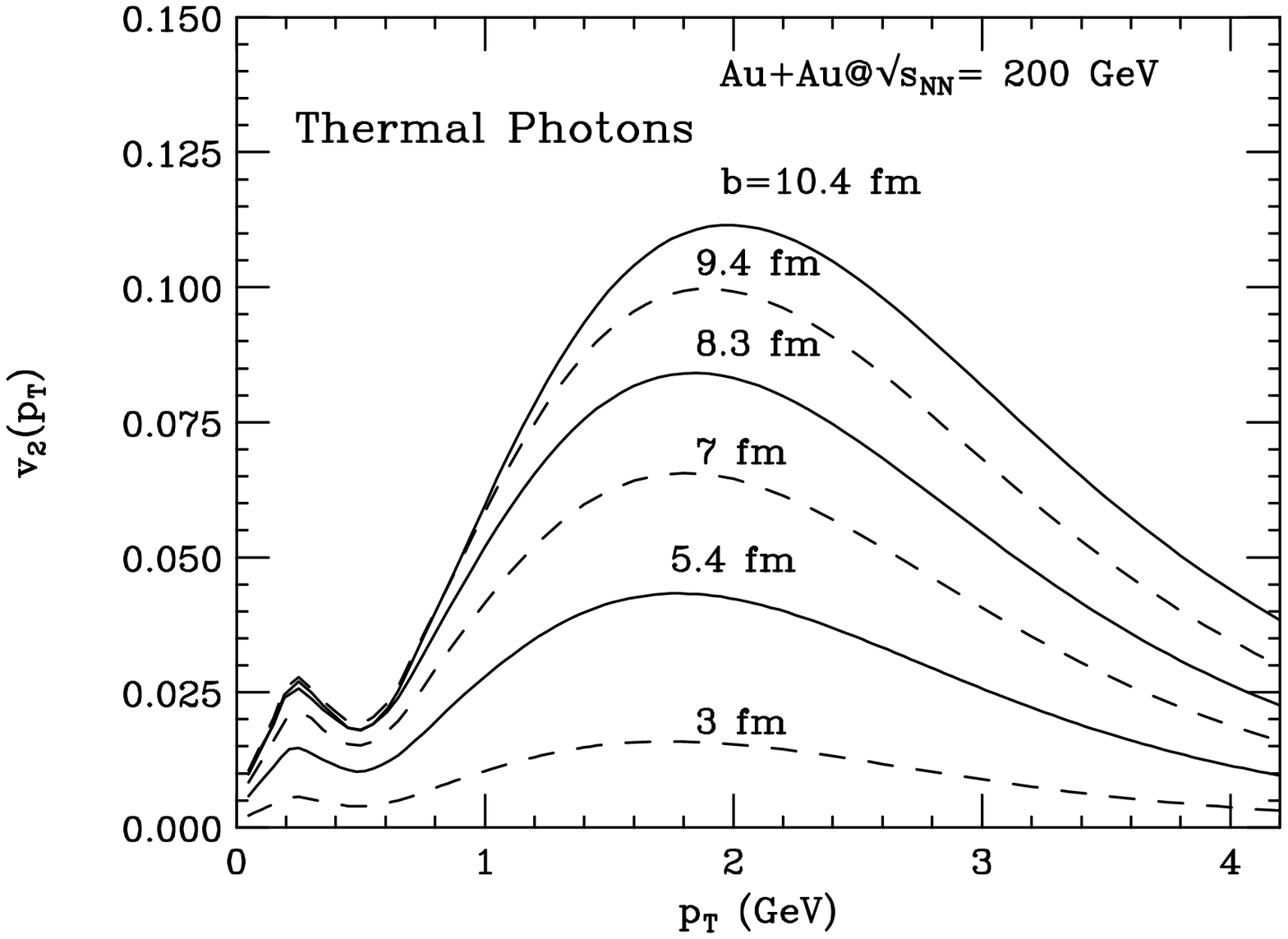}
\caption{Left panel: $v_2(p_T)$ for thermal photons along with the 
$v_2$ for $\pi$ and $\rho$ mesons. Contributions the from quark matter  
and  the hadronic matter are shown separately with the sum contribution. 
Right panel: Thermal photon $v_2$ for different centrality bins [taken 
from Ref.~\cite{CFHS}].}
\label{phot_v2_pt}   
\end{figure}

Ideal hydrodynamic model has successfully predicted the elliptic flow 
for different hadronic species upto a $p_T$ value of 1.5(2.3) GeV for 
mesons (baryons)~\cite{QGP3} at RHIC. Experimental data show that, 
$v_2(p_T)$ for hadrons saturates beyond that $p_T$ range. However, 
ideal hydrodynamics predicted $v_2(p_T)$ rises monotonically with $p_T$. 
The saturation at higher $p_T$ is is often explained as due to viscosity,
neglected in the discussion here.

Fig.~\ref{phot_v2_pt} shows the $p_T$ dependent $v_2$ for thermal 
photons~\cite{CFHS} for a typical impact 
parameter $b = 7$ fm. Contributions from quark matter (QM) and  hadronic 
matter (HM) along with the sum of the two are shown separately (left panel 
of Fig.~\ref{phot_v2_pt}). One can see that, $v_2(HM)$ rises monotonically 
with $p_T$, similar to the hadronic $v_2$ predicted by 
hydrodynamics. On the other hand, $v_2(QM)$ is very small at high $p_T$ or 
early times, as very little flow is generated by that time. $v_2(QM)$ 
rises for smaller values of $p_T$ and after attaining a peak value around 
$1.5 - 2.0$ GeV, it tends to 0 as $p_T$ $ \rightarrow 0$. The total flow or 
$v_2$(QM+HM) tracks the $v_2(QM)$ at high $p_T$ in-spite of very large 
$v_2(HM)$, as the yield of photons from hadronic phase is very small at 
high $p_T$. It is well known and mentioned earlier that, for photon 
energy  larger than the rest masses of the photon emitting  particles, 
the photon production cross-section peaks when the photon momentum and 
momentum of photon emitting particles become almost identical~\cite{wong}. 
Thus, at high $p_T$ photon $v_2$ reflects the anisotropies of the quarks 
and anti-quarks at early times. Also, the collision induced conversion 
of vector mesons $(\rho)$ starts dominating the photon production for 
$p_T \ge 0.4$ GeV, which gives rise to a structure at the transition 
point in the photon $v_2$ curve~\cite{CFHS}. As HM contribution dominates 
the $p_T$ spectrum and hydrodynamics is well applicable around that $p_T$ 
range, the structure is expected to survive in the experimental result also.

Recently Turbide {\it et al.}~\cite{TGFH} have shown that, total contribution 
to photon $v_2$ from prompt fragmentation (small $+ \ ve \ v_2$) and 
jets-conversion (small $- \ ve \ v_2$) photons is very small, almost 
equal to zero. Also, we know that the prompt photons from Compton and 
annihilation processes do not contribute to elliptic flow as their 
emission is not subjected to collectivity and has azimuthal symmetry. 
Thus, at low and intermediate 
$p_T$ range, $v_2$ from thermal photons plays a dominant role in deciding 
the nature of the direct photon $v_2$.
\begin{figure}[t]
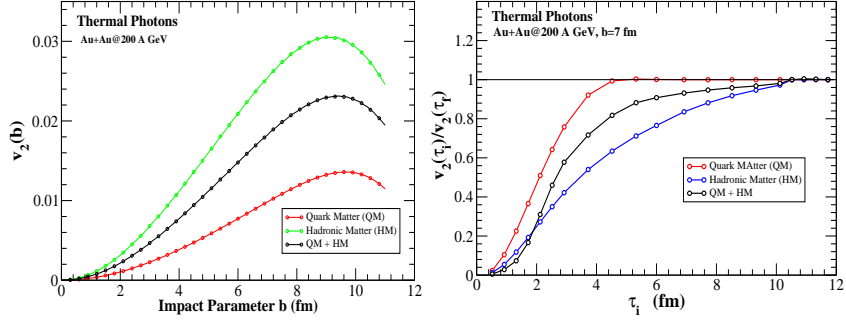

\centering
\includegraphics[height=4.2cm, width=5.50cm]{v_2_b.eps}
\includegraphics[height=4.2cm, width=5.50cm]{v_2_t.eps}
\caption{ Left panel: $p_T$ integrated elliptic flow of thermal photons 
from different phases at RHIC. Right panel: Time evolution of elliptic 
flow from different phases, normalized by the final value at the time of 
freeze-out [taken from Ref.~\cite{CHS}].}
\label{evolution}   
\end{figure}

For central collision of two spherical  nuclei, the spatial eccentricity 
($\epsilon$) of the overlapping zone is zero, thus the flow co-efficient 
$v_2$ is also zero. The $v_2(p_T)$ for thermal photons for different 
centrality bins are shown in the right panel of Fig.~\ref{phot_v2_pt}, 
where the value of $v_2$ rises from central towards peripheral collisions. 
Now, the elliptic flow as well as spatial eccentricity both rise with 
higher values of impact parameter. However, the ratio of the two remains 
independent of impact parameter up to a very large value of b~\cite{CHS}. 
The $p_T$ integrated $v_2$ for thermal photons from different phases as 
a function of collision centrality shows an interesting behavior (left 
panel of Fig.~\ref{evolution}). The $v_2(b)$ rises with b until 
for very peripheral collisions. For these the system size itself becomes 
very small to generate enough pressure gradient and elliptic flow, and 
as a result $v_2(b)$ decreases. The $p_T$ dependent as well as $p_T$ 
integrated time evolution results~\cite{CHS} of thermal photon spectra 
and elliptic flow show explicitly the gradual build up of spectra and 
$v_2$ with time very well. The  photon  $v_2$ from QGP phase saturates 
within about 4-5 fm/$c$ (right panel of Fig.\ref{evolution}), 
whereas $v_2$ from hadronic phase is not very significant at early 
times and it saturates much later. 

\subsection{Thermal Dilepton $v_2$}
\label{sec:4}

\begin{figure}
\centering
\includegraphics[height=4.2cm, width=5.50cm]{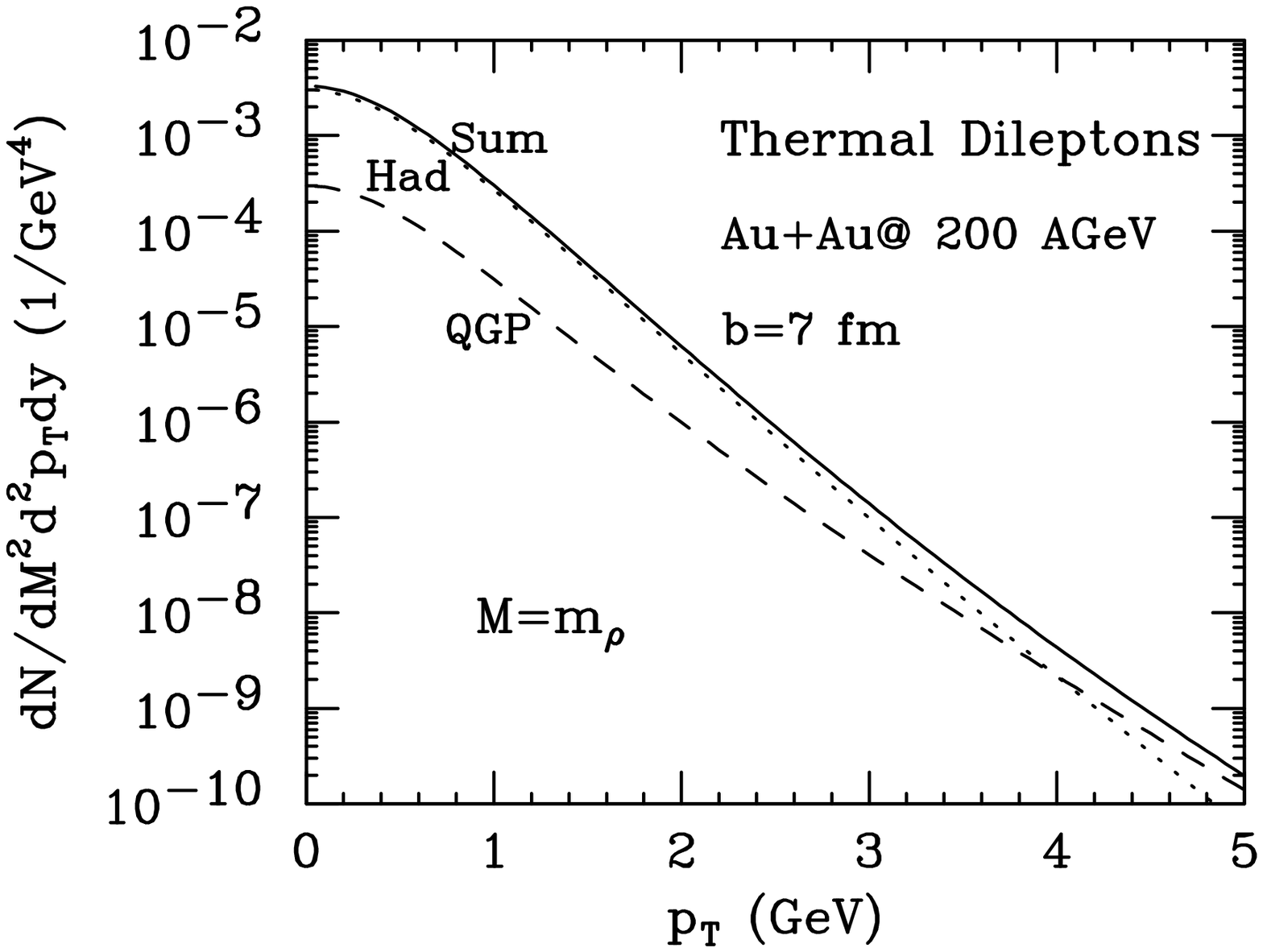}
\includegraphics[height=4.2cm, width=5.50cm]{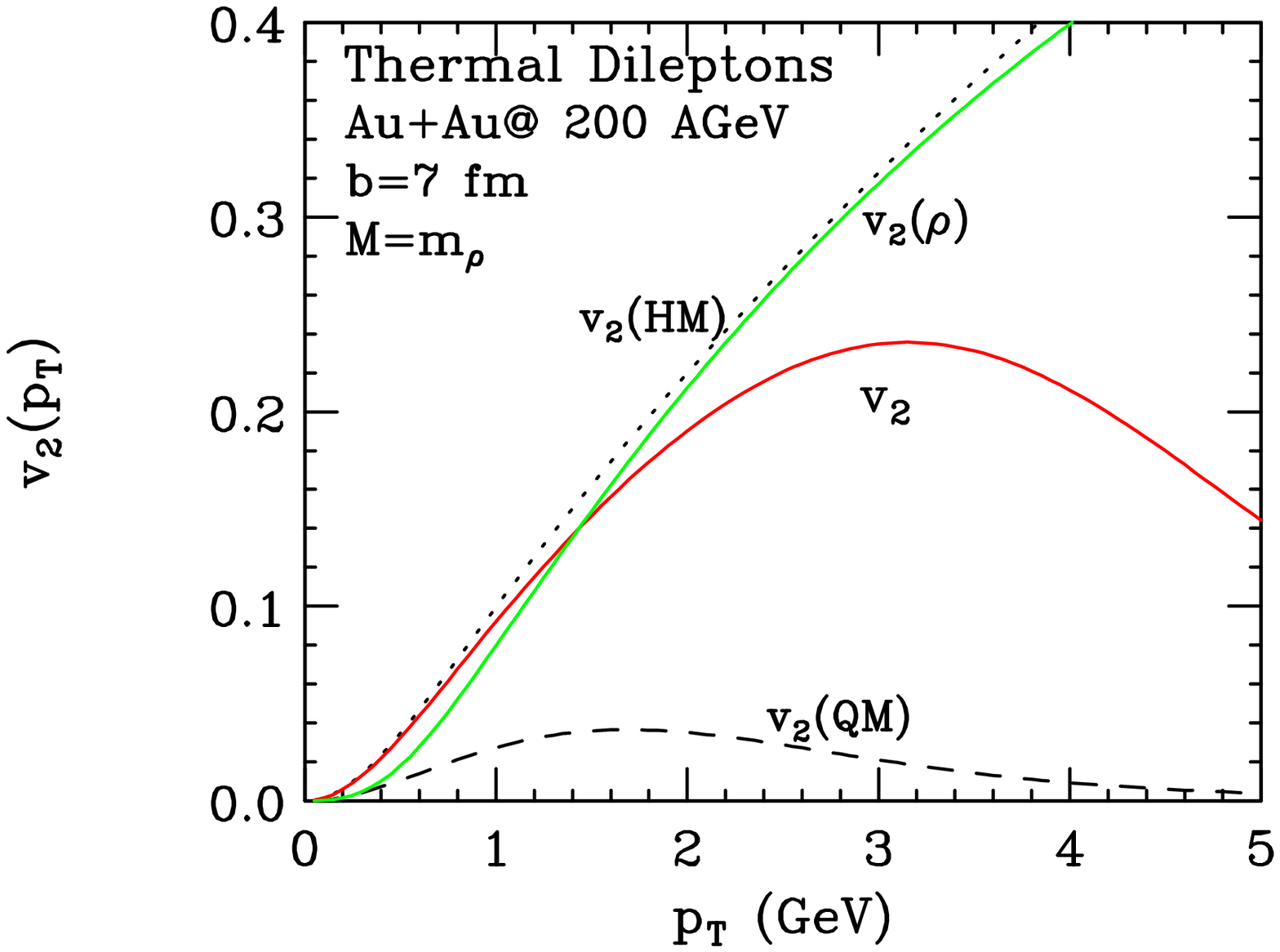}
\caption{$p_T$ spectra [let panel] and $v_2$ [right panel] for thermal 
dileptons at $M=m_{\rho}$. $v_2(\rho)$ is also plotted in the $v_2$ curve  
for comparison~\cite{dil}.}
\label{rho_dn_v2}   
\end{figure}

Elliptic flow  of thermal dileptons is another very interesting 
and illustrative observable which gives information of the different 
stages of heavy ion collision depending on invariant mass $M$ and 
transverse momentum $p_T$~\cite{dil}. At the $\rho$ and $\phi$ 
masses, the $p_T$ spectra and $v_2(p_T)$ show complete dominance 
of hadronic phase. At these resonance masses radiation from quark 
matter becomes significant only for very large $p_T ( \ge 4)$ GeV 
(left panel of Fig.~\ref{rho_dn_v2}).  The $v_2(p_T)$ at $M = m_\rho$  
also shows similar nature as spectra and the total $v_2$ is almost 
similar to hadronic $v_2$  upto a large $p_T$. Right panel of 
Fig.~\ref{rho_dn_v2} shows dilepton $v_2$ at rho mass along with 
$v_2$ of rho meson where, $v_2(HM)$ tracks $v_2(\rho)$. Also, the 
hadronic $v_2$ is little smaller than the $v_2(HM)$ for dileptons. 
The effective temperature of dilepton emission is a little larger 
than for hadrons, thus the spectra of the later is boosted by a somewhat 
larger radial flow~\cite{dil}, which results in smaller elliptic flow.
\begin{figure}
\centering
\includegraphics[height=4.2cm, width=5.50cm]{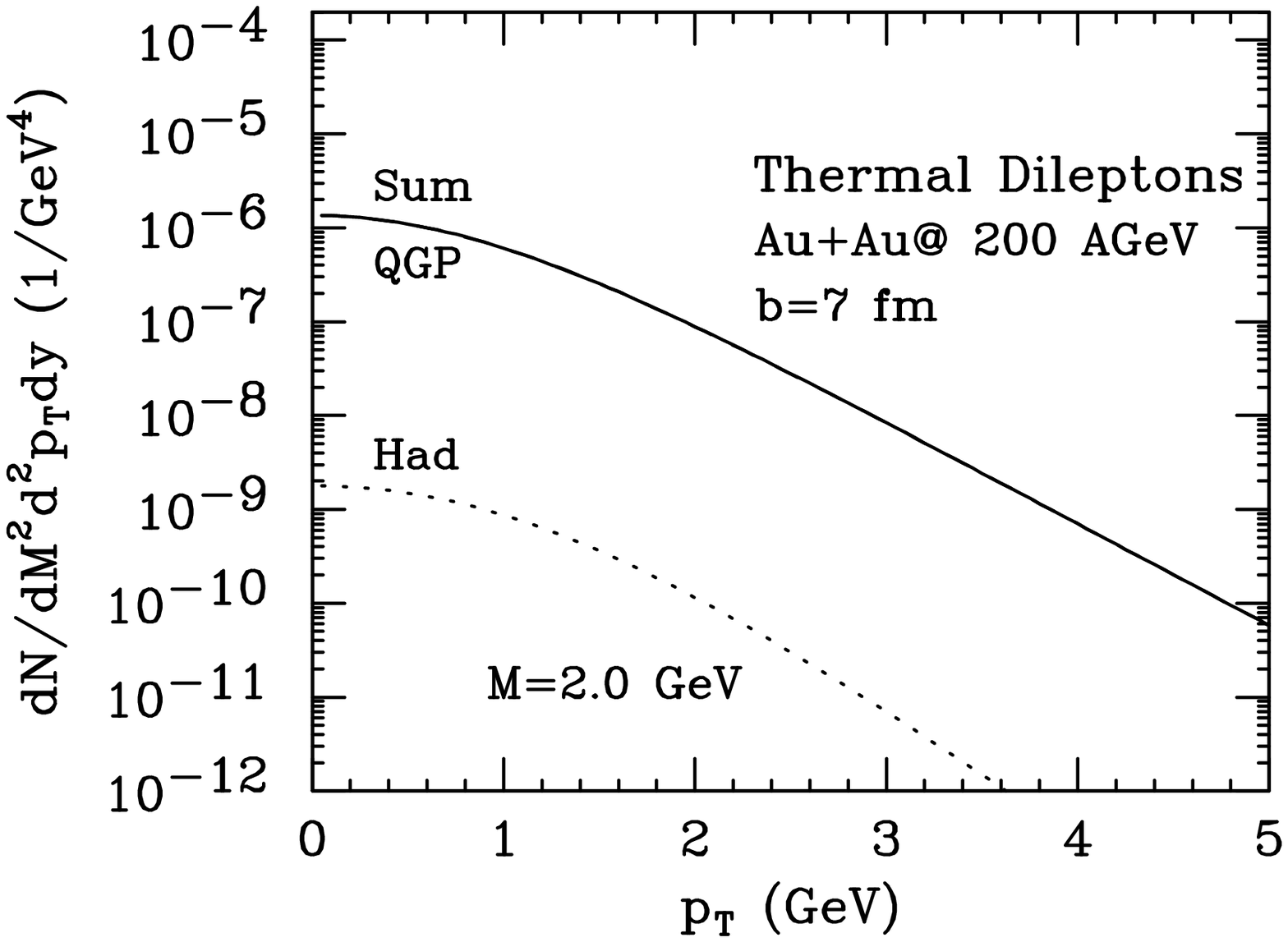}
\includegraphics[height=4.2cm, width=5.50cm]{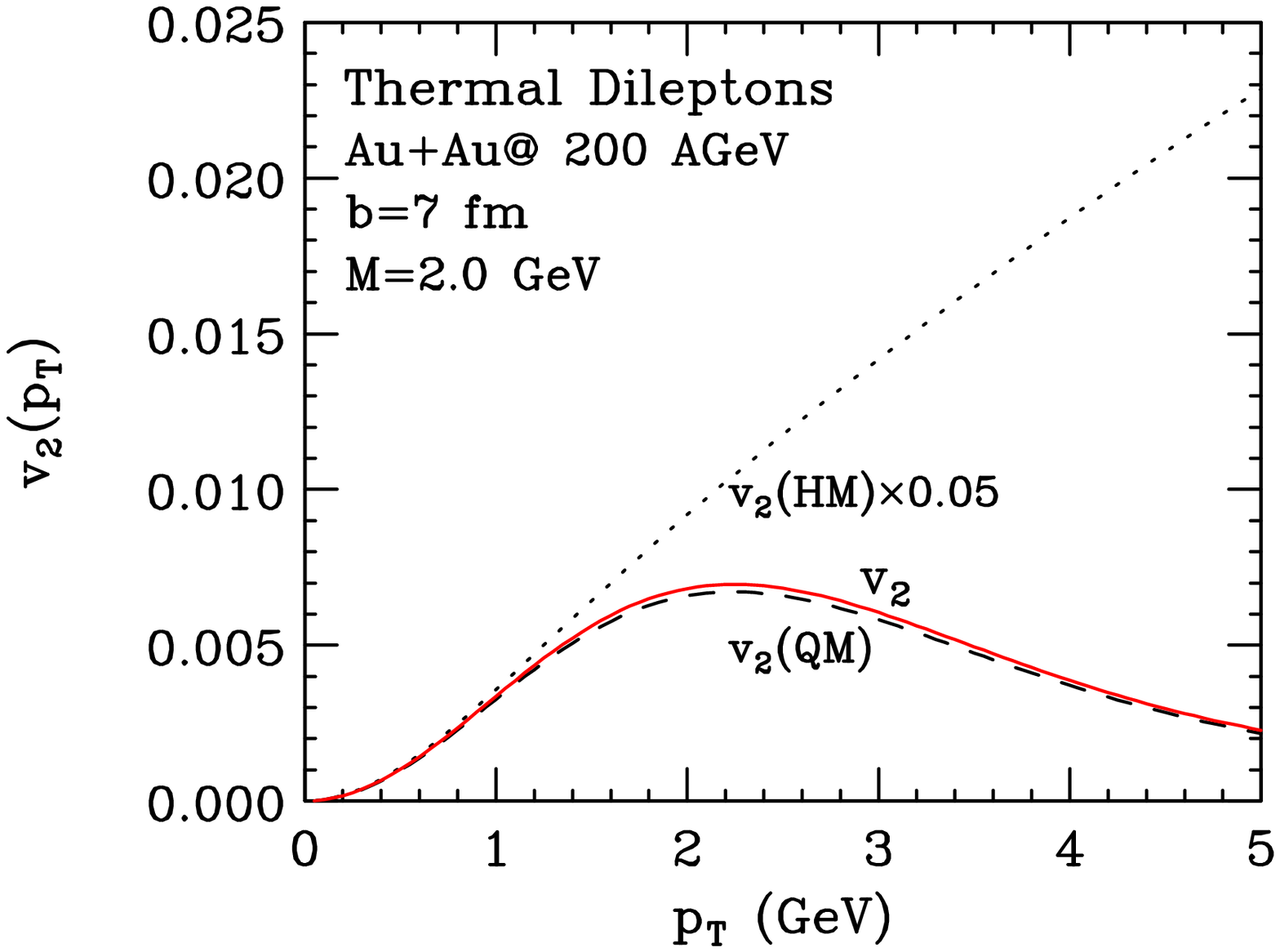}
\caption{$p_T$ spectra [let panel] and $v_2$ [right panel] for thermal 
dileptons at $M=2$ GeV~\cite{dil}.}
\label{2_dn_v2}    
\end{figure}

A totally different situation emerges for the $p_T$ spectrum 
and $v_2$ for dilepton invariant mass $M > M_{\phi}$. For $M = 2$ 
GeV, the dilepton $v_2$ is  similar to $v_2(QM)$ inspite of the 20 
times larger  $v_2(HM)$, as hadronic dileptons are very few compared 
to QGP radiation at this M value (Fig.~\ref{2_dn_v2}).

\begin{figure}
\centering
\includegraphics[height=4.2cm, width=5.50cm]{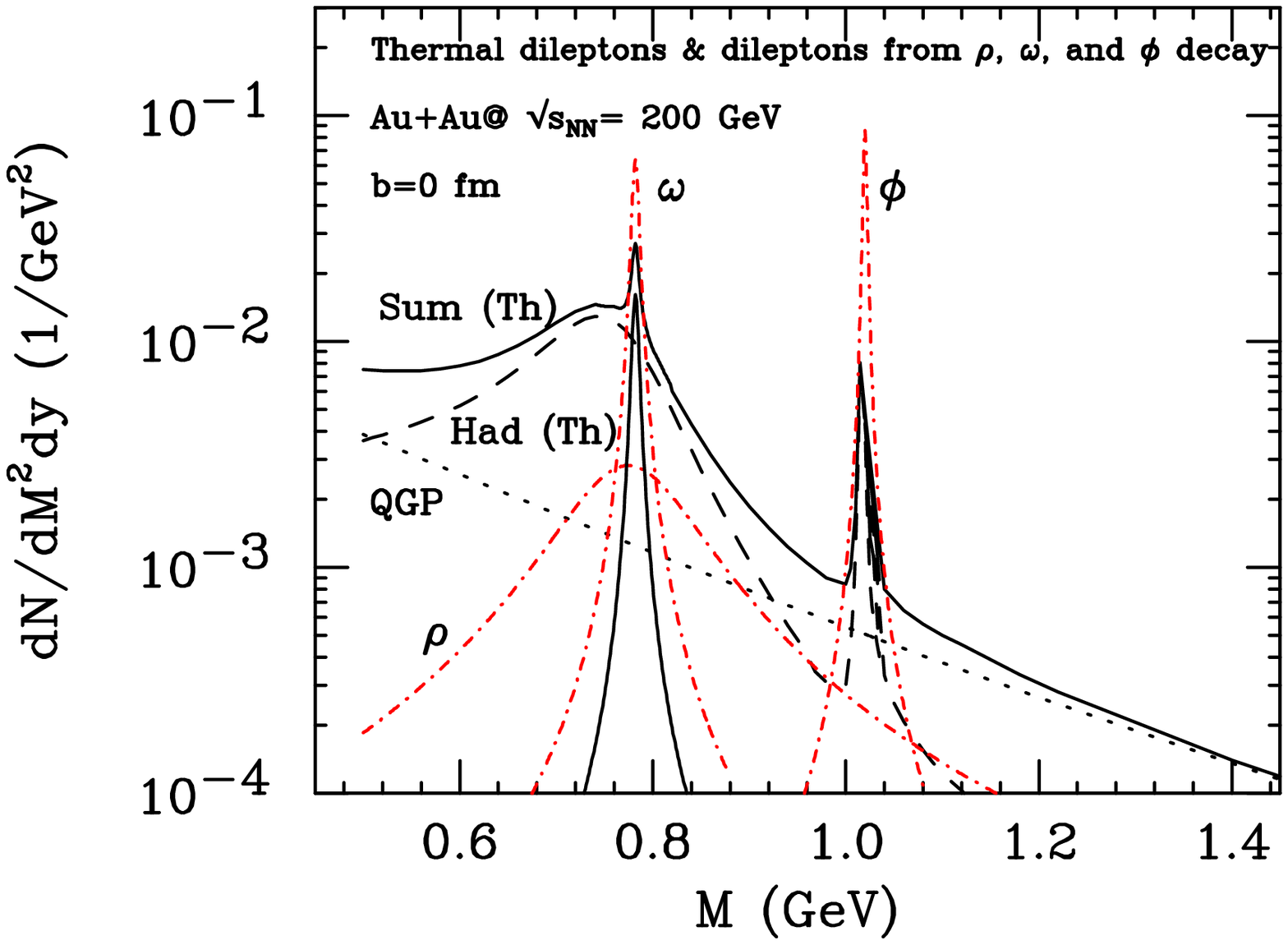}
\includegraphics[height=4.2cm, width=5.50cm]{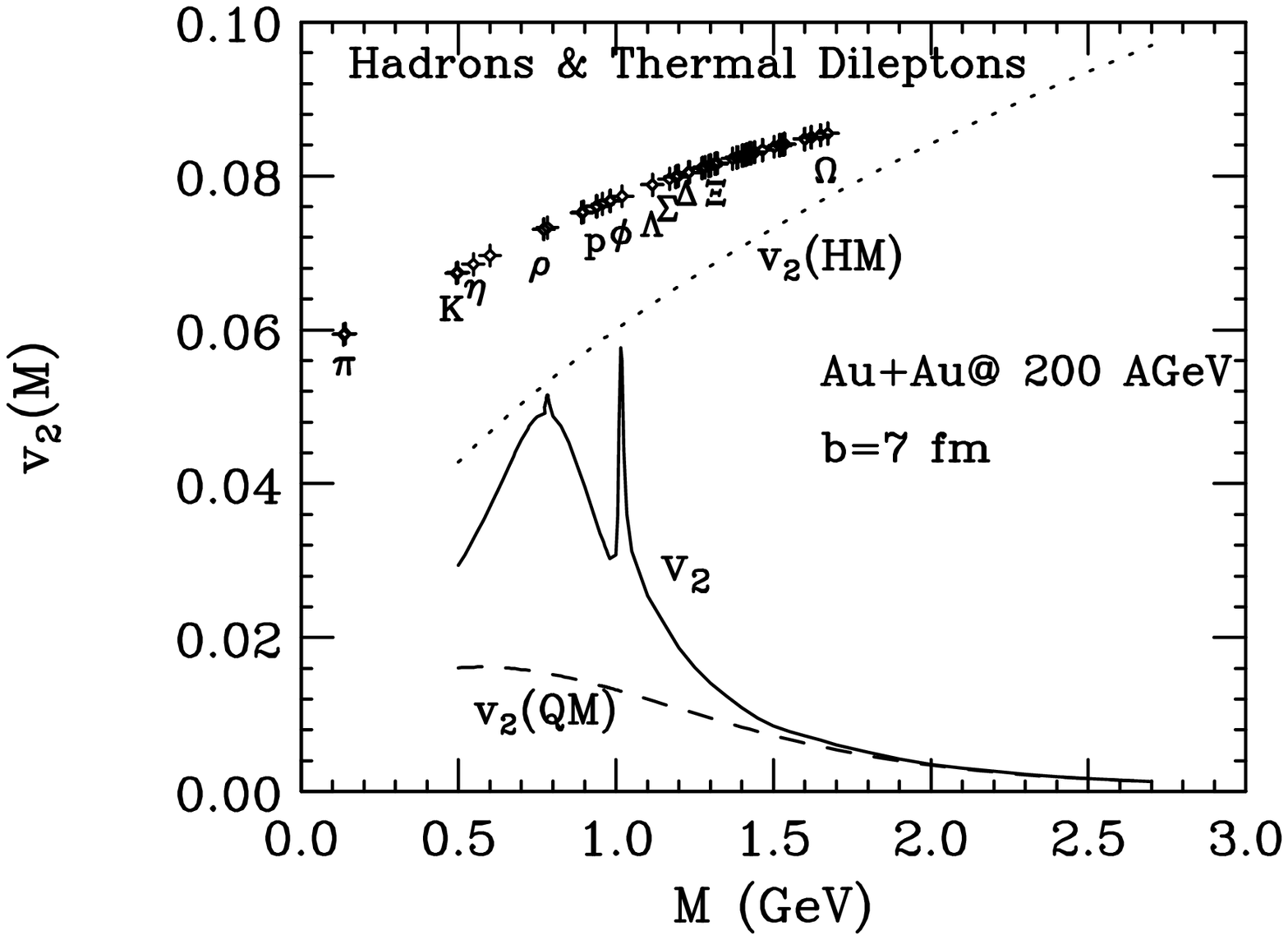}
\caption{Left panel: The mass spectra of thermal 
dileptons from a hydrodynamical simulation of central $200A$ GeV $Au+Au$ 
collision. Quark and hadronic matter contributions are shown 
separately. Right panel: $p_T$ integrated elliptic flow parameter 
for dileptons and various hadrons~\cite{dil}.}
\label{dil_spec_v2}     
\end{figure}

The $p_T$ integrated spectra and $v_2$ as a function of invariant mass 
$M$ show well defined peaks at the resonance masses ($\rho, \ \omega$ 
and $\phi$), as can be seen Fig.~\ref{dil_spec_v2}. For $M\le M_{\phi}$ 
the dilepton spectrum is totally dominated by hadronic phase and above 
that it is dominated by contribution from QGP. $v_2(QM)$ is very small 
at large $M$ and 
rises for smaller $M$. It shows a  nature similar to the thermal photon $v_2$. 
The $M$ dependent $v_2(HM)$ is significant only at the resonance masses 
and beyond $\phi$ mass its contribution to total $v_2$ is negligible. 
Thus, for large  values of $M$ or at early times flow comes from QGP 
phase and its value is very small, whereas at later times or at low 
$M$ values it is  from HM, which is very strong. The measurement of 
flow parameter at high $M$ and/or high $p_T$ values can be very useful 
to reveal a pure QGP signature. 

\subsection{Elliptic flow of decay photons}

We have already discussed that most of the produced photons in  heavy 
ion collisions are from the 2-$\gamma$ decay of $\pi^0$ and $\eta$ mesons 
and the direct photons contain a small fraction of the  inclusive 
photon spectrum. Thus it is very interesting to know the nature of 
$v_2$ of decay photons from pion and $\eta$ decays. The momentum 
distribution of decay photons from $\pi^0$ decay in an invariant 
form can be expressed as~\cite{cahn},
\begin{equation}
k_0\frac{dN}{d^3k}(p, k)=\frac{1}{\pi}\, \delta(p \cdot k-\frac{1}{2}
m^2)\, ,
\label{dec}
\end{equation}
\begin{figure}
\centering
\includegraphics[height=4.2cm, width=5.5cm]{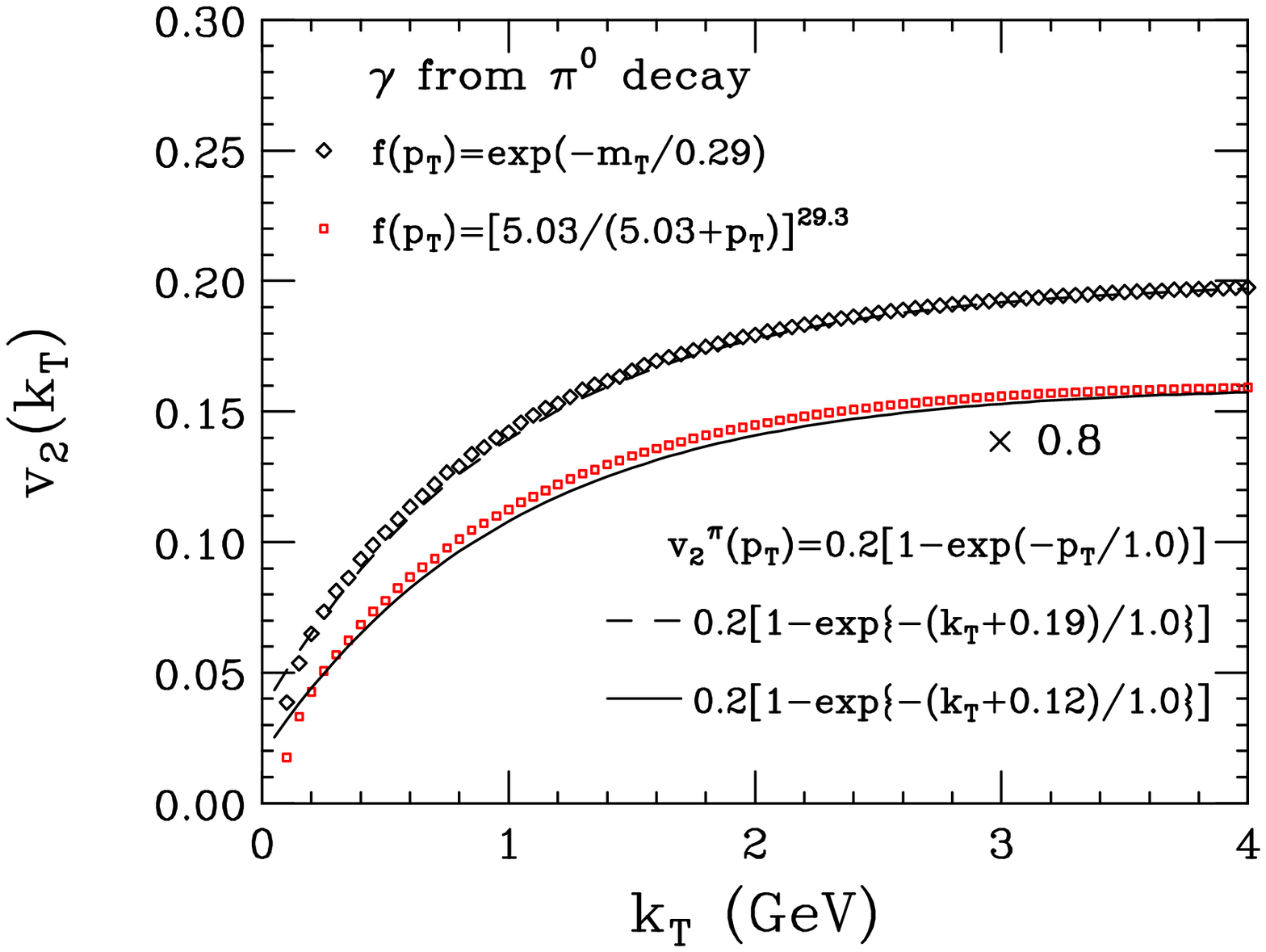}
\includegraphics[height=4.2cm, width=5.5cm]{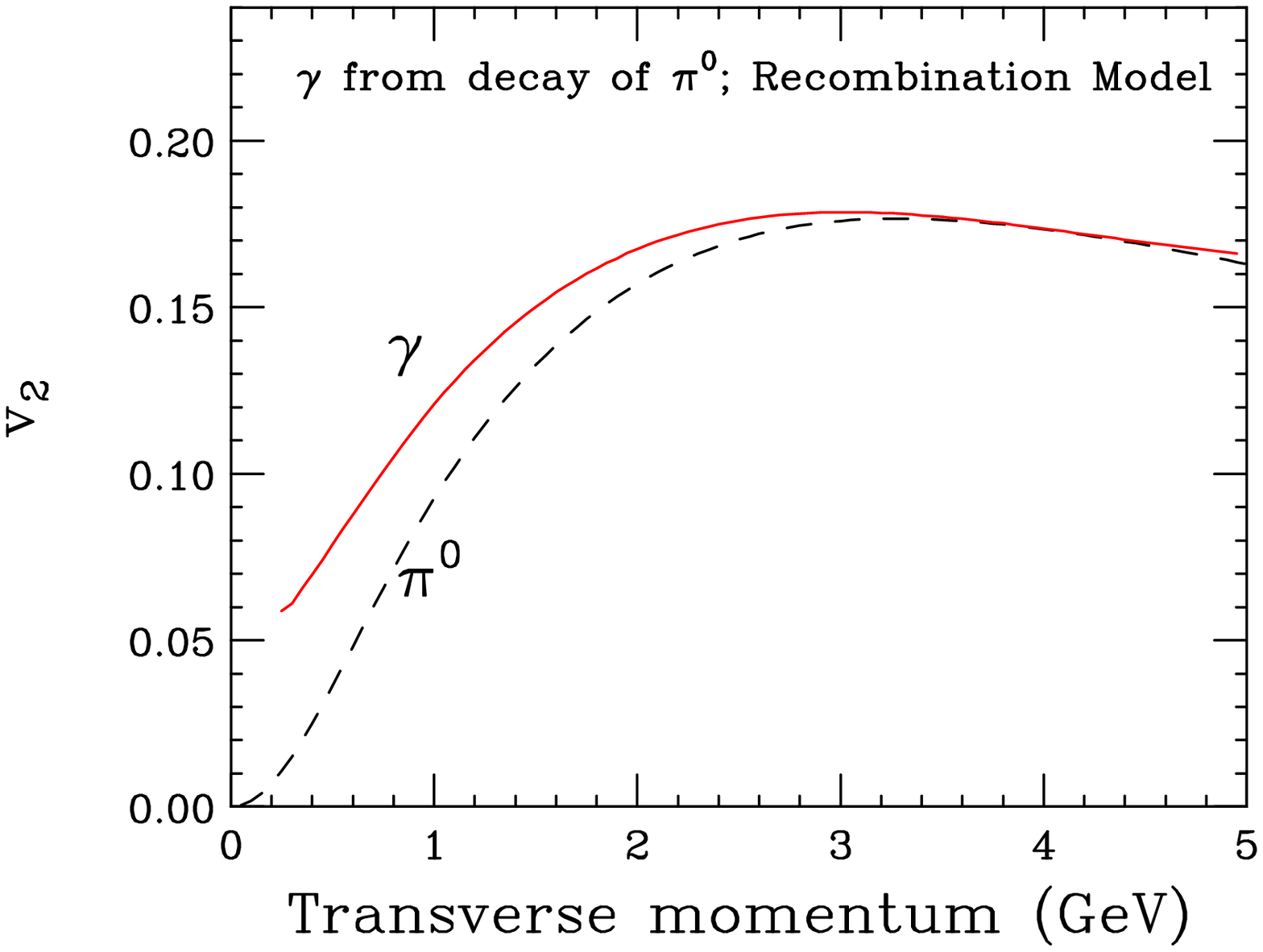}
\caption{Left panel: Elliptic flow of photons from $\pi^0$ decay at 
mid-rapidity. $f(p_T)$ stands for the momentum distribution of the 
$\pi^0$. Right panel: Elliptic flow parameters for photons from 
decay of $\pi^0$ obtained using the recombination model [taken from 
Ref.~\cite{decay}].}
\label{layek}
\end{figure}
where p and k are the 4-momentum of the pion and photons and m is 
the pion mass. Thus the Lorentz invariant cross-section of the decay 
photons using the decay kinematics from the Eqn.~(\ref{dec}) is,
\begin{equation}
k_0\frac{d\sigma}{d^3k}=\int \frac{d^3p}{E}\left(E \frac{d\sigma}{d^3p} 
\right)\frac{1}{\pi} \delta (p \cdot k -\frac{1}{2} m^2) \, .
\label{dsig}
\end{equation}
Layek {\it et al.}~\cite{decay} have calculated elliptic flow of 
decay photons considering several azimuthally asymmetric pion 
distributions (Fig.~\ref{layek}). They have shown that $k_T$ 
dependent $v_2$ of decay photons closely follows the $v_2(p_T)$ 
of $\pi^0$ evaluated at $p_T \sim k_T + \delta$ (where $\delta \sim 
0.1 - 0.2$ GeV). Similar results were obtained for decay photons 
from $\eta$ mesons also. This study could be useful in identifying 
additional sources of photons as the $v_2$ of $\pi^0$ is similar 
to that of $\pi^+$ and $\pi^-$. Also by using the property of 
quark number scaling or the recombination model, the decay photon 
$v_2$ can help to estimate the $v_2$ of constituent partons in the 
$\pi^0$ or $\eta$ mesons (right panel of Fig.~\ref{layek}). 
See Ref.~\cite{decay} for details.

\subsubsection{Experimental measurement of direct photon $v_2$}

PHENIX has measured direct photons and its $v_2$ by subtracting 
$v_2$ of decay photons (2-$\gamma$ decay of $\pi^0$ and $\eta$ 
mesons) from inclusive photon $v_2$ using appropriate weight 
factor~\cite{Sakaguchi}. The procedure followed by them to 
estimate direct photon $v_2$ is as follows:
\begin{equation}
v_2^{\rm dir.} = \frac {R \times v_2^{ \rm incl.} - v_2^{\rm bkgd.}} {R - 1}
\end{equation}
where R is the direct photon excess over hadron decays defined as,
\begin{equation}
R= \frac {(\gamma/\pi^0)_{\rm incl.}} {(\gamma/\pi^0)_{\rm bkgd.}} \nonumber
\end{equation}
\begin{figure}
\centering
\includegraphics[height=7.5cm, width=10.50cm]{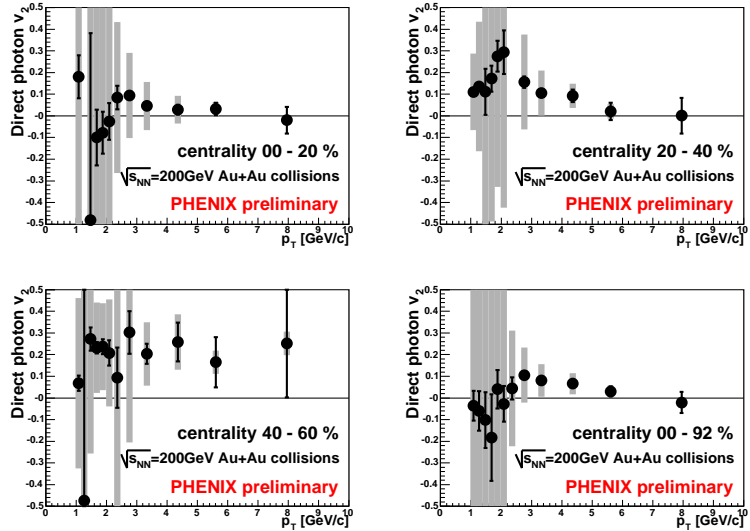}
\caption{ Elliptic Flow of direct photons from $200 A$ GeV $Au+Au$ 
collision at RHIC~\cite{miki}.}
\label{phenix_v2}
\end{figure}
The factor $R$ is measured from spectral analysis. The inclusive and 
hadron decay $v_2$ are measured using reaction plane method where the 
background photons are measured using a Monte Carlo simulation. We have 
already seen that the photon $v_2$ shows different signs and/or 
magnitude depending on the production procedure of photons in heavy 
ion collisions. Thus experimental measurement of photon $v_2$ can 
be a very powerful tool to disentangle the intermix contributions 
from different sources in different $p_T$ ranges. However, the 
experimentally measured preliminary PHENIX data still contain a large 
systematic error (Fig.~\ref{phenix_v2}), which need to be reduced 
before making any specific conclusion about direct photon $v_2$. 

\section{Photon tagged jets}

Jets are hard phenomenon characterized by a large momentum transfer 
between partons and are characterized by several hadrons in a small 
angle around a leading particle as a result of jet fragmentation. 
The hadrons surrounding a jet are known as associated particles of the jet. 

One of the key observables at RHIC energy is the strong suppression of  
leading hadron yield at high values of transverse momentum compared to 
expectations based on $p+p$ or $d+A$ collisions at same collision energy. 
This is the celebrated phenomenon of `jet quenching'. It is assumed that the 
hard partons lose a large fraction of their energy when passing 
through the strongly interacting medium. A variety of qualitatively
different models are available  using collisional energy loss and 
radiative energy loss mechanisms to describe the nuclear modification 
factor $R_{AA}$, defined as,
\begin{equation}
R_{AA}(p_T,y)= \frac {d^2N_{AA}/dp_T dy} {T_{AA}(b) d^2 \sigma^{NN}/dp_T dy}
\end{equation}
Here, $T_{AA}$ is the nuclear overlap function and $\sigma_{NN}$ is 
the nucleon-nucleon cross-section. The major problem for studying 
jet properties arises from the fact that, conventional calorimetric 
study can not measure the jet energy loss very accurately. Also it 
is very difficult to directly measure the modification of jet 
fragmentation function and jet production cross-section.

The `jet quenching' in heavy ion collisions can be studied by 
measuring the $p_T$ distributions of charged hadrons in opposite 
direction of a tagged direct photon. Quark-gluon Compton scattering 
and quark anti-quark annihilation process are the main mechanisms 
by which direct photons are produced at very high $p_T$, and jets 
are produced in the opposite direction of these photons. From 
momentum conservation law, the initial transverse energy of the 
produced jets are equal to that of the produced photons, i.e, 
$E_{\gamma} = E_{\rm jet}$. Thus uncertainties regarding the jet 
cross-section can be avoided by tagging a direct photon in the 
direction opposite to the jet (see. Ref.~\cite{WHS}).

Medium modified parton fragmentation function $D_{h/a}(z)$ 
( z is fractional momenta of the hadrons) 
is used to study the jet energy loss. The differential $p_T$ 
distribution of hadrons from jet fragmentation in the kinematical 
region $(\Delta\phi,\Delta y$) using $D_{h/a}(z)$ can be written as:
\begin{equation}
\frac{dN^{\rm jet}_{\rm ch}} {dyd^2p_T} = \sum_{r,h} r_a(E_T^{\gamma}) 
{\frac {D_{h/a}(p_T/E_T)} {p_T E_T}} {\frac {C(\Delta y \Delta \phi)} 
{\Delta y \Delta \phi}},
\end{equation}
where, $ C(\Delta y \Delta \phi)=\int_{|y| \le \Delta y/2} dy 
\int_{|\phi -\bar{\phi}_\gamma| \le \Delta \phi/2} d \phi  
f ( y -|\phi -\bar{\phi}_\gamma|)$  is an overall factor and 
$f(y,\phi)$ is the hadron profile around the jet axis and 
$r_a(E_T^{\gamma})$ is the fractional production cross section 
of a typical jet associated with the direct photons. Thus, comparison 
between extracted fragmentation function in $AA$ and $pp$ collisions 
can be used to determine the jet energy loss as well as the 
interaction mean free path in the dense matter produced in high 
energy heavy ion collisions. Thorston Renk~\cite{Renk_jet} has 
shown that by gamma hadron correlation measurement, the averaged 
probability distributions for quarks are accessible experimentally 
and he has also explained an analysis procedure capable of 
distinguishing between different energy loss scenarios leading 
to the same nuclear suppression factor. 
\begin{figure}
\centering
\includegraphics[height=4.2cm, width=5.50cm]{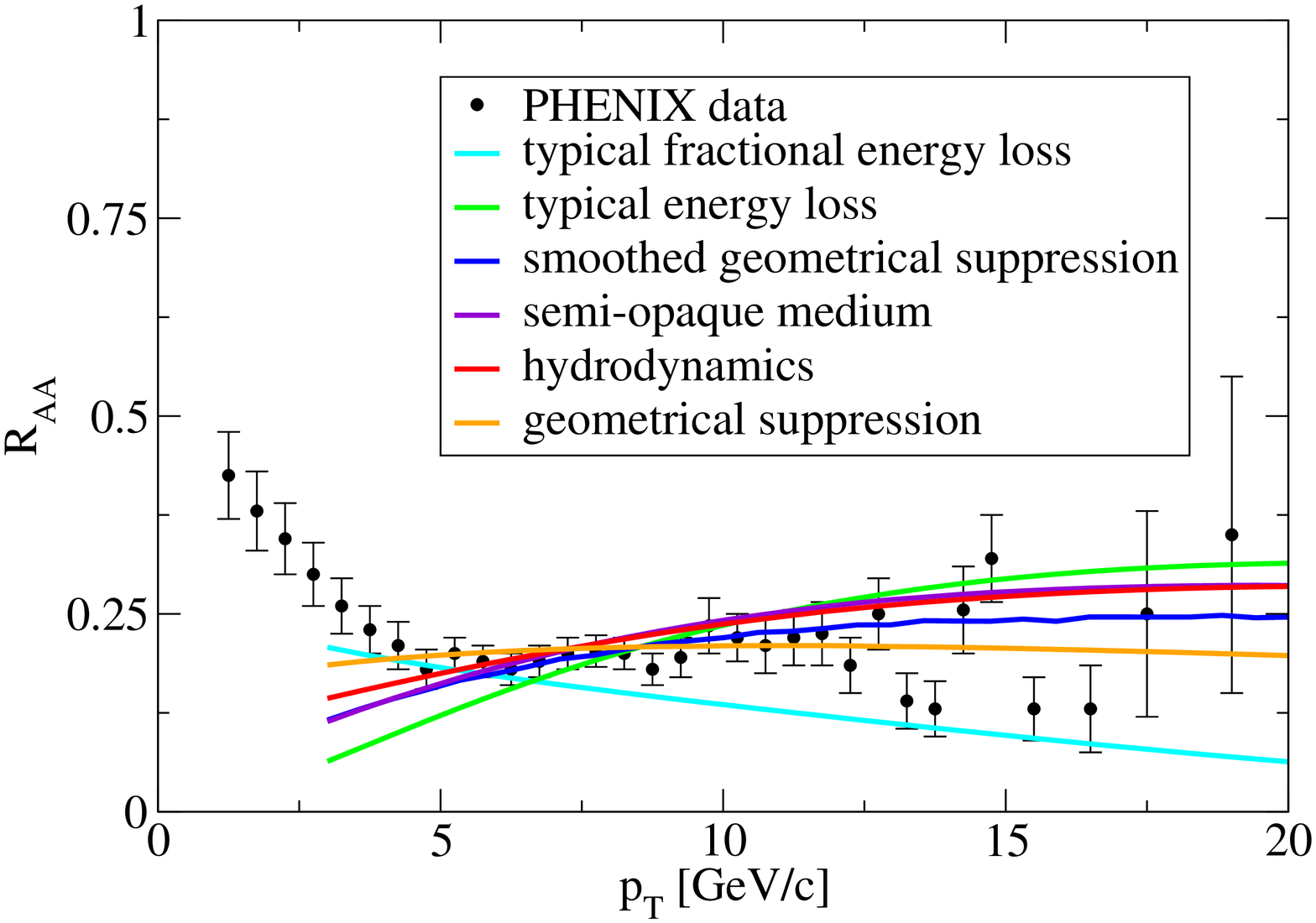}
\includegraphics[height=4.2cm, width=5.50cm]{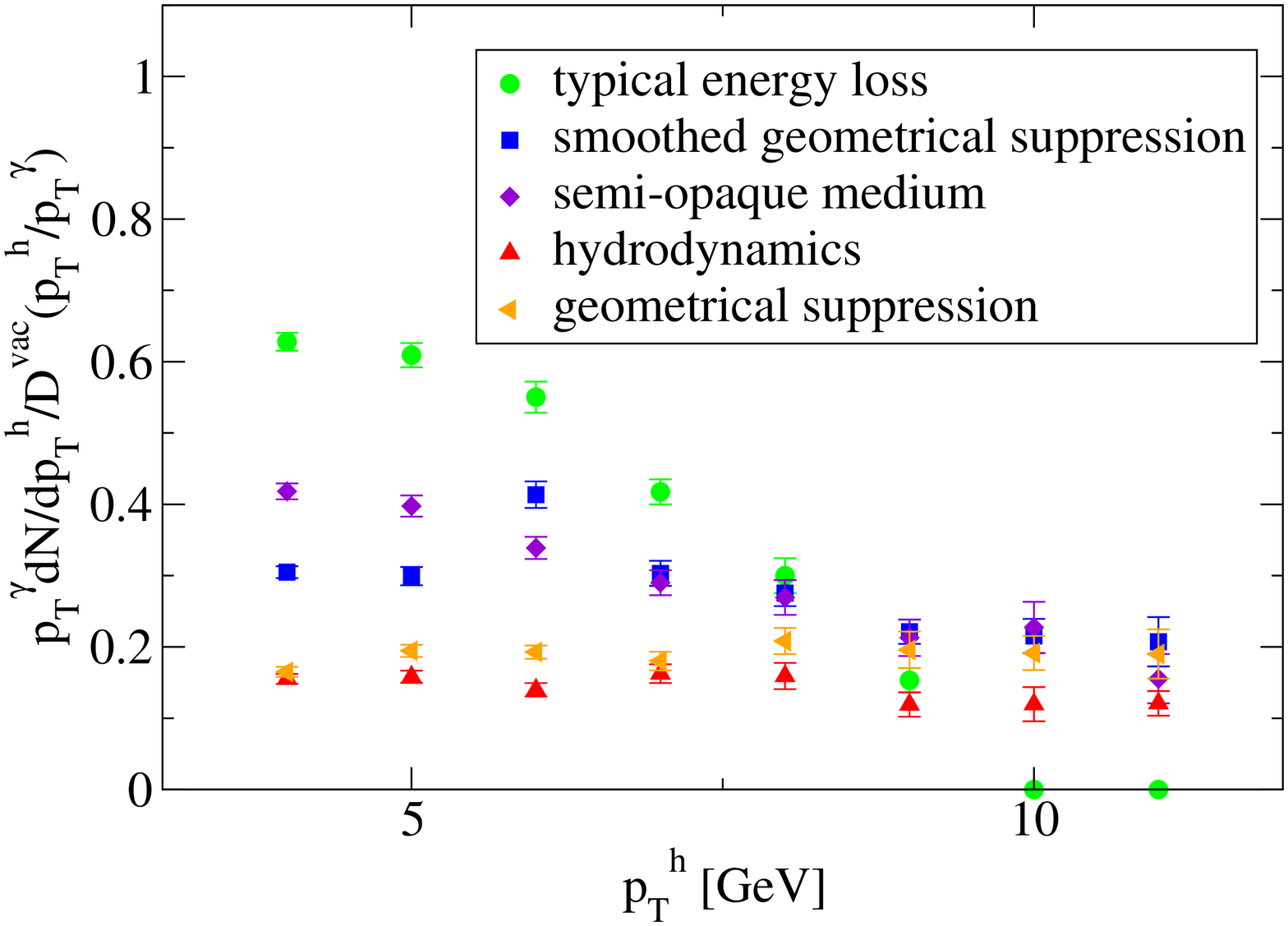}
\caption{ Left panel: Nuclear suppression factor $R_{AA}$ for 
different toy models and hydrodynamical simulation compared with 
PHENIX data. Right panel: Momentum spectrum of hard hadrons 
correlated back to back with a photon trigger normalized to the 
expectation of geometrical absorption (taken from 
Ref~\cite{Renk_jet}).}
\label{renk_1_2} 
\end{figure}
\subsubsection{Isolating the bremsstrahlung photons}

In order to measure direct photon cross section by suppressing the 
hadron decay background or in particular to isolate bremsstrahlung 
photons from accompanying hadrons, an interesting method known as `isolation 
cut' can be used successfully. The basic assumption for performing the 
cut is that the hadronic energy in a cone around the photon is less 
than a certain fraction of the photon energy, i.e, a photon is 
considered as isolated if the combined energy of the accompanying 
hadrons is less than $\epsilon E_\gamma$ ( where, $E_\gamma $ is 
the photon energy) inside a cone of half opening angle $\delta$ 
around the photon. The parameter $\epsilon$ is very small ( $\sim 
0.1$) and is called the energy resolution parameter. The cone around 
the tagged photon is known as isolation cone. The cone opening can 
be related to the radius $R$ of a circle centered around the photon 
in the centre of mass system, where R is defined in terms of rapidity 
$\eta$ and azimuthal angle $\phi$ as,
\begin{equation}
R \ge \sqrt { {\Delta \eta}^2 + {\Delta \phi}^2}
\end{equation}
For small rapidities, the half opening angle $\delta$ equals the 
radius $R$. This method was successfully implemented into the 
theoretical study of isolated prompt photon production considering 
fragmentation contribution also in next-to-leading order (NLO) by Gordon 
and Vogelsang~\cite{Vogel}. A very good accuracy of this method over 
a wide range of isolation parameters  for prompt photon production 
was demonstrated in their calculation. Results from RHIC and LHC for this 
would be very valuable.

\begin{figure}
\centering
\includegraphics[height=4.2cm, width=5.50cm]{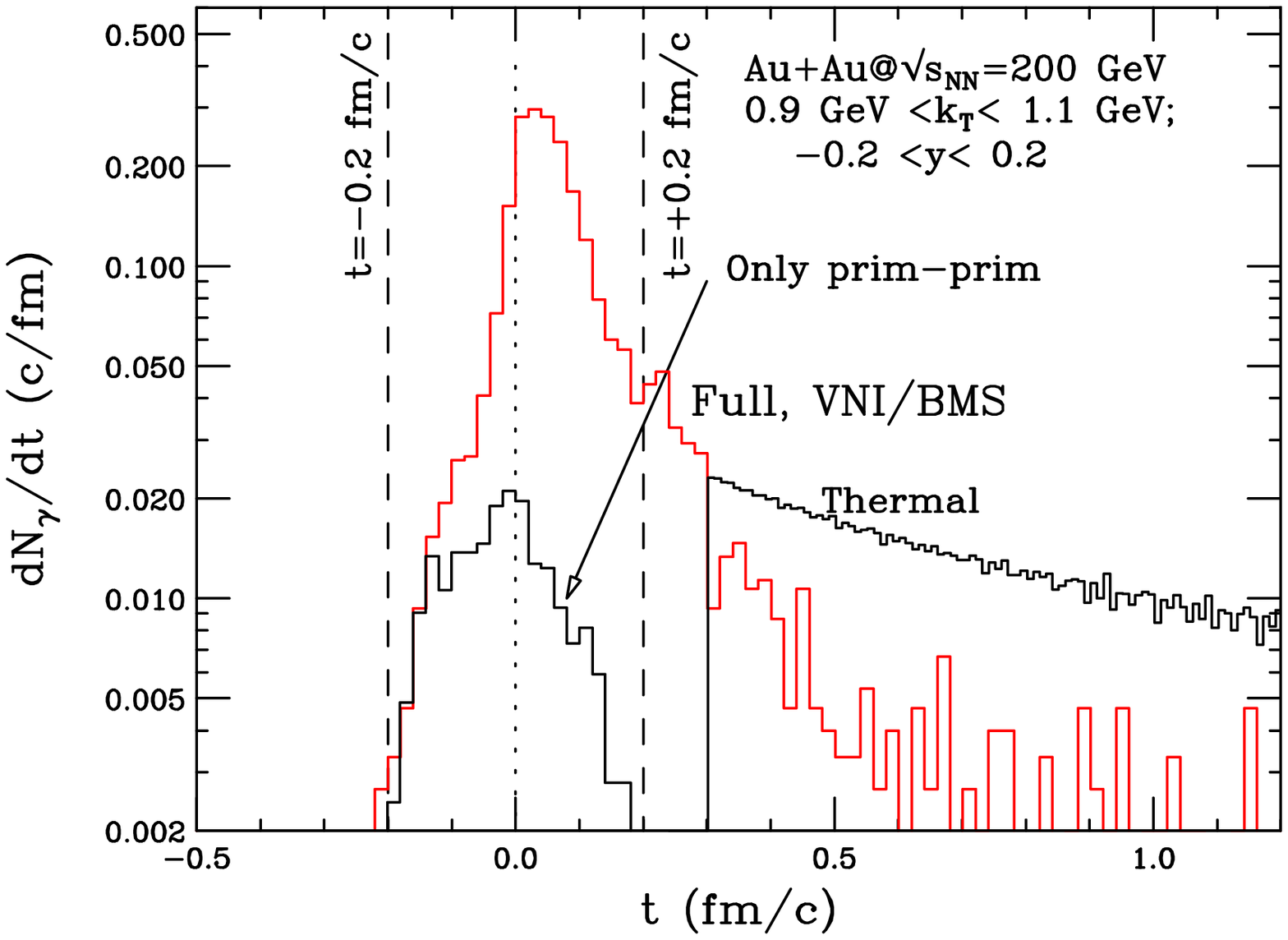}
\includegraphics[height=4.2cm, width=5.50cm]{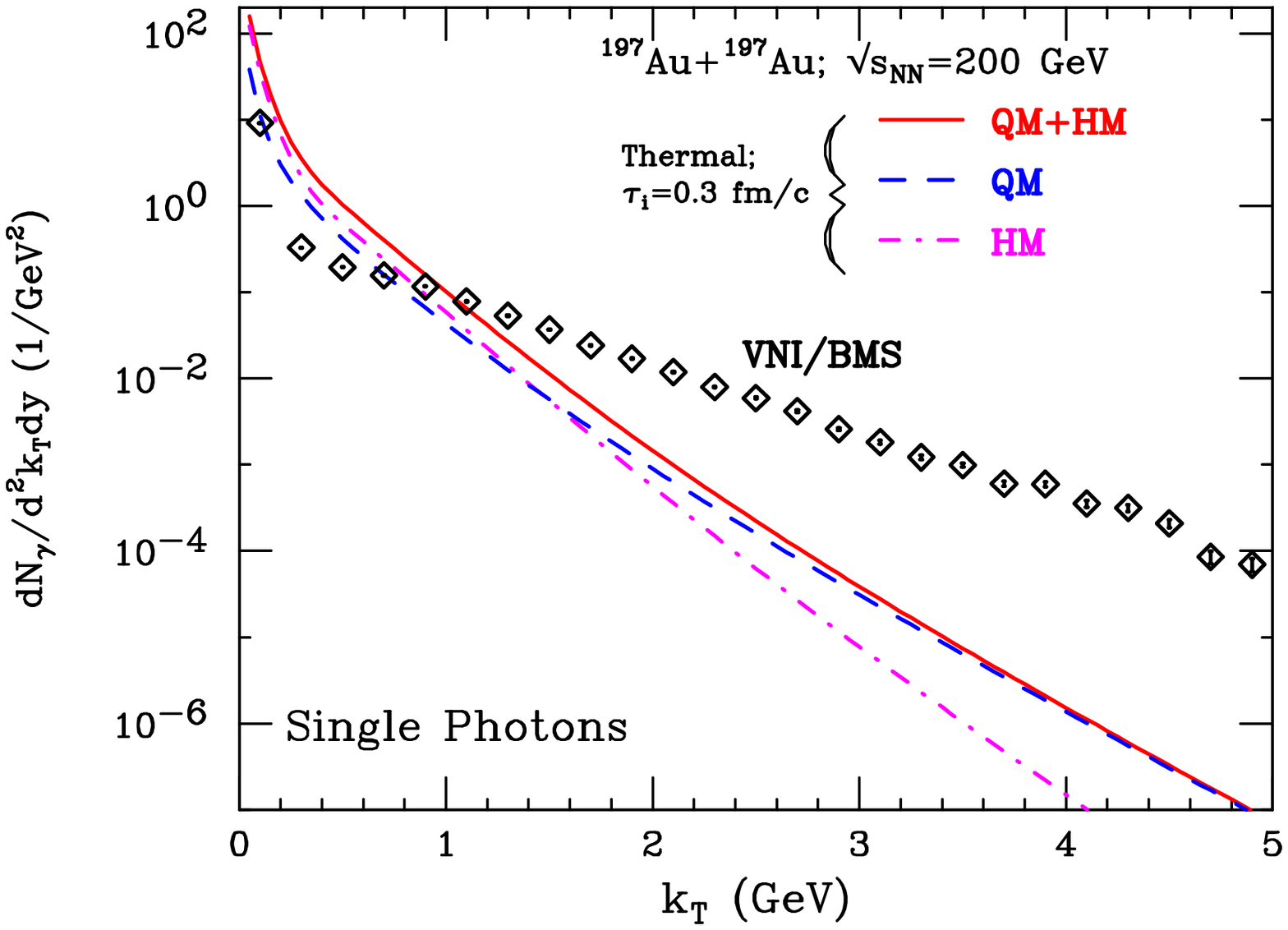}
\caption{ Left panel: The production rate (per event) of hard photons 
in a central collision of gold nuclei at $\sqrt {s_{NN}} = 200$ GeV 
as a function of time in the centre of mass system. Right panel:
Spectrum of photons from various sources. QM and HM denote quark 
matter and hadronic matter contribution respectively  (figures 
are from Ref.~\cite{prl_inter}.}
\label{interfero} 
\end{figure}

\subsubsection{Dilepton tagged jets }

In the study of photon tagged jets, the main problem arises from the 
jet pair production background, where a leading $\pi^0$ in the jet is 
misidentified as a photon. For an event having a huge background 
`isolation cut' is not a very efficient mechanism in the low $p_T$ 
range to study photon tagged jets in heavy ion collisions and it is 
useful only for $pp$ collisions. In the high $p_T$ range although the 
background contribution related problems are reduced, a substantial 
problem in photon isolation is created by small opening angle. On 
the contrary jets tagged by dileptons are not affected 
by background and can be used to observe $p_T$ imbalance, a signal of 
medium induced partonic energy loss. As we have mentioned earlier, 
for dileptons not only $p_T$ but also invariant mass $M$ is another 
equally important parameter which can be used accordingly to study 
dilepton tagged jets in the medium. At high $p_T$ and high $M$, the 
dilepton yield is much lower compared to low $p_T$ and low $M$ range, 
however the relative background contribution is also lower in that 
range. 

At very high transverse momentum Drell-Yan process ($h_1 + h_2 
\rightarrow l^+ l^- +X$) dominates the dilepton production from QGP 
phase. Srivastava {\it et al.}~\cite{dil_tagged} have estimated the 
results for dilepton tagged jets by studying  
Drell-Yan process at NLO in relativistic heavy ion collisions at 
RHIC and LHC energies. They have also shown that correlated charm and 
bottom decay are unimportant as background for dileptons having large 
transverse momentum or in the kinematical region of interest for jet 
quenching. Lokhtin {\it et al.}~\cite{lok} have studied the dimuon + 
jet production (including both $\gamma^*/Z \rightarrow \mu^+ \mu^- $ 
modes) at LHC energy and have shown the $p_T$ imbalance between 
$\mu^+ \mu^-$ pair and a leading particle in a jet is clearly visible 
even for moderate energy loss. It is directly related to absolute value 
of partonic energy loss and almost insensitive to the angular spectrum 
of emitted gluons and to experimental jet energy resolution.

\begin{figure}
\centering
\includegraphics[height=8.2cm, width=5.50cm]{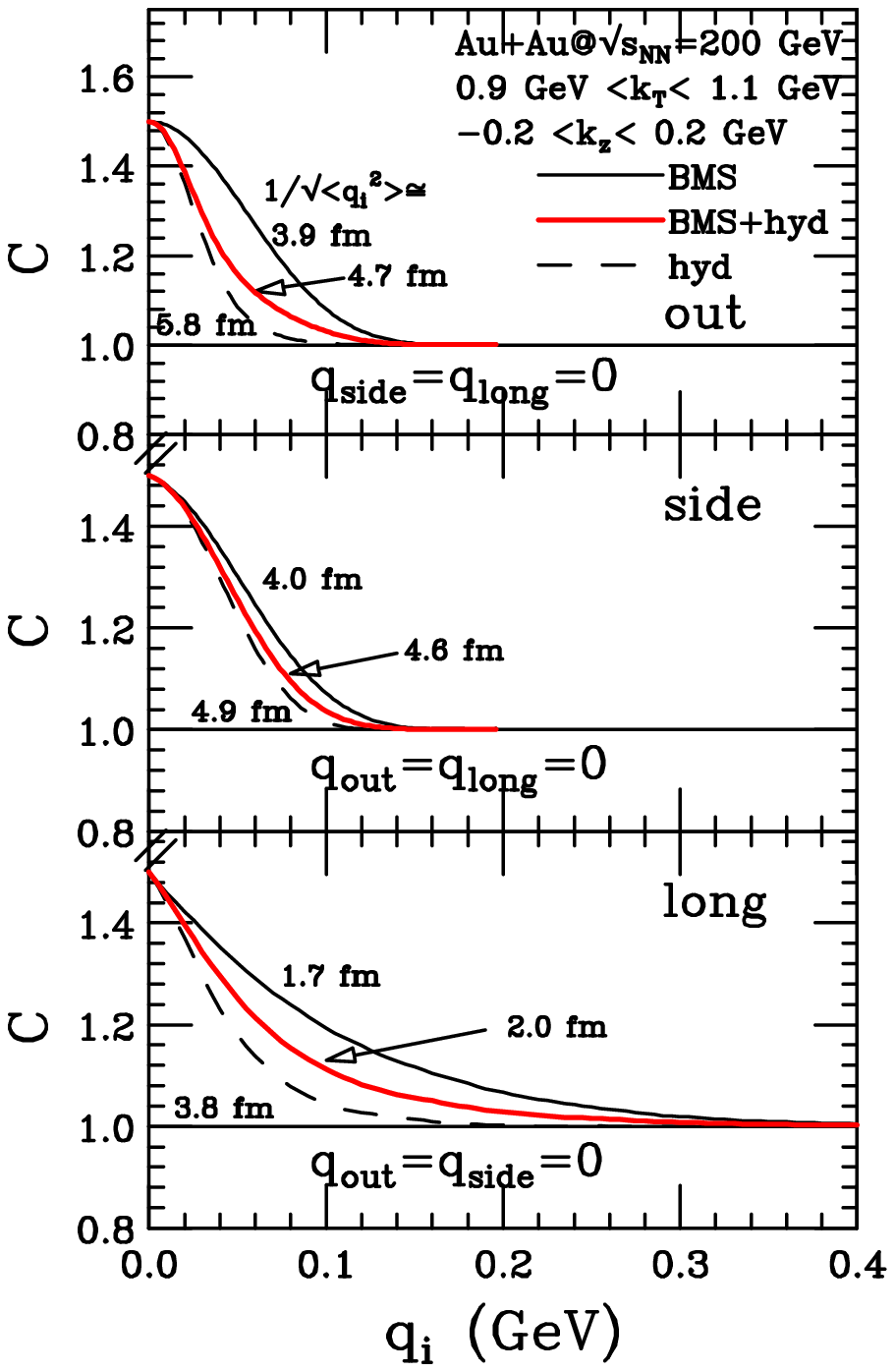}
\includegraphics[height=8.2cm, width=5.50cm]{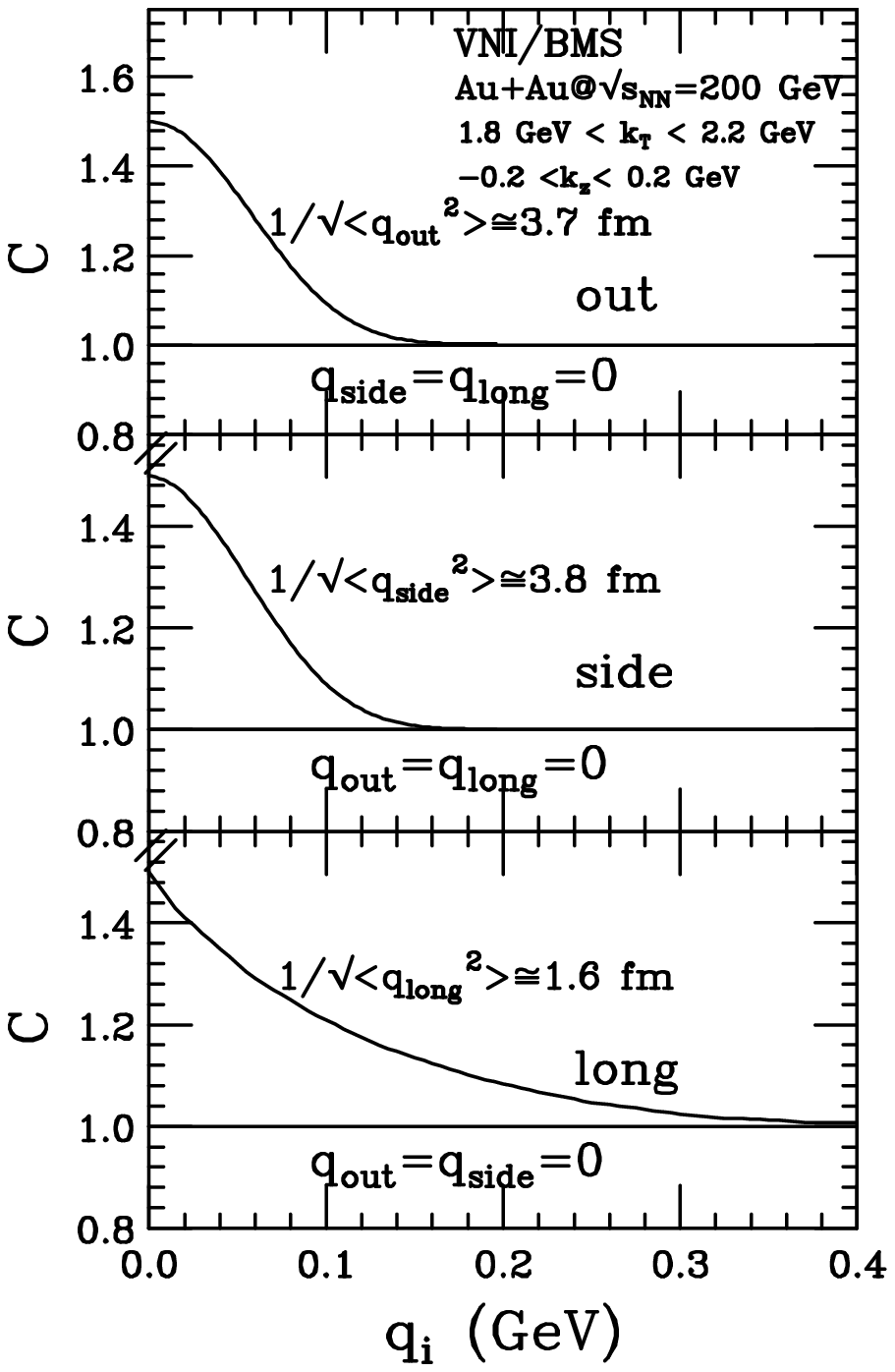}
\caption{ Left panel: The outward, side-ward and longitudinal 
correlations of direct photon predicted by PCM at $K_T = 2$ GeV, 
inclusion of thermal photons changes the results only marginally. 
Right panel: Intensity correlation of photons at 1 GeV, considering 
only PCM(BMS), only thermal (hyd), and all PCM+ thermal photons 
(BMS+hyd) (see Ref.~\cite{prl_inter}).}
\label{HBT}      
\end{figure}

\section{Intensity interferometry of thermal photons}

The quantum statistical interference between identical  particles 
emitted in relativistic heavy ion collisions, provide valuable 
insight about the shape and size of the particle emitting source. 
We know that direct photons emitted from different stages of the 
collision dominate the $p_T$ spectra depending on the range of 
transverse momentum range. Thus, one can extract space time 
dimension of the system at different stages of the collision by 
measuring the correlation radii for photons at different transverse momentum. 

Two particle correlation function $C(\vec{q},\vec{K})$ for photons 
having momenta $\vec{k_1}$ and $\vec{k_2}$, emitted from a completely 
chaotic source can be written as,
\begin{equation}
C(\vec{q},\vec{K}) =  1 + {\frac {1}{2}} {\frac {| \int d^4 x S(x,
\vec{K}) e^{iq.x}|^2} {\int d^4 x S(x,\vec{k_1})\int d^4 x S(x,\vec{k_2})}}
\end{equation}
In the above equation the factor 1/2 appears for averaging over spin 
and $S(x,\vec{K})$ is the space time density function defined as,
\begin{equation}
E \frac{dN}{d^3K} = \int d^4 x  S(x,\vec{K})
\end{equation}
where, 
\begin{equation}
\vec{q}= \vec{k_1} - \vec{k_2},  \vec{K}= (\vec{k_1} +\vec{k_2})/2
\end{equation}
The correlation function $C(\vec{q},\vec{K})$ can be expressed in 
terms of outward, side-ward and longitudinal momentum difference 
and radii ( $q_{\rm out}$, $q_{\rm side}$, $q_{\rm long}$ and 
$R_{\rm out}$, $R_{\rm side}$, $R_{\rm long}$ respectively )as,

\begin{equation}
C(q_{\rm out},q_{\rm side},q_{\rm long}) \sim 1+ {\frac{1}{2}} 
\exp[-(q_{\rm out}^2 R_{\rm out}^2 +q_{\rm side}^2 R_{\rm side}^2 
+ q_{\rm long}^2 R_{\rm long}^2)/2 ]
\end{equation}
Photon four momentum in terms of transverse momentum $k_T$, rapidity 
$y$ and azimuthal angle $\psi$ can be expressed as
\begin{equation}
k^{\mu}= (k_{T} \cosh y, k_{T} \cos \psi, k_{T} \sin \psi, k_{T} \sinh y) 
\end{equation}
and
\begin{eqnarray}
q_{\rm long}& =&|k_{1z} - k_{2z}|= |k_{1T} \sinh {y_1} - k_{2T} \sinh {y_2}| 
\nonumber\\
q_{\rm out}& =& {\bf{q_T}.k_T}/k_T, \nonumber\\  
q_{\rm side}& =&| {\bf{q_T}} - q_{\rm out}{\bf{k_T}}/k_{T}|\nonumber
\end{eqnarray} 
Photon interferometry in the QGP phase has been investigated by 
several theoretical groups and the first experimental results on 
direct photon was obtained by WA98~\cite{phot_wa98} collaboration. 
Bass, M\"uller, Srivastava~\cite{prl_inter}, have calculated two body 
quantum correlation of high energy photon using Parton Cascade Model 
(PCM) and ideal hydrodynamic model for central $200A$ GeV $Au+Au$
collision at RHIC. They have shown that one can differentiate between 
the direct photons from early per-equilibrium stage and the same from 
later QGP and hadronic gas stages depending on features 
of the correlation function. Left panel of Fig.~\ref{interfero} shows 
that about 88\%  of the total photons having transverse momenta ($ 1 
\pm 0.1$) GeV (produced by hard parton scattering in PCM~\cite{TGF}) 
are emitted 
within a time period of 0.3 fm/$c$ at rapidity $y=0$. Right panel of the 
same figure shows that at $k_T = 1$ GeV, the contribution of 
pre-equilibrium phase and thermal photons are similar to the total 
photon yield, whereas at $k_T =2$ GeV  hard photons from PCM outshine 
the thermal ones by an order of magnitude. The emission time in PCM 
is very small and hydrodynamic calculation with $\tau_0 = 0.3$ fm/$c$ allows 
a smooth continuation of the emission rate. Intensity correlation for 
$k_T \ge 2$ GeV (Fig.~\ref{interfero}) reveals a pre-thermal photon 
dominated small size source of brief duration.  On the contrary, for 
$k_T = 1$ GeV we can see much larger radii for an extended source and 
suppression of pre-thermal contribution over thermal.
\begin{figure}
\centering
\includegraphics[height=4.2cm, width=5.50cm]{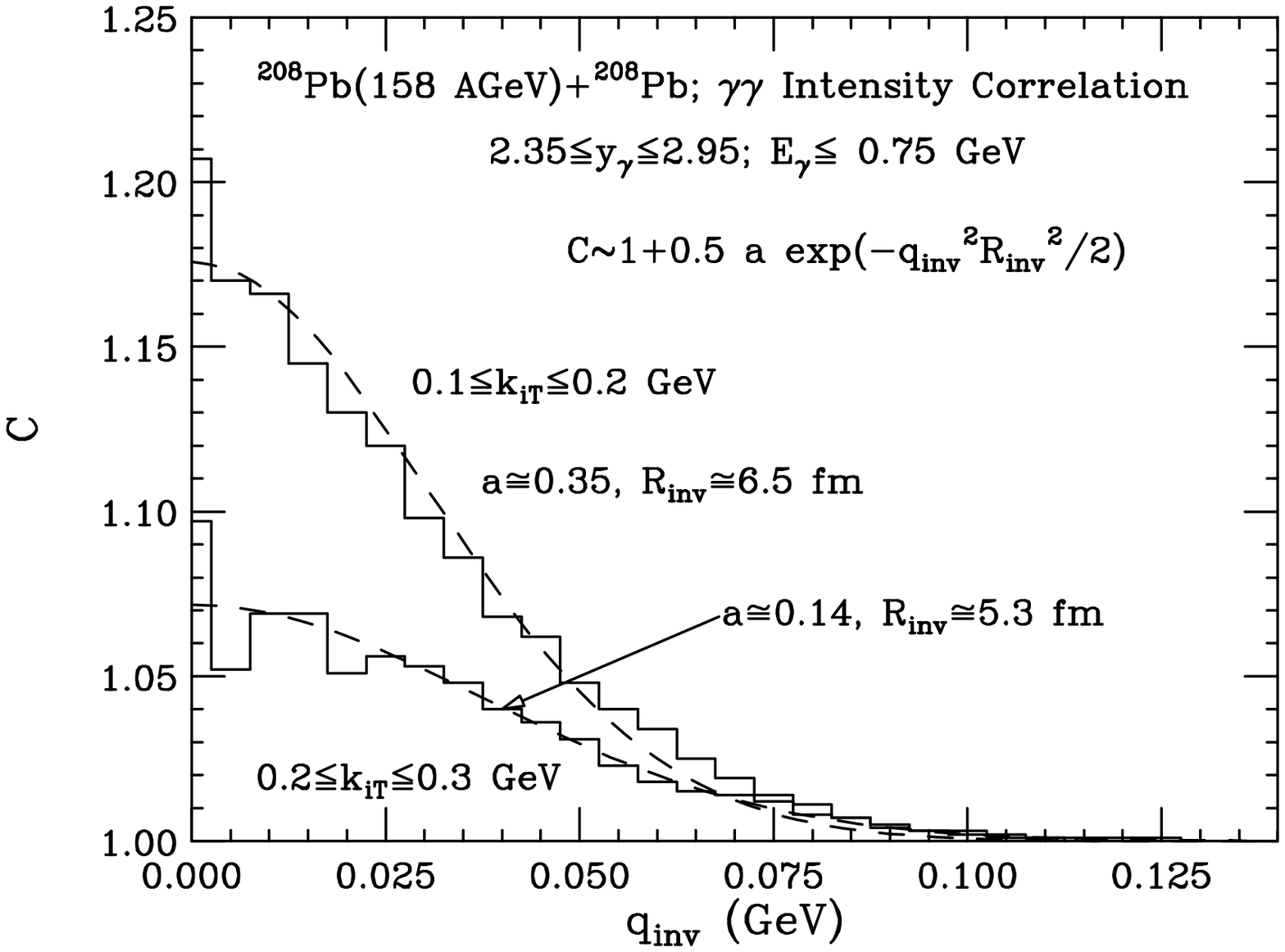}
\includegraphics[height=4.2cm, width=5.50cm]{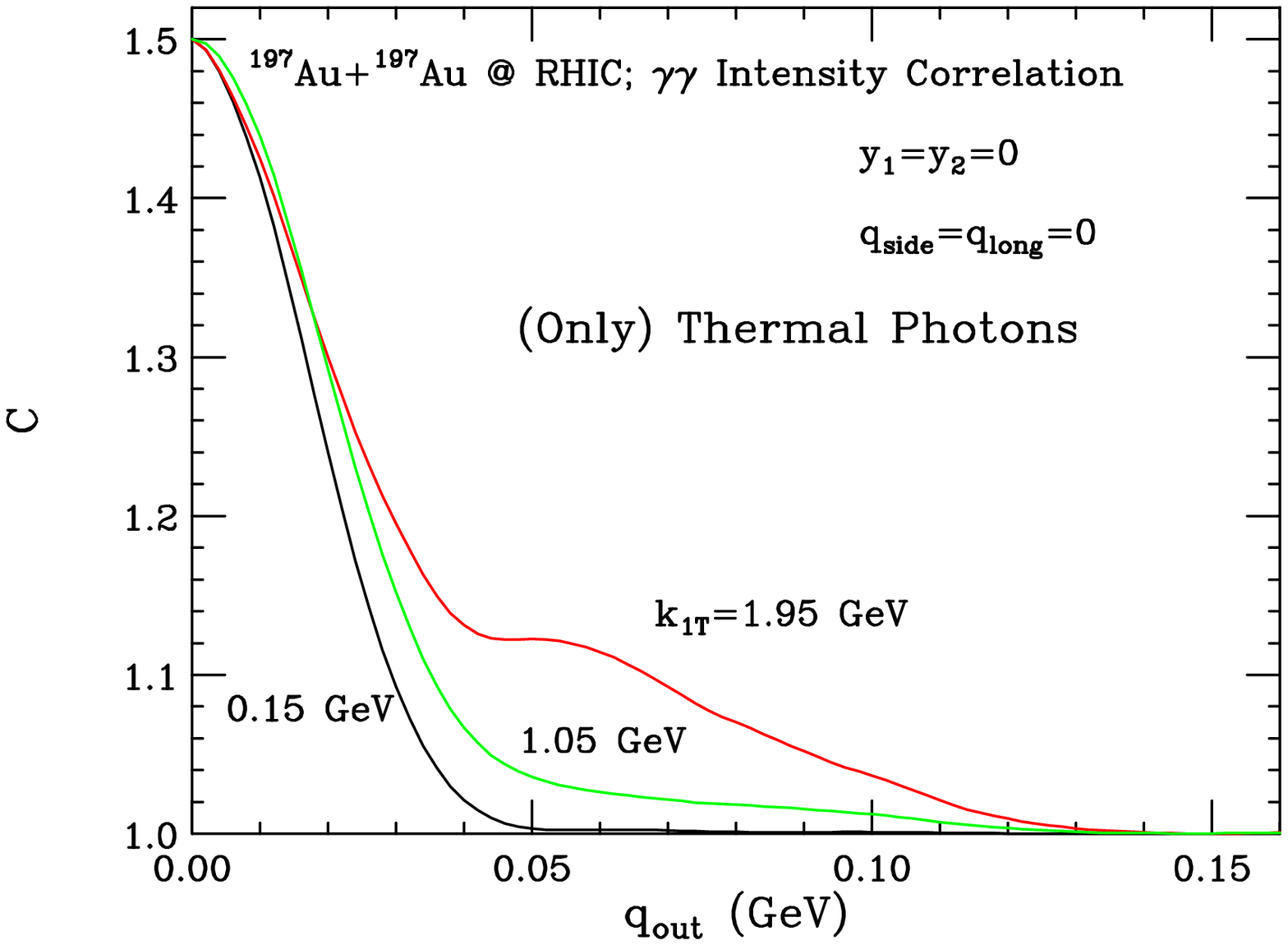}
\includegraphics[height=4.2cm, width=5.50cm]{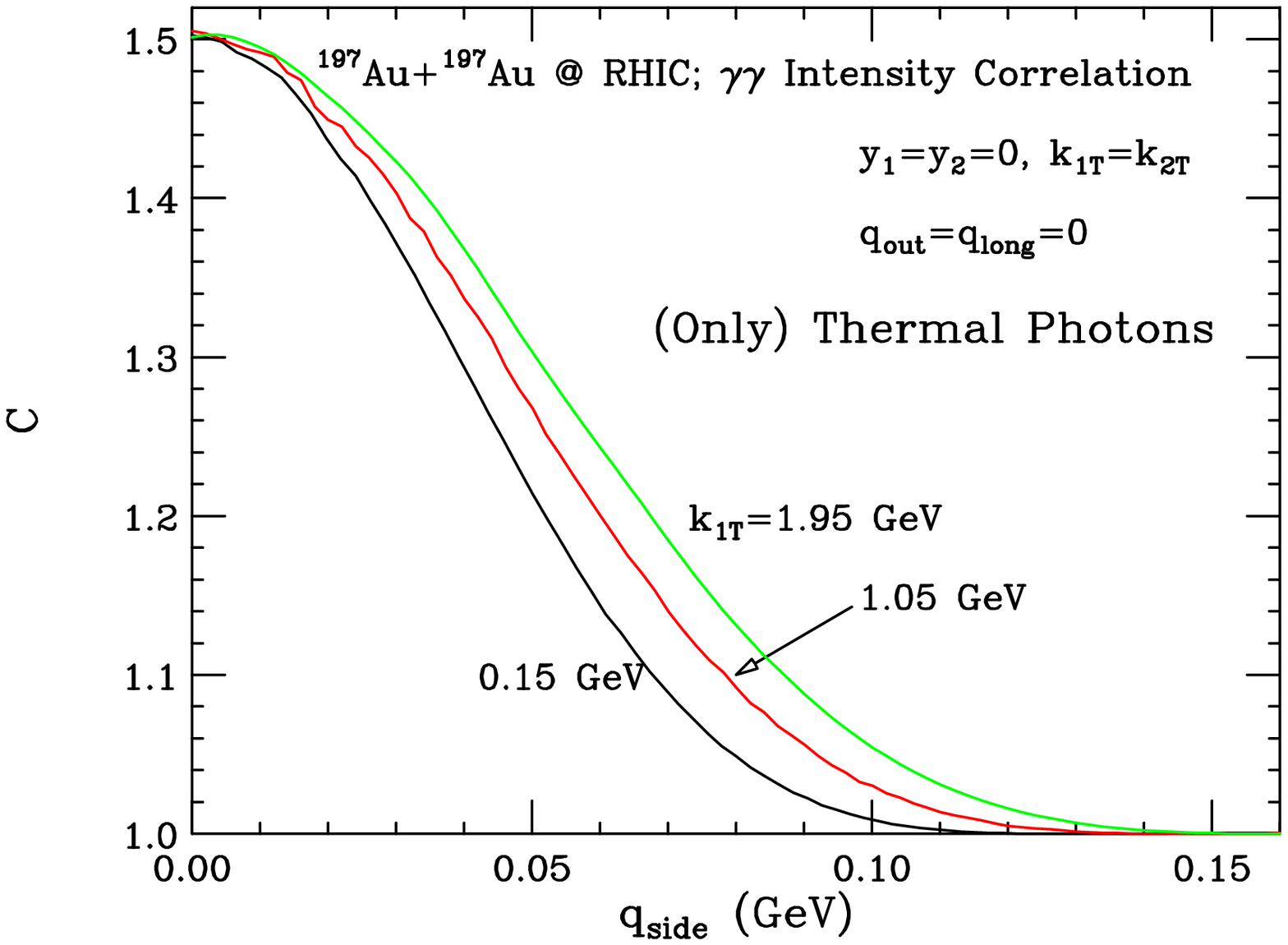}
\includegraphics[height=4.2cm, width=5.50cm]{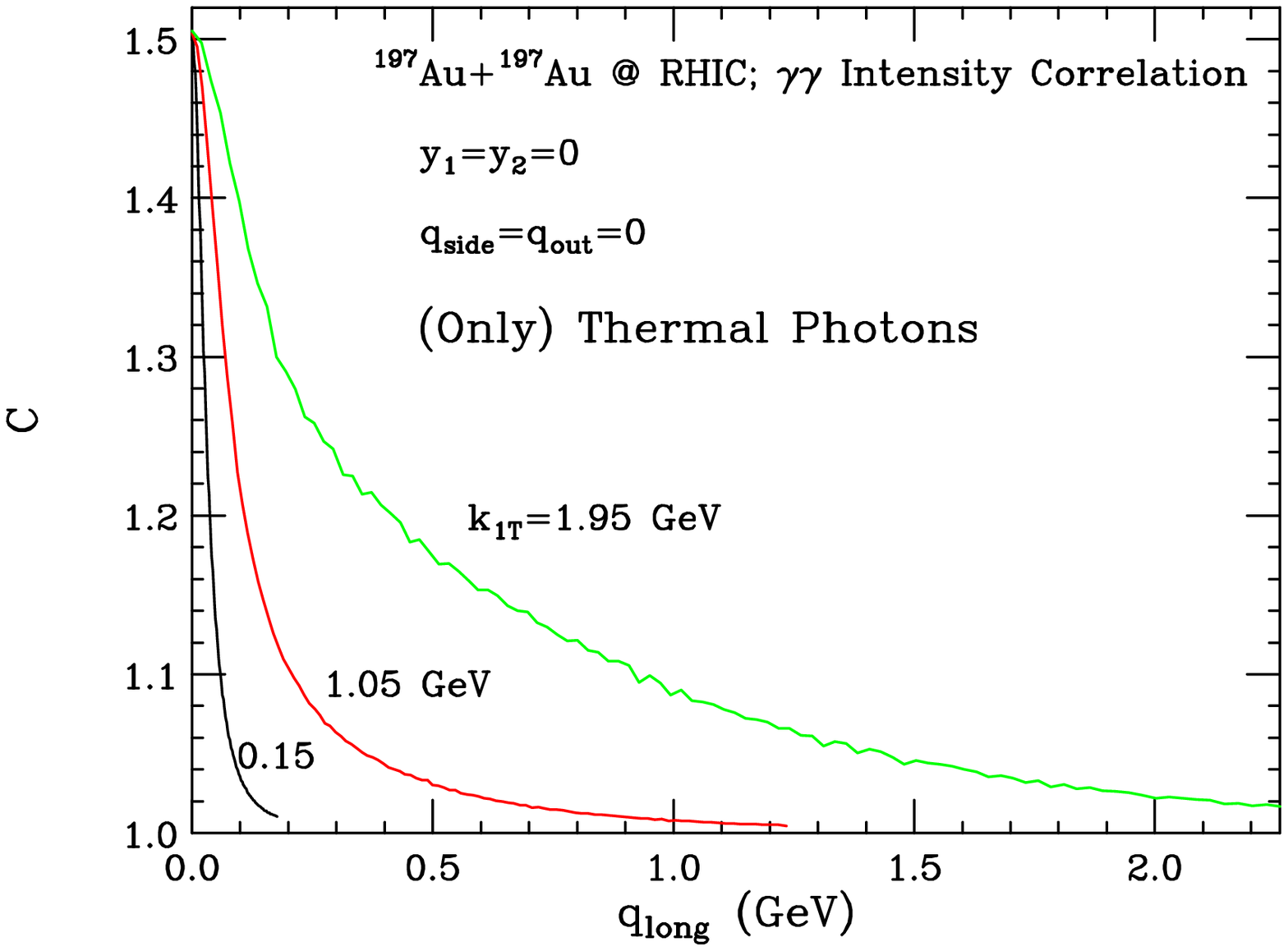}
\caption{One dimensional correlation function for the kinematic 
window used in WA98 experiment, assuming a fully source and emitting 
only single photon. Outward side-ward and longitudinal correlation 
function for thermal photons in central collision at RHIC energy. 
(taken from Ref.~\cite{DKS_WA98}).}
\label{intensity}    
\end{figure}

Intensity interferometry of thermal photons having transverse 
momentum $k_T \sim 0.1 - 2.0$  GeV~\cite{DKS_WA98} provide an 
accurate information about the temporal and spatial structure 
of the interacting medium. 

In reference~\cite{DKS_WA98} WA98 data is compared with theoretical 
results and prediction are given for RHIC and LHC energies. One 
dimensional correlation function in terms of $q_{inv}$ (invariant 
momentum difference) for different $k_T$ and rapidity windows along 
with WA98 data is shown in figure~\ref{intensity}. Theoretical 
results are well fitted in the form $C = \ 1 \ + \ 0.5 \ a \exp[-q_{\rm 
inv}^2 R_{\rm inv}^2/2]$ and are in reasonable agreement with the 
experimental data. At RHIC energy, contribution from quark matter 
increases and as a result the two source aspect in the outward 
correlation radii become more clear. Similar results are obtained 
for LHC energies. It is found that the transverse momentum dependence 
of the different radii are  quite different from the corresponding 
results for pions and do not decrease as $1/\sqrt{m_T}$. For SPS, 
RHIC, and LHC energies the longitudinal correlations show similar 
values, which can be explained as a result of boost invariance of 
the flow pattern.  

\section{Epilogue}

We have tried to give a reasonably complete introduction to the 
exciting possibilities provided by radiation of direct photons 
and dileptons for the study of the dynamics of relativistic heavy 
ion collisions.

While low mass dileptons provide insights into the medium modification 
of vector mesons, those having intermediate masses carry signatures 
of thermal radiation from the quark gluon plasma. We have, due to 
lack of space, left out the discussion of correlated decay of charm 
and bottom mesons which give a large contribution to dileptons. These 
are also important as they carry valuable information about the extent 
of thermalization, elliptic flow, and energy-loss exhibited by heavy 
quarks.

We have discussed sources of direct photons, consisting of prompt photons, 
thermal 
radiation from quark and hadronic matter and those due to passage of 
jets through QGP. We have also seen that photons carry information 
about the initial temperature, evolution of elliptic flow, and size 
of the system. High statistics data at RHIC and LHC along with photon 
or dilepton tagged jets will go a long way in seeing that we realize 
the full potential of electromagnetic probes of quark gluon plasma.

\section{Acknowledgment}
This write up is based on the notes taken by RC and LB from the lectures 
given by  DKS at the QGP Winter School 2008 at Jaipur. We thank the 
organizers for their warm hospitality. The work discussed here has benefited 
from discussions and collaborations with colleagues over many 
years and from many countries, and we take this opportunity to thank all of 
them and also to apologize to those whose work may not have found mention 
in these introductory lectures as this write up is not intended to be a 
review.

Finally we thank the organizers for their patience and for giving us 
extra time to complete this write up. 

%
%

%


\printindex
\end{document}